\documentclass[ip,amsmath,amsfonts]{jpsj3}
\usepackage{txfonts}
\usepackage{bm}
\usepackage{braket}
\usepackage{widetext}
\usepackage{amsmath}
\usepackage{amsfonts}
\usepackage{graphicx}

\title{Majorana Fermions and Topology in Superconductors}

\author{Masatoshi Sato$^1$ and Satoshi Fujimoto$^2$}
\inst{$^1$Yukawa Institute for Theoretical Physics, Kyoto University, Kyoto
606-8502, Japan \\ 
$^2$Department of Materials Engineering Science, 
Osaka University, Toyonaka, Osaka 560-8531, Japan
} %\\

\abst{Topological superconductors are novel classes of quantum condensed phases, characterized
by topologically nontrivial structures of Cooper pairing states. 
On the surfaces of samples and in vortex cores of topological superconductors, Majorana fermions, which are particles
identified with their own anti-particles,
appear as Bogoliubov quasiparticles. The existence and stability of Majorana fermions are ensured by bulk topological invariants constrained by the symmetries of the systems. Majorana fermions in topological superconductors obey a new type of quantum statistics referred to as non-Abelian statistics, which is distinct from bose and fermi statistics, and can be 
 utilized for application to topological quantum computation. 
Also, Majorana fermions give rise to various exotic phenomena such as ``fractionalization", non-local correlation, and ``teleportation".   A pedagogical review of these subjects is presented.
We also discuss interaction effects on topological classification of superconductors, and
the basic properties of Weyl superconductors.
 }

\kword{topological superconductor, unconventional superconductivity, Majorana fermion, non-Abelian statistics, Andreev bound state, quantum computation, Weyl superconductor}

\begin{document}
\maketitle

\section{Introduction}

Topology in condensed matter physics has a long history.
It plays an important role in the classification of topological defects in condensed matter systems, such as vortices, dislocations, and disclinations, i.e.
non-trivial textures in real space configurations.
In 1982, a milestone was achieved by Thouless, Kohmoto, den Nijs and
Nightingale, who found an intimate relation between topological invariants and the Hall conductivity  
in the quantum Hall effect.\cite{thouless, kohmoto}
This was the first example of topological nontriviality realized in bulk quantum solid state systems, which is the origin of the notion of topological phases.
In contrast to topological textures in real space mentioned above, topological phases
are characterized by nontriviality in the Hilbert space of quantum states.
In the last decade, remarkable advances have been achieved in this direction.
%in the field of topological phases in condensed matter physics. 
After pioneering works by Haldane\cite{haldane} and Volovik\cite{volovik} for the quantum Hall effect state and the superfluid $^3$He, respectively, Kane-Mele's celebrated papers initiated the exploration of topological phases
in band insulators.\cite{kane-mele1,kane-mele2}
Since then, 
the notion of topological phases has been extended to various other systems,
including superconductors, magnets, and correlated electron
systems.\cite{hasan-kane,zhang, tanaka-review}
%In the topological phases, there are topologically nontrivial structures in their many-body Hilbert space.
For topological superconductors, a nontrivial structure arises from the phase winding of superconducting order parameters
in momentum space. This can be regarded as a natural extension of a vortex of the superconducting order %, i.e. a well-established topological defect, 
to momentum space.
One of the most important consequence of such topologically nontrivial structures in superconductors is the existence of Majorana fermions, which are 
zero-energy Bogoliubov quasiparticles.\cite{read-green,beenakker-review,alicea-review,review-flensberg}
% composed of the equal-weight superposition of an electron and a hole.
Because of particle-hole symmetry of superconducting states, a zero-energy single-particle state must be the equal-weight superposition of 
an electron and a hole. This implies that the Hermitian conjugate of this state is the same as itself; i.e. a particle is identical to an anti-particle,
which is a signature of a Majorana fermion.
In topological superconductors, the Majorana zero-energy state is realized as an Andreev bound state at the surfaces of samples
and in vortex cores. An important point here is that the Majorana zero-energy state is protected by the bulk topological non-triviality 
of the Hilbert space (the momentum space), and is not affected by extrinsic factors such as conditions of surfaces, impurities, and crystal imperfection.
This is in contrast to Andreev zero-energy bound state realized at the surface of $d$-wave superconductors, which is sensitive to the direction of the surface.

%In this paper, we present a pedagogical review on recent remarkable development of this field.
%, and discuss various exotic phenomena associated with Majorana fermions in topologically non-trivial superconductors.
Majorana fermions in topological superconductors give rise to various exotic phenomena.
In this review, we discuss several representative phenomena: (i)
non-Abelian statistics, (ii) ``fractionalization'' and the $4\pi$-periodic Josephson effect, (iii) nonlocal correlation, and (iv) thermal responses. 
Non-Abelian statistics is a novel quantum statistics distinct from Fermi and Bose statistics.\cite{moore-read,nayak-wilczek,ivanov,stone-chung}
Thus, Majorana fermions in this context are not usual fermions.
Also, they are different from Abelian anyons in the fractional quantum Hall effect, which are characterized by
fractional charge.
The most important feature of non-Abelian statistics is that the exchange operations of particles
are not commutative, and the state depends on the order of the exchange operations.
This peculiar property stems from the topological degeneracy associated with Majorana fermions. 
It has been proposed that particles obeying non-Abelian statistics can be utilized for the realization of fault-tolerant quantum
computation, which is expected to be dramatically robust against decoherence from the environment. This scheme is called topological quantum computation.\cite{TQC-rev} 
Because of the possible technological application in the future as well as the importance of
its fundamental concept,
the non-Abelian statistics is one of the most intriguing features of Majorana fermions. However, its experimental realization and detection are still the most important open issue. 

Majorana fermions in topological superconductors behave as if they emerge from the splitting of electrons into two parts.\cite{semenoff}
In fact, in the second quantized language, an electron field is complex, while a Majorana field is real, and two Majorana
real fields can be combined into a complex fermion field for an electron (or a hole).
Thus, the emergence of Majorana fermions in topological superconductors can be regarded as the "fractionalization" of electrons.
Actually, the fractionalized character of Majorana fermions appears as the $4\pi$-periodic Josephson effect.\cite{kitaev-majorana-chain,yakovenko}
In this effect, the Josephson tunnel current mediated via two Majorana fermions
carries charge $e$, not $2e$: i.e., a Cooper pair with charge $2e$ splits into two Cooper pairs of Majorana fermions
with charge $e$.
The fractionalized character also results in the non-local correlation of two Majorana fermions.\cite{bolech,tewari,nilsson,fu-majorana}
 Two spatially well separated Majorana fermions exhibit a certain type of long-range correlation which is independent of the distance between them, and a phenomenon similar
to teleportation can occur.
These distinct features of Majorana fermions, fractionalization and non-local correlation, have not yet been experimentally established.

As mentioned before, a Majorana fermion in topological superconductors is realized as the equal-weight 
superposition of an electron and a hole. Thus, its topological character does not appear
in charge transport. However, since energy is conserved, heat transport phenomena can be
used for the characterization of topological responses. In fact, the quantum thermal Hall effect can occur
in a topological superconductor with broken time reversal symmetry. The heat current is carried by Majorana surface states.
At sufficiently low temperatures, the $T$-linear coefficient of the thermal Hall conductivity is quantized, reflecting
the total number of chiral Majorana modes carrying the heat current.

The organization of this paper is as follows. 
In Sect. \ref{sec:top}, we introduce the basic concepts of topological superconductors, and 
in Sect. \ref{sec:symmetry} we discuss the classification of topological phases of superconductors based on
the symmetry of systems, and present topological invariants characterizing distinct topological phases.
We also discuss the classification of topological defects such as a vortex which plays an important role in superconductors.
In Sect. \ref{sec:realization}, we explain the fundamental properties of topological superconductors, 
particularly focusing on the case with broken time-reversal symmetry, for which typical features
such as Majorana surface states emerge. We also present various scenarios for the realization of topological superconductors in materials.
In Sect. \ref{sec:materials}, we overview candidate materials of topological superconductors, mentioning
the current state of experimental researches. From Sect. \ref{sec:non-abelian} to Sect. \ref{sec:thermal}, we 
discuss exotic phenomena caused by Majorana fermions.
In Sect. \ref{sec:non-abelian}, we review the non-Abelian statistics of Majorana fermions in topological superconductors.
%As mentioned above, the non-Abelian statistics is one of the most dramatic features of Majorana fermions, though its experimental realization and detection are still the most important open issue. 
We introduce the basic ideas of the non-Abelian statistics of Majorana fermions, and discuss 
possible experimental detection schemes for this intriguing phenomenon.
In Sect. \ref{sec:4pi-josephson}, we consider  the $4\pi$-periodic Josephson effect, and 
in Sect. \ref{sec:non-local}, we overview the non-local correlation effects of Majorana fermions such as "teleportation".
%The non-local correlation may be understood as a result of splitting an electron into two Majorana fermions.
%In fact, the fractionalized character of Majorana fermions also appears as the the $4\pi$-Josephson effect: i.e.
%the Josephson current mediated via Majorana fermions carries charge $e$, not $2e$.
In Sect. \ref{sec:thermal}, the thermal responses of Majorana fermions are discussed.
%In section \ref{sec:nanowire}, we briefly mention about recent experimental exploration of Majorana zero energy states
%for nanowire systems, and related theoretical works evaluating the experimental results critically.
In Sect. \ref{sec:TQC}, we present an elementary introduction to topological quantum computation utilizing Majorana fermions.
In Sect. \ref{sec:interaction}, the recent development of our understanding of interaction effects in topological superconductors is reviewed. 
In Sect. \ref{sec:weyl},  we overview the basic properties of Weyl superconductivity, which is another topological phase of superconductors,
characterized by the existence of Weyl fermions as bulk gapless quasiparticles. The chiral anomaly associated with Weyl fermions gives rise to various exotic phenomena.

\section{Topological Superconductors}
\label{sec:top}

\subsection{Andreev bound states}
\label{sec:topABS}

%\begin{widetext}
\begin{table*}
\begin{center}
\caption{Various Andreev bound states and corresponding topological
 numbers.} 
\begin{tabular}{c||c|c|c|c}
\hline\hline
Energy dispersion & Chiral& Helical& Conical &Flat\\ 
 &($E=c k_y$) &($E=\pm c k_y$) & ($E=\pm c\sqrt{k_x^2+k_y^2}$)& ($E=0$)\\ \hline
Topological number & TKNN number & Kane-Mele's $Z_2$ number & 3D winding number & 1D
		 winding number \\ \hline
Related materials & Sr$_2$RuO$_4$ & Noncentrosymmetric SCs & $^3$He-B &
High-$T_{\rm c}$ cuprates \\ \hline
\end{tabular} 
\label{table:1}
\end{center}
\end{table*}
%\end{widetext}
%In particular, the bulk-boundary correspondence%
%which ensures the existence of gapless surface states by bulk 
%topological numbers.

\begin{figure}
\includegraphics[width=8.5cm]{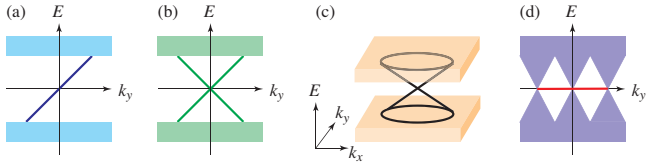}
\caption{(Color online) Edge and surface states in topological superconductors. (a)
Chiral Majorana edge mode. (b) Helical Majorana edge mode. (c) Helical
surface Majorana fermion. (d) Flat edge mode.}
\label{fig:edge_mode}
\end{figure}

Through the study of unconventional superconductivity, it has been recognized
that some unconventional superconductors may support surface
bound states called Andreev bound state.\cite{kashiwaya-review}
For instance, chiral $p$-wave superconductors, which are considered to be realized
in Sr$_2$RuO$_4$, host Andreev bound states with a linear dispersion, as
illustrated in Fig. \ref{fig:edge_mode}(a).
%Also, high-$T_{\rm c}$ cuprates have surface Andreev bound states with flat
%dispersion. 
Before the recent study of topological superconductors, however, such Andreev bound
states had been understood, on a case-by-case basis, as a result of
interference between quasiparticles due to the Andreev reflection. 
Recent development of topological superconductors provides a unified
topological viewpoint for these Andreev bound states, which  
has been obtained by using  
analogy between the Andreev bound states and edge states in quantum
Hall effects.\cite{halperin1}
Indeed, it has been known that quantum Hall states also have gapless edge
states similar to the Andreev bound states in chiral $p$-wave
superconductors.\cite{volovik,read-green}

For quantum Hall states, the existence of edge states has been
understood as a consequence of the intrinsic topology of the
system.\cite{hatsugai} 
First, for a bulk quantum Hall state without a boundary, the Hall
conductance $\sigma_{xy}$ is given  by a topological number $\nu_{\rm
TKNN}$ called
the Thouless-Kohmoto-den Nijs- Nightingale (TKNN) number (or
the first Chern number in mathematics),\cite{thouless, kohmoto} 
\begin{eqnarray}
\sigma_{xy}=\frac{e^2}{h}\nu_{\rm TKNN}, 
\end{eqnarray}
where $e$ is the unit charge of an electron and $h$ is the Planck constant.
On the other hand, if the quantum Hall state has a boundary, the Hall
conductance is given by
\begin{eqnarray}
\sigma_{xy}=\frac{e^2}{h}N, 
\end{eqnarray}
with $N$ the number of edge states on the boundary.\cite{halperin1}
Because these two different equations express the same quantum
Hall effect,  the bulk topological number $\nu_{\rm TKNN}$ should be the same as $N$,
\begin{eqnarray}
\nu_{\rm TKNN}=N. 
\end{eqnarray}
Thus, we can understand the existence of edge states, i.e., a nonzero value
of $N$,  as being a result of a nonzero value of $\nu_{\rm TKNN}$.
%relate the topological number $C$ to the number of edges
%state $N$. 
Generally, such a relation between bulk topological numbers and gapless
boundary states is called ``bulk-boundary correspondence''.

Remembering the similarity between Andreev bound states and edge states
in quantum Hall states,
one can naturally expect a similar correspondence for the Andreev bound
states. 
Actually, even for chiral $p$-wave superconductors, we can define the
TKNN number for the bulk systems without boundaries, which yields 
$|\nu_{\rm TKNN}|=1$
(or $|\nu_{\rm TKNN}|=2$ if the spin degrees of freedom are taken into account).
Similarly to quantum Hall states, the existence of the edge state in
chiral $p$-wave superconductors can be explained by the intrinsic
topology of the bulk systems.\cite{volovik,read-green}  

Depending on the symmetry of the system, Andreev bound states may
have different energy dispersions from that in
Fig. \ref{fig:edge_mode}(a).   
From the viewpoint of the bulk-boundary correspondence, these
differences result from differences in the corresponding
topological numbers.
For instance, Andreev bound states with a flat dispersion exist on
the $(110)$ surface of high-$T_{\rm c}$ cuprates
(Fig. \ref{fig:edge_mode}(d)).\cite{hu,tanaka-kashiwaya,ryu-hatsugai}
In this case, the corresponding topological number is not the TKNN
number but the one-dimensional (1D) winding number.\cite{sato-tanaka}
Furthermore, on surfaces of the superfluid $^3$He-B, 
Andreev bound states appear with conical dispersion
(Fig. \ref{fig:edge_mode}(c)),\cite{hara-nagai} whose corresponding topological number is the 
three-dimensional (3D) winding number.\cite{grinevich, schnyder, sato2009} 
We summarize the relation between various Andreev bound states and the
corresponding topological numbers in Table \ref{table:1}.

Superconductors with nonzero bulk topological numbers are called
topological superconductors.
The existence of gapless Andreev bound states on their surfaces is one of the pieces of 
direct evidence for topological superconductivity.

\subsection{Majorana fermions}
\label{subsec:Majorana}

As mentioned above, there is a similarity between Andreev bound states
and edge states in quantum Hall states. 
What is the difference between them?

\begin{figure}[b]
\begin{center}
\includegraphics[width=2cm]{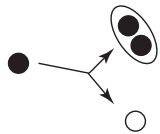}
\end{center}
\caption{An electron (black circle) becomes a hole (white circle) by
the formation of a Cooper pair.}
\label{fig:Cooper}
\end{figure}

The most important difference is that Andreev bound states
do not carry definite charges, whereas edge states in quantum Hall states
do. 
In superconducting states, an electron can become a hole by the
formation of a Cooper pair
( see Fig.\ref{fig:Cooper} ).
Therefore, fermionic excitations in the superconducting state are
naturally expressed as a superposition of an electron $c_\sigma({\bm
x})$ and a hole $c^{\dagger}_\sigma({\bm x})$,
\begin{eqnarray}
\Psi({\bm x})=
\left(
\begin{array}{c}
c_{\uparrow}({\bm x}) \\
c_{\downarrow}({\bm x})\\
c_{\uparrow}^{\dagger}({\bm x})\\
c_{\downarrow}^{\dagger}({\bm x})
\end{array}
\right). 
\label{eq:nambu}
\end{eqnarray} 
Since an electron and a hole have an opposite charge, 
the quasiparticle cannot have a
definite charge.

The wave function in Eq. (\ref{eq:nambu}) satisfies
\begin{eqnarray}
\Psi^*({\bm x})=\tau_x \Psi({\bm x}), 
\label{eq:Majorana_condition}
\end{eqnarray}
where $\tau_x$ is the Pauli matrix in the Nambu space.
This equation implies that the quasiparticle $\Psi({\bm x})$ is essentially
the same as its antiparticle $\Psi^{*}({\bm x})$. Such a self-conjugate
property is not seen in ordinary fermions. 
This self-conjugate condition is called the Majorana condition because
a class of fermions named Majorana fermions satisfies this condition.

Majorana fermions are Dirac fermions satisfying the self-conjugate
property. Originally, they were introduced as an elementary particle.\cite{majorana}
As discussed in the above, all quasiparticles in
superconductors, i.e., 
even those in conventional $s$-wave superconductors, satisfy the Majorana
condition.
%%% amended by Fujimoto on 2016 April 15.
%, but they are not Majorana fermions in general since most of
%them are not Dirac fermions.    
Furthermore, surface Andreev bound states have linear dispersions, and
thus their effective Hamiltonian is given by the massless Dirac
Hamiltonian. 
Hence, Andreev bound states can naturally be considered as Majorana
fermions in condensed matter physics.\cite{read-green, stone-roy}  
%%%%%%%%%%%%%%%%%%%%%%%%

\section{Symmetry and Topology}
\label{sec:symmetry}

In this section, we discuss the classification of topological superconductors based on the symmetry of systems.\cite{schnyder,kitaev-class,ryu-furusaki,altland}
As will be seen below, particle-hole symmetry and time-reversal symmetry play crucial roles.
We also present topological invariants which characterize distinct topological phases.
In Sects. \ref{subsec:PHS} and \ref{subsec:TRS}, we discuss generic spinful cases, 
and consider the case with spin-rotation symmetry in Sect. \ref{subsec:spin-rot} and 
the spinless (or fully spin-polarized) case in Sect. \ref{subsec:spinless}.
The case with additional crystal symmetry such as mirror reflection symmetry
is discussed in Sects. \ref{subsec:TCS} and \ref{subsec:Ising}.

\begin{table*}[t]
\begin{center}
\caption{Symmetry of superconductors (SCs) and the relevant topological numbers. The third
 to fifth columns indicate the absence (0) or presence ($\pm 1$) of
 time-reversal symmetry (TRS), particle-hole symmetry (PHS), and chiral symmetry (CS), respectively, where $\pm$ denotes the sign of ${\cal
 T}^2$ and ${\cal C}^2$. The topological invariants, ${\bm
 Z}_2^{(\gamma_{\rm geom})}$, ${\bm Z}^{\rm (TKNN)}$, ${\bm
 Z}_2^{(\gamma_{{\rm geom}/2})}$, 
${\bm Z}^{\rm (3dW)}$, and ${\bm Z}^{\rm (1dW)}$ are given by
 Eqs. (\ref{eq:Z2_PHS}), (\ref{eq:TKNN}), (\ref{eq:Z2_DIII}), (\ref{eq:3dW}), and (\ref{eq:1dW}), respectively.
${\bm Z}_2^{\rm (KM)}$ is the Kane-Mele's ${\bm Z}_2$ number.}
\begin{tabular}{c c c c c c c c}
\hline\hline
&AZ class & TRS & PHS & CS  & $d=1$ & $d=2$ & $d=3$\\ 
\hline
Spinful or Spinless SC  & D  & 0 & +1 & 0 & ${\bm Z}_2^{(\gamma_{\rm geom})}$ & ${\bm Z}^{\rm (TKNN)}$ & 0\\ 
Spinful SC with TRS & DIII & -1 & +1 & 1 & ${\bm Z}_2^{(\gamma_{{\rm geom}/2})}$ & ${\bm
			 Z}^{\rm (KM)}_2$ & ${\bm Z}^{\rm (3dW)}$\\
Spinful SC with SU(2)-SRS  & C & 0 & -1 & 0 & 0 & $2 {\bm Z}^{\rm (TKNN)}$ & 0\\
Spinful SC with SU(2)-SRS+TRS &CI & +1 & -1 & 1 & 0 & 0 & $2{\bm Z}^{\rm (3dW)}$\\

Spinless SC with TRS &BDI & +1 & +1 & 1 & ${\bm Z}^{\rm (1dW)}$ & 0 & 0\\
\hline
\end{tabular} 
\label{table:Periodic}
\end{center}
\end{table*}

\subsection{BdG Hamiltonian}
Generally, the Hamiltonian of electrons in superconductors is given by  
\begin{eqnarray}
{\cal H}=\sum_{\alpha\beta{\bm k}}
{\cal E}_{\alpha\beta}({\bm k})c^{\dagger}_{{\bm k}\alpha}c_{{\bm k}\beta}
+\frac{1}{2}\sum_{\alpha\beta{\bm k}}\left(
\Delta_{\alpha\beta}({\bm k})
c^{\dagger}_{{\bm k}\alpha}c^{\dagger}_{{-\bm k}\beta}
+{\rm h.c.}\right),
\end{eqnarray}
where $c_{{\bm k}\alpha}$ ($c^{\dagger}_{{\bm k}\alpha}$)
is the annihilation (creation) operator of an electron with momentum ${\bm
k}$, and the suffix $\alpha$ labels internal degrees of freedom for the fermion
such as spin, orbit, and so forth.
The first term on the right-hand side is the Hamiltonian of electrons in the
normal state, and the second term appears in the superconducting state because of the formation of Cooper
pairs with a gap function $\Delta_{\alpha\beta}({\bm k})$.
Here the anticommutation relation of $c_{{\bm k}\alpha}^{\dagger}$
yields $\Delta({\bm k})=-\Delta^t(-{\bm k})$. 
The above Hamiltonian is called the Bogoliubov-de Gennes (BdG)
Hamiltonian.

The BdG Hamiltonian can be written in the following matrix form:
\begin{eqnarray}
{\cal H}=\frac{1}{2}\sum_{{\bm k}}\Psi^{\dagger}({\bm k}){\cal H}({\bm
 k})\Psi({\bm k}), 
\label{eq:BdG_matrix}
\end{eqnarray}
with 
\begin{eqnarray}
{\cal H}({\bm k})=\left(
\begin{array}{cc}
{\cal E}({\bm k})& \Delta({\bm k})\\
\Delta^{\dagger}({\bm k})& -{\cal E}^t(-{\bm k})
\end{array}
\right)_{\alpha\beta}
,\quad
\Psi({\bm k})=\left(
\begin{array}{c}
c_{{\bm k}\alpha}\\
c^{\dagger}_{-{\bm k}\alpha}
\end{array}
\right), 
\label{eq:BdG_matrix3}
\end{eqnarray}
where the summation over the index $\alpha$ is implicit in Eq. (\ref{eq:BdG_matrix}).   
We have neglected the constant term caused by the anticommutation relation
between $c_{{\bm k}\alpha}$ and $c_{{\bm k}\alpha}^{\dagger}$ since it
merely shifts the origin of the energy.
Performing the Fourier transformation, 
\begin{eqnarray}
\Psi({\bm k})=\frac{1}{\sqrt{V}}
\sum_{\bm x}e^{i{\bm k}{\bm x}}\Psi({\bm x}),
\quad 
{\cal H}({\bm k})=%\frac{1}{\sqrt{V}}
\sum_{\bm x}e^{i{\bm k}{\bm x}}{\cal H}({\bm x}),
\end{eqnarray}
we also have the BdG Hamiltonian
in the coordinate space,
\begin{eqnarray}
{\cal H}=\frac{1}{2}\sum_{{\bm x}{\bm y}}\Psi^{\dagger}({\bm x}){\cal
 H}({\bm x}-{\bm y})\Psi({\bm y}) 
\label{eq:BdG_matrix2}
\end{eqnarray} 
This form indicates that  fermionic excitations of
a superconductor are given as a superposition of an electron and hole,
$\Psi({\bm x})$ in Eq.(\ref{eq:nambu}),
as mentioned in Sect. \ref{subsec:Majorana}.
The quasi-particle spectrum in the superconducting state is obtained by solving
the eigen equation of the BdG Hamiltonian
\begin{eqnarray}
{\cal H}({\bm k})|u_n({\bm k})\rangle=E_n({\bm k})|u_n({\bm k})\rangle. 
\label{eq:BdG_equation}
\end{eqnarray}

%When the system supports topological defects, ${\cal H}({\bm x}-{\bm
%y})$ in Eq.(\ref{}) is replaced by ${\cal H}({\bm x}-{\bm y},{\bm R})$
%with ${\bm R}=({\bm x}+{\bm y})/2$.

\subsection{Particle-hole symmetry (charge-conjugation symmetry)}
\label{subsec:PHS}

As discussed in Sect.\ref{subsec:Majorana}, the wave function $\Psi({\bm
x})$ satisfies the self-conjugate condition in 
Eq.(\ref{eq:Majorana_condition}). 
Correspondingly, the BdG Hamiltonian in Eq.(\ref{eq:BdG_matrix2}) should
be invariant 
under the replacement of $\Psi^{\dagger}({\bm x})$ and $\Psi({\bm
k})$ with their conjugates $\Psi^t({\bm x})\tau^t_x$ and $\tau_x
\Psi^*({\bm x})$, respectively.
Because ${\cal H}$ transforms as
\begin{eqnarray}
&&{\cal H}\rightarrow
\frac{1}{2}\sum_{{\bm x}, {\bm y}}\Psi^{t}({\bm
 x})\tau_x^t {\cal H}({\bm x}-{\bm y})\tau_x\Psi^{*}({\bm y})
\nonumber\\
&&= -\frac{1}{2}\sum_{{\bm x},{\bm y}}\Psi^{\dagger}({\bm x})\left[
\tau^t_x 
{\cal H}({\bm y}-{\bm x})\tau_x
\right]^t
\Psi({\bm y})
%\nonumber\\
%&&
+\frac{1}{2}\sum_{\bm x}{\rm tr}{\cal H}({\bm x})
\nonumber\\
\label{eq:BdG_PH}
\end{eqnarray}
under the replacement, this requirement imposes a constraint on ${\cal
H}({\bm x})$,
\begin{eqnarray}
\tau_x^t {\cal H}({\bm x})\tau_x=-{\cal H}^t(-{\bm x}).
\label{eq:PHS_coordinate}
\end{eqnarray}
Here note that the last term in Eq.(\ref{eq:BdG_PH}) vanishes
if ${\cal H}({\bm x})$ satisfies Eq.(\ref{eq:PHS_coordinate}).
In the momentum space, the constraint is given as
\begin{eqnarray}
\tau_x^t {\cal H}({\bm k})\tau_x=-{\cal H}^t(-{\bm k}).
\label{eq:PHS_momentum}
\end{eqnarray}
Using the hermiticity of ${\cal H}({\bm k})$, Eqs.(\ref{eq:PHS_coordinate}) and (\ref{eq:PHS_momentum}) are also
written as
\begin{eqnarray}
{\cal C}{\cal H}({\bm x}){\cal C}^{-1}=-{\cal H}({\bm x}),  
\quad
{\cal C}{\cal H}({\bm k}){\cal C}^{-1}=-{\cal H}(-{\bm k}),  
\label{eq:PHS}
\end{eqnarray}
where ${\cal C}=\tau_x K$ with the complex conjugate operator $K$. 
The constraint on the BdG Hamiltonian in Eq. (\ref{eq:PHS}) is
called as the particle-hole symmetry. 
We can verify directly that ${\cal H}({\bm k})$ in Eq.(\ref{eq:BdG_matrix3}) actually has the
particle-hole symmetry.
This symmetry is one of the fundamental symmetries in the
Altland-Zirnbauer classification scheme,\cite{altland} and 
systems supporting particle-hole symmetry are classified as class D.

From the particle-hole symmetry, one can show that a positive energy
state is always paired with a negative energy state.
%obtain a positive energy state
%by a negative energy state
%The particle-hole symmetry imposes a constraint on the quasiparticle spectrum.
%that a positive energy quasiparticle
%state is paired with a negative one.
Indeed, from a solution $|u_n({\bm k})\rangle$ of Eq.(\ref{eq:BdG_equation}) with a
positive (negative) energy $E_n({\bm k})>0$ [$E_n({\bm k})<0$], one can obtain a solution with
momentum $-{\bm k}$ as
%$|u_{-n}({\bm k})\rangle$ with a negative energy $E_{-n}({\bm
%k})=-E_n({\bm k})\rangle$ as
\begin{eqnarray}
|u_{-n}(-{\bm k})\rangle ={\cal C}|u_n({\bm k})\rangle,
\end{eqnarray}
which has a negative (positive) energy $-E_n({\bm
k})$. 

In one dimension, the paired structure of the
spectrum enables us to
define a topological number.\cite{Qi-Hughes-Zhang}
The paired structure implies that the gauge field constructed from
negative-energy states, 
$
{\mathcal A}^{(-)}(k)=i\sum_{E_n(k)<0}\langle u_n(k)|\partial_k u_n(k)\rangle, 
$
is not independent of that constructed from positive-energy states,
$
{\mathcal A}^{(+)}(k)=i\sum_{E_n(k)>0}\langle u_n(k)|\partial_k u_n(k)\rangle, 
$
which leads to the relation ${\mathcal A}^{(-)}(k)={\mathcal A}^{(+)}(-k)$.
Therefore, the geometrical phase of $A^{(-)}(k)$ along the 1D
Brillouin zone (BZ),
\begin{eqnarray}
\gamma_{\rm geom}=\oint_{\rm BZ} dk {\mathcal A}^{(-)}(k), 
\label{eq:Z2_PHS}
\end{eqnarray}
is recast into
\begin{eqnarray}
\gamma_{\rm geom}&=&\frac{1}{2}\oint_{\rm BZ} dk \left[
{\mathcal A}^{(+)}(k)+{\mathcal A}^{(-)}(k)
\right]
\nonumber\\
&=&\frac{i}{2}\oint_{\rm BZ} dk 
\sum_{n}\langle u_n(k)|\partial_k u_n(k)\rangle 
\nonumber\\
&=& \frac{i}{2}\oint_{\rm BZ}dk {\rm tr}\left[U^{-1}(k)\partial_k U(k)\right]
\nonumber\\
&=& \frac{i}{2}\oint_{\rm BZ}dk \partial_k\ln\left[{\rm det}U(k)\right], 
\end{eqnarray}
where $U_{mn}(k)$ is given by the $m$-th
component of $|u_n(k)\rangle$. 
The U(1) phase of ${\rm det}U(k)$ contributes to the line
integral, and thus 
the uniqueness of ${\rm det}U(k)$ in the BZ yields $\gamma_{\rm geom}=\pi
N$ with an integer $N$.
Finally, taking into account the $2\pi$ ambiguity of $\gamma_{\rm geom}$, we have two
physically different values of $\gamma_{\rm geom}$, i.e., $\gamma_{\rm geom}=0$ and $\pi$, 
which define the ${\bm Z}_2$ topological invariant for 1D
superconductors.  
The system is topologically trivial (non-trivial) if $\gamma_{\rm
geom}=0$ ($\gamma_{\rm geom}=\pi$).

As mentioned in Sect.\ref{sec:topABS}, in two dimensions, class D superconductors
are topologically characterized by the TKNN integer,
\begin{eqnarray}
\nu_{\rm TKNN}=\frac{1}{2\pi}\int_{\rm BZ}d^2k {\cal F}_{xy}^{(-)}({\bm k}), 
\label{eq:TKNN}
\end{eqnarray} 
where ${\cal F}_{xy}^{(-)}({\bm k})$ is the field strength of the gauge field 
${\cal A}_\mu^{(-)}({\bm k})=i\sum_{E_n({\bm k})<0}\langle u_n({\bm
k})|\partial_{k_{\mu}} u_n({\bm k})\rangle$.
On the other hand, no topologically nontrivial class D superconductor
exists in three dimensions.

%\subsection{1D ${\bm Z}_2$ invariant}
%The particle-hole symmetry 

\subsection{Time-reversal symmetry and chiral symmetry}
\label{subsec:TRS}

Similarly to topological insulators, the presence of time-reversal symmetry also
enriches the topological structure in superconductors. 
%Eigenstates of the BdG Hamiltonians form the Kramers doublet
%$(|u_{2n-1}(k)\rangle)$ 
%Spin $1/2$ fermions with the time-reversal symmetry ${\cal T}$ form 
%Kramers doublet 
The quasiparticle state $|u({\bm k})\rangle$ has a Kramers partner ${\cal
T}|u(-{\bm k})\rangle$ with time-reversal operator ${\cal T}$, and 
we can define various topological numbers, 
using the Kramers degeneracy.
The Kane-Mele's
${\bm Z}_2$ 
invariant for quantum spin
Hall states also characterizes the topology of two-dimensional (2D)
time-reversal invariant
superconductors,\cite{sato-fujimoto1,tanaka-yokoyama-balatsky-nagaosa, nakosai1} but the
coexistence of
particle-hole symmetry gives other topological numbers as well.

In one dimension, we can generalize the ${\bm Z}_2$ topological
invariant introduced in Sect. \ref{subsec:PHS}.\cite{sato2010} 
Although $\gamma_{\rm geom}$ in Eq.(\ref{eq:Z2_PHS}) itself is always trivial 
because each component of the Kramers doublet equally contributes to $\gamma_{\rm geom}$, 
one can avoid it by taking only one state for each Kramers pair to
calculate the geometric phase.
Namely, for the Kramers pair $\left(|u_{2n-1}(k)\rangle,
|u_{2n}(k)\rangle\equiv {\cal T}|u_{2n-1}(-k)\rangle\right)$,
the geometrical phase $\gamma_{{\rm geom}/2}$ of $|u_{2n-1}(k)\rangle$,
\begin{eqnarray}
\gamma_{{\rm geom}/2}=\oint_{\rm BZ} {\mathcal A}_{1/2}^{(-)}(k),
\label{eq:Z2_DIII}
\end{eqnarray}
with
\begin{eqnarray}
{\mathcal A}_{1/2}^{(-)}(k)=i\sum_{E_{2n-1}(k)<0}\langle u_{2n-1}(k)|\partial_k u_{2n-1}(k)\rangle 
\end{eqnarray}
is quantized as $\gamma_{{\rm geom}/2}=0,\pi$ $({\rm mod }~ 2\pi )$.
Therefore, $\gamma_{{\rm geom}/2}$ defines a ${\bm Z}_2$ topological invariant in
one dimension.
That is, when $\gamma_{{\rm geom}/2}=\pi \,({\rm mod}~  2\pi )$, the system is topologically non-trivial.  
Here note that the geometrical phase $\gamma'_{{\rm geom}/2}$ of
the remaining state $|u_{2n}(k)\rangle$ is equal to $\gamma_{{\rm geom}/2}$ up the $2\pi$-ambiguity,
and thus it does not define an independent topological invariant.

Because of the time-reversal symmetry ${\cal T}$ and the
particle-hole symmetry ${\cal C}$, time-reversal BdG Hamiltonians have
so-called chiral symmetry,
\begin{eqnarray}
\Gamma {\cal H}({\bm k}) \Gamma^{-1}=-{\cal H}({\bm k}), 
\end{eqnarray}
where $\Gamma=i{\cal T}{\cal C}$ is a unitary operator with $\Gamma^2=1$.
Using the chiral symmetry $\Gamma$, one can define a topological
invariant as the winding number in
three dimensions,\cite{schnyder} 
\begin{align}
w_{\rm 3D}
=\frac{1}{48\pi^2}\int_{\rm BZ}  d^3k \epsilon^{\mu\nu\lambda}{\rm tr}
\left[\Gamma{\cal H}^{-1}(\partial_{k_{\mu}}{\cal H})
{\cal H}^{-1}(\partial_{k_{\nu}}{\cal H})
{\cal H}^{-1}(\partial_{k_{\lambda}}{\cal H})
\right],
\label{eq:3dW}
\end{align}
where ${\rm BZ}$ is the 3D BZ.
This topological invariant characterizes 3D time-reversal
invariant topological superconductors/superfluids such as the $^3$He B-phase.

Using the chiral symmetry, we can also define the topological invariant\cite{wen-zee}
\begin{eqnarray}
w_{\rm 1D}=\frac{i}{4\pi}\oint_C  dk_{\mu}{\rm tr}
\left[\Gamma{\cal H}^{-1}\partial_{k_{\mu}}{\cal H}
\right],
\label{eq:windingnumber_1D}
\end{eqnarray}
with a closed loop $C$ in the BZ.
Whereas $w_{\rm 1D}$ is found to be zero for 1D
time-reversal invariant superconductors, 
it can be nonzero when $C$ encloses a line (point)
node of superconductors in three (two) dimensions.\cite{sato2006, beri, mizuno,
yada,schnyder-ryu1, sato-tanaka}
For instance, nodes in high-$T_{\rm c}$ cuprates have nonzero $w_{\rm
1D}$. The 1D winding number $w_{\rm 1D}$ also ensures the
existence of flat band Andreev bound state on the surface of the
cuprates.\cite{sato-tanaka}

%In one-dimensions, the particle-hole symmetry 
%If $C$ is invariant under the particle-hole symmetry, 
%so it cannot 
%Thus, in one-dimension, $w_{\rm 1D}$ always vanishes.
%In general, $w_{\rm 1D}$ can be nonzero 
%On the other hand, $w_{\rm 3D}$ does not have such a restriction.
%Three-dimensional time-reversal invaraint superconductors
%are topologically characterized by $w_{\rm 3D}$. 

In the Altland-Zirnbauer scheme, systems having ${\cal T}$ and ${\cal C}$
are classified as class DIII.

%$\Gamma{\cal H}^{-1}({\bm k})\partial_{\mu}{\cal H}({\bm k})$
%In the diagonal basis of $\Gamma$,
%\begin{eqnarray}
%\Gamma=\left(
%\begin{array}{cc}
%{\bm 1}_{N\times N} & 0 \\
%0 & -{\bm 1}_{N\times N}
%\end{array}
%\right), 
%\end{eqnarray}
%the chiral symmetry implies that ${\cal H}({\bm k})$ takes the 
%off-diagnal form,
%\begin{eqnarray}
%{\cal H}({\bm k})=
%\left(
%\begin{array}{cc}
%0 & h({\bm k})\\
%h^{\dagger}({\bm k}) & 0
%\end{array}
%\right).
%\end{eqnarray}
%The one-form $h^{-1}({\bm k})\partial_{\mu}h({\bm k})$

\subsection{Spin-rotation symmetry}
\label{subsec:spin-rot}

Most superconductors host spin-singlet Cooper pairs, and 
their spin-orbit interaction can often be
neglected without changing their qualitative properties.
Such superconductors have, in effect, the SU(2) spin-rotation symmetry.

The SU(2) spin-rotation symmetry implies
\begin{eqnarray}
\left[
\left(
\begin{array}{cc}
e^{i{\bm s}\cdot{\bm \theta}/2} & 0 \\
0 & e^{-i{\bm s}^*\cdot{\bm \theta}/2}
\end{array}
\right),
{\cal H}({\bm k})\right]=0
\end{eqnarray}
with the Pauli matrix ${\bm s}=(s_x,s_y,s_z)$ in the spin space and the
SU(2) rotation angle ${\bm \theta}$.
By taking the derivative with respect to ${\bm \theta}$, 
the above equation reduces to the commutation relations
\begin{eqnarray}
[J_i, {\cal H}({\bm k})]=0,
\quad
J_i=\left(
\begin{array}{cc}
s_i & 0\\
0 & -s_i^*
\end{array}
\right), 
\quad 
(i=1,2,3).
\end{eqnarray}
From $[J_z, {\cal H}({\bm k})]=0$,  
${\cal H}({\bm k})$ is block-diagonal in the diagonal basis
of $J_z$. Each block-diagonal subsector has a definite eigenvalue of $J_z$,
and thus no mixing is allowed between different subsectors.
Furthermore, 
using the other commutation relations, $[J_{x,y}, {\cal H}({\bm k})]=0$,
one also finds that the $J_z=-1$ subsector is essentially the same as the
$J_z=1$ subsector. Therefore, it is enough to consider only the $J_z=1$
sector.
Actually, the BdG Hamiltonian (\ref{eq:BdG_matrix}) is rewritten in
terms of the following reduced Hamiltonian in the $J_z=1$ sector,
\begin{eqnarray}
{\cal H}=\sum_{{\bm k}} \phi^{\dagger}({\bm k})h({\bm k})\phi({\bm k}),
\end{eqnarray}
with
\begin{eqnarray}
h({\bm k})
=\left(
\begin{array}{cc}
\epsilon({\bm k}) & \psi({\bm k})\\
\psi^{\dagger}({\bm k}) & -\epsilon^{t}(-{\bm k})
\end{array}
\right),
\quad 
\phi({\bm k})
=\left(
\begin{array}{c}
c_{{\bm k}\uparrow}\\
c_{-{\bm k}\downarrow}^{\dagger}
\end{array}
\right), 
\end{eqnarray}
where $\epsilon({\bm k})$ and $\psi({\bm k})$ are defined as
\begin{eqnarray}
{\cal E}({\bm k})=\epsilon({\bm k})s_0,
\quad 
\epsilon^{\dagger}({\bm k})=\epsilon({\bm k}),
\nonumber\\
\Delta({\bm k})=i\psi({\bm k})s_y,
\quad
\psi^t({\bm k})=\psi(-{\bm k}).
\end{eqnarray}

We should note here that the reduced Hamiltonian $h({\bm k})$ does not
have the original
particle-hole symmetry: Because of the
anticommutation relation
$\{{\cal C}, J_i\}=0$, the particle-hole symmetry maps a state $|u({\bm
k})\rangle$ with $J_z=1$ to ${\cal C}|u(-{\bm k})\rangle$ with a
different eigenvalue, i.e., $J_z=-1$,
\begin{eqnarray}
J_z {\cal C}|u(-{\bm k})\rangle=-{\cal C}J_z|u(-{\bm
 k})\rangle=-{\cal C}|u(-{\bm k})\rangle.
\end{eqnarray}
Thus, the $J_z=1$ sector cannot retain the original particle-hole symmetry.

Although $h({\bm k})$ cannot retain the original particle-hole
symmetry, it has its own alternative symmetry.
%Since $J_x$ interchanges the $J_z=1$ sector with
%the $J_z=-1$ one, 
Because of the commutation relation $[J_x{\cal C}, J_z]=0$,
the combination of $J_x$ and ${\cal C}$
maps the $J_z=1$ sector to itself, which defines the ``spin-rotation
particle-hole symmetry'' ${\cal C}_J\equiv J_x{\cal C}$ in the $J_z=1$ sector. 
(One can also consider $J_y{\cal C}$ as a combined particle-hole
symmetry, but it ultimately defines the same symmetry as $J_x{\cal C}$
in the $J_z=1$ sector).
In terms of the reduced Hamiltonian $h({\bm k})$, the spin-rotation
particle-hole symmetry is given by
\begin{eqnarray}
{\cal C}_J h({\bm k}){\cal C}_J^{-1}=-h(-{\bm k}),
\end{eqnarray}
with ${\cal C}_J=i\tau_y K$.
While the original particle-hole symmetry obeys
${\cal C}^2=1$, the spin-rotation particle-hole symmetry satisfies
${\cal C}_J^2=(J_x{\cal C})^2=-1$.
In the Altland-Zirnbauer scheme, systems having such particle-hole
symmetry are classified as class C.

If ${\cal H}({\bm k})$ is time-reversal invariant, so is the reduced
Hamiltonian $h({\bm k})$, but in a manner similar to the particle-hole
symmetry, its
time-reversal transformation is
different from the original one:
Since the original time-reversal transformation flips the spin of
electrons,  $h({\bm k})$ defined in the $J_z=1$ sector does
not retain the original time-reversal symmetry. 
%However, in a manner similar to the particle-hole symmetry, 
Instead, 
the $J_z=1$
sector has the ``spin-rotation time-reversal symmetry'' ${\cal T}_J\equiv
J_x{\cal T}$ as a combination of $J_x$ and ${\cal T}$. 
Correspondingly, $h({\bm k})$ satisfies
\begin{eqnarray}
{\cal T}_J h({\bm k}) {\cal T}_J^{-1}=h(-{\bm k}),
\end{eqnarray}
with ${\cal T}_J=K$. 
Different from the original time-reversal symmetry
${\cal T}$, which satisfies ${\cal T}^2=-1$, the spin-rotation
time-reversal symmetry ${\cal T}_J$ obeys ${\cal T}_J^2=1$.
Systems with ${\cal C}_J$ and ${\cal T}_J$ are classified as class CI.

Topological numbers for class C and class CI systems are summarized in
Table \ref{table:Periodic}. Note that only even numbers are possible for
the topological numbers.
From the bulk-boundary correspondence, 
this means that only even numbers of surface gapless states are possible
in these classes of topological superconductors.
Therefore, to realize locally unpaired Majorana fermions in topological
superconductors, we need the strong spin-orbit interaction or
spin-triplet Cooper pairs which break the SU(2) spin-rotation symmetry.

\subsection{Spinless superconductors}
\label{subsec:spinless}

As discussed in the previous section, the SU(2) spin-rotation symmetry must be
broken in order to realize locally unpaired Majorana fermions.
An example of systems with broken SU(2) spin-rotation symmetry is
a fully spin-polarized electron system.    
Fully spin-polarized electrons can be regarded as spinless
electrons, since their spin degrees of freedom are completely locked in a
particular direction.  
Superconducting states of fully spin-polarized electrons are called
as spinless superconductors.

%In spinless superconductors, the SU(2) spin-rotation symmertry is
%completely broken: For electrons with fully polarized spin in the
%$z$-direction, the spin rotation symmertry along the $x$ and $y$-direction is
%obviously broken.  Furthermore, the $J_z$
%spin-rotation symmetry is spontaneously broken by Cooper
%pairs. Indeed, only the equal-spin pairing $\langle c_{{\bm k},\uparrow}c_{-{\b%m
%k},\uparrow}\rangle$ is possible in the spinless superconductors, so 
%the $J_z$ spin-rotation symmetry must be broken.

Retaining only up-spin electrons and up-spin holes in Eq.(\ref{eq:BdG_matrix}), one can obtain
the BdG Hamiltonian for spinless superconductors,
\begin{eqnarray}
{\cal H}=\frac{1}{2}\sum_{{\bm k}}
\psi_{\uparrow}^{\dagger}({\bm k})h_{\uparrow}({\bm k})\psi_{\uparrow}({\bm k}),
\quad
\psi_{\uparrow}({\bm k})=
\left(
\begin{array}{c}
c_{\uparrow}({\bm k})\\
c_{\uparrow}^{\dagger}({\bm k})
\end{array}
\right) 
\end{eqnarray}
with
\begin{eqnarray}
h_{\uparrow}({\bm k})=
\left(
\begin{array}{cc}
{\cal E}_{\uparrow\uparrow}({\bm k}) & \Delta_{\uparrow\uparrow}({\bm k})\\
\Delta^{\dagger}_{\uparrow\uparrow}({\bm k}) &-{\cal
 E}^t_{\uparrow\uparrow}(-{\bm k})   
\end{array}
\right).
\end{eqnarray}
Here ${\cal E}_{\uparrow\uparrow}({\bm k})={\cal E}^{\dagger}_{\uparrow\uparrow}({\bm k})
$ and $\Delta_{\uparrow\uparrow}^t({\bm
k})=-\Delta_{\uparrow\uparrow}(-{\bm k})$.
As well as the original BdG Hamiltonian, the spinless BdG Hamiltonian
$h_{\uparrow}({\bm k})$ has particle-hole symmetry,
\begin{eqnarray}
{\cal C}h_{\uparrow}({\bm k}){\cal C}^{-1}=-h_{\uparrow}(-{\bm k}), 
\end{eqnarray}
and thus it is classified as class D.

Spinless superconductors cannot be invariant under ${\cal T}$
since the ordinary time-reversal operation flips the spin of electrons, 
Nevertheless, they may be invariant under a combination of the spin-flip
and the time-reversal, which we denote ${\cal T}_{\rm s}$.
In term of the spinless BdG Hamiltonian, the
combined time-reversal symmetry reads
\begin{eqnarray}
{\cal T}_{\rm s} h_{\uparrow}({\bm k}){\cal T}_{\rm s}^{-1}=h_{\uparrow}(-{\bm k}), 
\end{eqnarray}
with ${\cal T}_{\rm s}=K$. ${\cal E}_{\uparrow\uparrow}({\bm k})$ and
$\Delta_{\uparrow\uparrow}({\bm k})$ should be real in order to have ${\cal
T}_{\rm s}$-invariance.
The time-reversal invariant spinless superconductors are characterized
by ${\cal T}_{\rm s}$ and ${\cal C}$ obeying ${\cal T}_{\rm s}^2=1$ and ${\cal
C}^2=1$, and are classified as class BDI in the Altland-Zirnbauer
classification.  
Class BDI can be topologically non-trivial in one dimension, but not
in two and three dimensions. 
Similarly to class DIII, class BDI has chiral symmetry as a combination of
${\cal T}_{\rm s}$ and ${\cal C}$,
\begin{eqnarray}
\Gamma_{\rm s} h_{\uparrow}({\bm k})\Gamma_{\rm
 s}^{-1}=-h_{\uparrow}({\bm k}),
\quad 
\Gamma_{\rm s}={\cal T}_{\rm s}{\cal C}. 
\end{eqnarray} 
The 1D topological number in class BDI is given
by
\begin{eqnarray}
w^{\rm BDI}_{\rm 1D}=\frac{i}{4\pi} \int_{\rm BZ}dk{\rm tr}\left[\Gamma_{\rm s}
h_{\uparrow}^{-1}\partial_k h_{\uparrow}\right].
\label{eq:1dW}
\end{eqnarray}

\subsection{Topological defects}
\label{subsec:topodefect}

In the presence of a defect such as a vortex, the BdG Hamiltonian depends
on the distance ${\bm R}$ from the defect, ${\cal H}({\bm k}, {\bm R})$,
as well as the momentum ${\bm k}$.
Considering the topology of ${\cal H}({\bm k}, {\bm R})$ away from the
defect, we can examine the topological stability of the defect and
gapless defect modes at the same time.\cite{teo-kane2,shiozaki-sato} 

For this purpose, consider a $D$-dimensional sphere $S^D$ 
surrounding a defect in $d$ dimensions. 
For instance, consider $D=1$ ($D=2$) dimensional sphere $S^1$ $(S^2)$ for a line
(point) defect in $d=3$ dimensions.
The BdG Hamiltonian ${\cal H}({\bm k}, {\bm R})$ is now defined on the
base space $({\bm k}, {\bm R})\in T^d\times S^D$ with the
$d$-dimensional BZ $T^d$.
Away from the defect, the system is gapped, so in a manner similar to
uniform superconductors, one can define various nontrivial topological
numbers for ${\cal H}({\bm k}, {\bm R})$.

Among the topological numbers for ${\cal H}({\bm k}, {\bm R})$, some do
not change their value even when one neglects either the
${\bm k}$-dependence  or ${\bm R}$-dependence of ${\cal H}({\bm k}, {\bm R})$. 
The former ${\bm k}$-independent topological numbers are the topological
numbers of the defect itself, which ensures the stability of the defect,
while the latter ${\bm R}$-independent numbers are simply the bulk topological numbers without the defect. 
From the other topological numbers, we can predict the existence of gapless modes
localized on the defect.

For instance, away from the core, a vortex in a 2D
superconductor is well described by the following BdG Hamiltonian,
\begin{eqnarray}
{\cal H}({\bm k}, {\bm R})=
\left(
\begin{array}{cc}
{\cal E}({\bm k}) 
& \Delta({\bm k})e^{i\theta}\\
\Delta^{\dagger}({\bm k})e^{-i\theta} 
& -{\cal E}^{-t}(-{\bm k})  
\end{array}
\right) 
\end{eqnarray}
with ${\bm k}=(k_x, k_y)$ and ${\bm R}=(R\cos\theta, R\sin\theta)$.
Here $\theta$ is the angle around the vortex.
The vortex can host a Majorana zero mode localized on the core, whose
existence is ensured by the topological invariant
\begin{eqnarray}
\nu=\frac{1}{4\pi^2}\int_{T^2\times S^1}\epsilon^{ijk}{\rm tr}
\left[{\cal A}_i\partial_j{\cal A}_k-i\frac{2}{3}{\cal A}_i{\cal
 A}_j{\cal A}_k\right]
\quad {\rm mod} 2,  
\label{eq:CS}
\end{eqnarray}
where the gauge field matrix $({\cal A}_i)_{mn}({\bm k}, \theta)$ is defined as
\begin{eqnarray}
({\cal A}_i)_{mn}({\bm k},\theta)=i\langle u_m({\bm k},
 \theta)|\partial_i u_n({\bm k},\theta)\rangle, 
\quad 
(i=k_x, k_y, \theta), 
\end{eqnarray}
with negative energy states $|u_n({\bm k},\theta)\rangle$ of ${\cal H}({\bm
k},\theta)$. 
Here the trace in Eq.(\ref{eq:CS}) is taken for all the negative-energy states.

A systematic classification of topological defects and their gapless
modes has been performed based on the $K$-theory.\cite{teo-kane2, shiozaki-sato} 

\subsection{Topological crystalline superconductors}
\label{subsec:TCS}

In addition to the particle-hole and time-reversal symmetries,
superconductors may have crystal symmetries.
Such material dependent symmetries also influence on their topological
properties.\cite{mizushima-sato,teo,ueno,chui,zhang1,fang,liu-law,benalcazar,mizushima-sato1,morimoto}

For instance, consider the mirror reflection symmetry
with respect to the $xy$-plane,
\begin{eqnarray}
{\cal M}_{xy}{\cal H}({\bm k}){\cal M}^{-1}_{xy}={\cal H}(k_x, k_y, -k_z), 
\end{eqnarray}
where ${\cal M}_{xy}$ is the mirror reflection operator. 
Since the BdG Hamiltonian ${\cal
H}(k_x, k_y, 0)$ on the mirror invariant plane at $k_z=0$ commutes with ${\cal
M}_{xy}$, ${\cal
H}(k_x, k_y, 0)$ is block diagonal in the diagonal basis of ${\cal
M}_{xy}$. 
For each block-diagonal subsector with a definite eigenvalue of
${\cal M}_{xy}$, the mirror Chern number can be introduced as the Chern number
of the subsector. 
When the mirror Chern number is nonzero, gapless states appears on
surfaces preserving the mirror reflection symmetry. 
The gapless states are stable provided that the mirror symmetry is retained.

Whereas a similar mirror-protected topological surface state exists in
insulators,\cite{CTI-fu, CTI-ex1, CTI-ex2, CTI-ex3}  superconductors have their own novel properties.\cite{ueno} 
In superconductors, two different mirror reflection symmetries
are possible for the same crystal structure.
A superconducting state retains the mirror reflection symmetry of the
normal state when the gap function is invariant under the mirror reflection,
but this is not the only possibility.
Even when the gap function changes its sign under mirror reflection, 
the superconducting state may support a mirror reflection symmetry as a
combination of the original mirror reflection and a U(1) gauge rotation
for the gap function. 
For the former (latter), the mirror-even (mirror-odd) gap function, ${\cal
M}_{xy}$ is given by ${\cal M}_{xy}^{+}=is_z\tau_z$ (${\cal
M}^-_{xy}=is_z\tau_0$). 

%Interestingly, the nature of the gapless state depends on the
%transformation law of the gap function under the mirror reflection:

The two different mirror reflection
symmetries realize different mirror-protected surface states.
For the mirror-even case, the mirror reflection operator ${\cal
M}_{xy}^+$ commutes with the charge conjugation operator ${\cal
C}$. Thus, by applying ${\cal C}$ to an eigenstate $|\lambda\rangle$ with the
eigenvalue $\lambda=\pm i$ of ${\cal M}_{xy}^+$, we have an eigenstate
with the opposite eigenvalue $-\lambda$. (Note that ${\cal C}$ is anti-unitary.)
Therefore, the particle-hole symmetry merely interchanges the two mirror
subsectors, so each mirror subsector does not have its own
particle-hole symmetry. 
As a result, the surface state in each mirror subsector is not
self-conjugate, and thus it is a Dirac fermion. 
On the other hand, for the mirror-odd case, ${\cal M}_{xy}^-$
anticommutes with ${\cal C}$. In this case, the charge conjugation operator
${\cal C}$ maps a mirror eigensector into itself, so each mirror
subsector has its own particle-hole symmetry. 
Therefore, a surface state or a vortex state in each mirror subsector can be
a self-conjugate Majorana fermion.
The non-Abelian statistics of the mirror protected Majorana fermions
is discussed in Ref. \citen{sato-yamakage-mizushima} 

Crystal symmetries other than mirror reflection symmetry also
protect topological states.
%Furthermore, to gapless modes localized on topological defects  
A complete classification of topological crystalline
insulators/superconductors and topological defects protected by
order-two space group symmetries is given in
Refs. \citeonline{shiozaki-sato,shiozaki-sato-gomi}. 

\subsection{Majorana Ising property}
\label{subsec:Ising}

Crystals may have an anti-unitary symmetry ${\cal T}_{\cal G}$ as a combination of
time-reversal symmetry ${\cal T}$ and crystal symmetry ${\cal G}$. 
Such an anti-unitary symmetry, which is called magnetic symmetry, can
survive even when both ${\cal T}$ and ${\cal G}$ are broken.
For example, the magnetic mirror reflection symmetry with respect to the
$xy$-plane is obtained by combining the mirror reflection ${\cal
M}_{xy}$ with the time-reversal symmetry ${\cal T}=is_y K$,
\begin{eqnarray}
{\cal T}_{{\cal M}_{xy}}{\cal H}({\bm k}){\cal
 T}_{{\cal M}_{xy}}^{-1} 
={\cal H}(-k_x, -k_y, k_z),
\end{eqnarray} 
with ${\cal T}_{{\cal M}_{xy}}={\cal T}{\cal M}_{xy}$.
When we apply a magnetic field normal to the $z$-direction, both the mirror
reflection and time-reversal symmetries are explicitly broken, but the magnetic
mirror reflection symmetry ${\cal T}_{{\cal M}_{xy}}$ remains.
As is shown below, 
the magnetic symmetry also stabilizes topological
phases,\cite{sato-fujimoto1, mizushima-sato, mizushima-sato1, fang} in which 
Majorana fermions show a unique anisotropic response to magnetic
fields.\cite{sato-fujimoto1, chung1, higashitani, shindou, mizushima-sato}

First, we note that 
the magnetic symmetry
${\cal T}_{\cal G}$ results in
a chiral crystal symmetry $\Gamma_{\cal G}$,
by combining it with the particle-hole symmetry.
Then, along a 1D line in the momentum subspace invariant under ${\cal G}$,
we can define a 1D winding number in a similar manner as
Eq.(\ref{eq:windingnumber_1D}).
As a result, we may have a gapless state ensured by the 1D
winding number.
For the magnetic mirror reflection symmetry ${\cal T}_{{\cal M}_{xy}}$, we
have the chiral mirror symmetry, 
\begin{eqnarray}
\Gamma_{{\cal M}_{xy}} {\cal H}({\bm k})\Gamma_{{\cal M}_{xy}}^{-1}
=-{\cal H}(k_x, k_y, -k_z),
\end{eqnarray}
with $\Gamma_{{\cal M}_{xy}}=e^{i\alpha}{\cal C}{\cal T}{\cal M}_{xy}$.
(Here the phase $\alpha$ is determined so that $\Gamma_{{\cal
M}_{xy}}^2=1$.)
Then, say, for a line with a fixed $k_x$ in the mirror invariant plane at
$k_z=0$, we can define the topological invariant 
\begin{eqnarray}
w_{\rm 1D}(k_x)=\frac{i}{4\pi}\oint d k_y 
{\rm tr}\left[\Gamma_{{\cal M}_{xy}}{\cal H}^{-1}(k_x, k_y, 0)
\partial_{k_y} {\cal
	 H}(k_x, k_y, 0)\right]. 
\end{eqnarray}
When $w_{\rm 1D}(k_x)\neq 0$, from the bulk-boundary correspondence,
$|w_{\rm 1D}(k_x)|$ zero modes exist on a surface normal to the $y$-axis.

One can show that the zero modes $|u_0\rangle$ protected by the magnetic
symmetry ${\cal T}_{\cal G}$ are eigenstates of the chiral crystal operator
$\Gamma_{\cal G}$.
At the same time, from the particle-hole symmetry, 
$|u_0\rangle$ satisfies
${\cal C}|u_0\rangle=|u_0\rangle$ by choosing its phase appropriately.
The latter condition implies that the zero modes are self-conjugate
Majorana fermions, and the former condition determines the spin and/or orbital
structure of the Majorana zero modes. 
In particular, for spinful systems with magnetic mirror reflection
or magnetic $C_2$ rotation, 
from these two conditions, one can show that the density operator of the
Majorana zero modes is identically zero, and the spin density operator of the
Majorana zero modes is nonzero only in a particular
direction.\cite{mizushima-sato, shiozaki-sato} 
Therefore, the Majorana zero modes can couple to a magnetic field only in a
particular direction, resulting in an anisotropic magnetic
response.

\section{Realization of Topological Superconductors Supporting Majorana Fermions}
\label{sec:realization}

In this section,we explain the basic properties of topological superconductors, 
particularly, focusing on class D systems with broken time-reversal symmetry, for which typical features
such as the existence of Majorana fermions are demonstrated.
We also discuss how the class D topological superconductors can be realized in real systems.

\subsection{Chiral $p$-wave superconductors}
\label{subsec:chiralp}

A prototype of topological superconductors supporting Majorana fermions is a
spinless chiral $p$-wave superconductor in two dimensions.\cite{read-green} 
The BdG Hamiltonian is 
\begin{eqnarray}
{\cal H}=\frac{1}{2}
\sum_{\bm k}
(c^{\dagger}_{{\bm k}}, c_{-{\bm k}})
{\cal H}({\bm k})
\left(
\begin{array}{c}
c_{\bm k}\\
c^{\dagger}_{-{\bm k}}
\end{array}
\right)
\end{eqnarray}
with 
\begin{eqnarray}
{\cal H}({\bm k})=\left(
\begin{array}{cc}
\frac{k_x^2+k_y^2}{2m}-\mu_0 & \Delta_0(k_x+ik_y)\\
\Delta_0(k_x-ik_y) & -\frac{k_x^2+k_y^2}{2m}+\mu_0
\end{array}
\right),
\label{eq:BdGchiralp}
\end{eqnarray}
($\Delta_0>0$).  Now, following the original argument by Read and Green,\cite{read-green} we consider the case
of small $\mu_0$.
Under this assumption, the Fermi momentum becomes small; thus, we can
neglect the ${\bm k}^2$ term in the Hamiltonian.
Equation (\ref{eq:BdGchiralp}) reduces to 
\begin{eqnarray}
{\cal H}({\bm k})\rightarrow\left(
\begin{array}{cc}
-\mu_0 & \Delta_0(k_x+ik_y)\\
\Delta_0(k_x-ik_y) & \mu_0
\end{array}
\right).
\label{eq:BdGchiralp2}
\end{eqnarray}
Using this simplified Hamiltonian, we first examine the edge state.
For this purpose, 
consider a 2D semi-infinite system that extends in the
positive $x$ direction with the boundary at $x=0$.
To realize such a system, 
we replace $k_i$ and $\mu_0$ in Eq.(\ref{eq:BdGchiralp2}) as follows;
\begin{eqnarray}
k_i\rightarrow -i\partial_i,
\quad
\mu_0\rightarrow \mu(x), 
\end{eqnarray}
with
\begin{eqnarray}
\mu(x)=\left\{
\begin{array}{rl}
-\mu_0 <0,&\mbox{for $x<0$} \\
\mu_0 >0,&\mbox{for $x>0$}
\end{array}
\right. .
\end{eqnarray}
The latter equation implies that no electron exists in the $x<0$ region; thus,
it effectively realizes a semi-infinite system with $x>0$. 
The edge state is examined by solving the BdG equation
\begin{eqnarray}
\left(
\begin{array}{cc}
-\mu(x) & \Delta_0[-i\partial_x+\partial_y]\\
\Delta_0[-i\partial_x-\partial_y] & \mu(x)
\end{array}
\right)
\left(
\begin{array}{c}
u \\
v
\end{array}
\right)=E
\left(
\begin{array}{c}
u \\
v
\end{array}
\right). 
\end{eqnarray}
We can find a solution localized at $x=0$ :
\begin{eqnarray}
\left(
\begin{array}{c}
u_{k_y} \\
v_{k_y}
\end{array}
\right)=
N
\left(
\begin{array}{c}
i^{1/2} \\
i^{-i/2}
\end{array}
\right)\exp\left[
ik_y y-\int_0^x dx'\frac{\mu(x')}{\Delta_0}\right], 
\end{eqnarray}
where $N$ is a real normalization constant. The energy spectrum of this
edge state is linear in the momentum $k_y$,
\begin{eqnarray}
E=\Delta_0 k_y. 
\end{eqnarray}

To see that the edge state is a Majorana fermion, consider the mode
expansion,
\begin{eqnarray}
\left(
\begin{array}{c}
c(x) \\
c^{\dagger}(x)
\end{array}
\right)=\sum_{k_y}\gamma_{k_y}
\left(
\begin{array}{c}
u_{k_y} \\
v_{k_y}
\end{array}
\right)+\cdots. 
\end{eqnarray} 
Then, from the relation $u_{k_y}=v_{-k_y}^*$, one finds that
$\gamma_{k_y}$ satisfies the Majorana condition in the momentum space,
\begin{eqnarray}
\gamma_{k_y}=\gamma^{\dagger}_{-k_y}. 
\end{eqnarray}
Also, for $\gamma_{k_y}$, the time-dependent BdG equation
\begin{eqnarray}
i\partial_t c(x)=[c(x), {\cal H}] 
\end{eqnarray}
reduces to the 1D Dirac equation in the momentum space,
\begin{eqnarray}
i\partial_t \gamma_{k_y}=\Delta_0 k_y \gamma_{k_y}. 
\end{eqnarray}
Thus, $\gamma_{k_y}$ is a Majorana fermion.

A vortex in this system also supports a Majorana zero
mode. The BdG Hamiltonian with a vortex is obtained by replacing
\begin{eqnarray}
k_i\rightarrow (-i\partial_i+eA_i)\equiv -i{\cal D}_i,
\quad
\Delta_0 \rightarrow \Delta e^{-i\theta},
\quad \mu_0\rightarrow \mu(r)
\nonumber\\
\end{eqnarray} 
in Eq.(\ref{eq:BdGchiralp2}).
Here $r$ is the distance from the vortex core, and $\theta$ is the angle
around the vortex.
$A_i$ is the gauge potential of the vortex flux which can be
approximated as
\begin{eqnarray}
A_i=-\partial_i\theta/2e. 
\end{eqnarray}
This gauge potential is singular at $r=0$, but the singularity can be
avoided by taking $\mu(r)$ as
\begin{eqnarray}
\mu(r)=\left\{
\begin{array}{rl}
-\mu_0<0, & \mbox{for $r\rightarrow 0$}\\
\mu_0>0, & \mbox{for $r\rightarrow\infty$ }
\end{array}
\right. .
\end{eqnarray}
Because the chemical potential is negative near the vortex core,
electrons cannot approach there; thus, the
singularity becomes harmless.
The zero mode is obtained by solving the BdG equation
\begin{eqnarray}
\left(
\begin{array}{cc}
-\mu(r) & \Delta e^{-i\theta}(-i{\cal D}_x+{\cal D}_y)
\\
\Delta (-i{\cal D}_x-{\cal D}_y)e^{i\theta}
& \mu(r)
\end{array}
\right)
\left(
\begin{array}{c}
u_0 \\
v_0
\end{array}
\right)=0, 
\end{eqnarray}
whose solution is
\begin{eqnarray}
\left(
\begin{array}{c}
u_0 \\
v_0
\end{array}
\right)=N \left(
\begin{array}{c}
i^{1/2} \\
i^{-1/2}
\end{array}
\right)
\exp\left[-\int_0^r dr'\frac{\mu(r')}{\Delta}\right], 
\end{eqnarray}
with a real normalization constant $N$. 
The quantum operator of the zero mode $\gamma$,
which is defined as the coefficient of the expansion
\begin{eqnarray}
\left(
\begin{array}{c}
c({\bm x})\\
c^{\dagger}({\bm x})
\end{array}
\right)=
\sqrt{2}\gamma\left(
\begin{array}{c}
u_0 \\
v_0
\end{array}
\right)+\cdots , 
\end{eqnarray}
satisfies the Majorana condition
\begin{eqnarray}
\gamma=\gamma^{\dagger} 
\label{eq:Majoranacond}
\end{eqnarray}
and the anticommutation relation
\begin{eqnarray}
\{\gamma^{\dagger}, \gamma\}=1/2.
\label{eq:Majoranaanticommu}
\end{eqnarray}
Here we have used  
\begin{eqnarray}
\gamma=\int dxdy\left[u_0^*({\bm x})c({\bm x})+
v_0^*({\bm x})c^{\dagger}({\bm x})
\right]/\sqrt{2} 
\end{eqnarray}
and
\begin{eqnarray}
\{c^{\dagger}({\bm x}), c({\bm x}')\}=\delta({\bm x}-{\bm x}'),
\nonumber\\
\{c({\bm x}), c({\bm x}')\}
=\{c^{\dagger}({\bm x}), c^{\dagger}({\bm x}')\}
=0,
\end{eqnarray}
to derive these relations.
Equations (\ref{eq:Majoranacond}) and (\ref{eq:Majoranaanticommu}) show that $\gamma$ is a Majorana
zero mode. 

In a similar manner, one can show that $N$ well-separated vortices
support $N$ Majorana zero modes satisfying
\begin{eqnarray}
\gamma_i^2=1,
\quad
\gamma_i\gamma_j=-\gamma_j\gamma_i,  
\end{eqnarray} 
for $i\neq j$.
Later, we will argue that this relation is an important ingredient in
realizing the non-Abelian statistics 

\subsection{Dirac systems}

Majorana zero modes and thus non-Abelian statistics can be realized
even in an $s$-wave superconducting state. 
This possibility was first pointed out in 2003.\cite{sato} 
Here we will review the idea of this paper.

The basic idea is to consider a 2D bulk Dirac fermion
system instead of an ordinary electron system. 
As was discussed in Sect.\ref{subsec:Majorana}, quasiparticles in superconducting states
may satisfy the Majorana condition.
Thus, if we consider bulk Dirac fermions, they automatically realize
Majorana fermions even in an $s$-wave superconducting state.    
In the Nambu representation with the basis $\Psi'({\bm x})=(c^{\uparrow}({\bm x}),
c_{\downarrow}({\bm x}), c^{\dagger}_{\downarrow}({\bm x}),
-c^{\dagger}_{\uparrow}({\bm x}))$, the BdG Hamiltonian of the Majorana
fermions is given by
\begin{eqnarray}
{\cal H}=\left(
\begin{array}{cc}
-i\sigma_i\partial_i & \Phi^*\\
\Phi & i\sigma_i\partial_i
\end{array}
\right),
\label{eq:BdGDirac} 
\end{eqnarray}
where $\sigma_{i=x,y}$ are the Pauli matrices, and $\Phi$ is an $s$-wave
gap function.
Note that the diagonal term is not the Hamiltonian for ordinary electrons 
but that for Dirac fermions.

Now we will show that a Majorana zero mode exists in a vortex of
this system.
When the system supports a vortex, $\Phi$ in Eq.(\ref{eq:BdGDirac}) becomes
\begin{eqnarray}
\Phi=\Phi(r)e^{i\theta},
\quad
\Phi(\infty)=\Phi_0 >0, 
\end{eqnarray} 
where $r$ is the distance from the vortex, and $\theta$ is the angle
around the vortex.
The BdG Hamiltonian for the zero mode
\begin{eqnarray}
\left(
\begin{array}{cc}
-i\sigma_i\partial_i & \Phi(r)^{-i\theta}\\
\Phi(r)e^{i\theta} & i\sigma_i\partial_i
\end{array}
\right)
\left(
\begin{array}{c}
{\bm u} \\
{\bm v}
\end{array}
\right) =0 
\end{eqnarray}
is easily solved and the solution is\cite{jackiw-rossi,callan-harvey} 
\begin{eqnarray}
{\bm u}
=N
\left(
\begin{array}{c}
0 \\
i^{1/2}
\end{array}
\right)\exp\left[-\int_0^r dr'\Phi(r')
\right],
\quad 
{\bm v}=i\sigma_y{\bm u}^{*}. 
\end{eqnarray}
The quantum operator $\gamma$ for this solution in the mode expansion
\begin{eqnarray}
\Psi'({\bm x})=\sqrt{2}\gamma 
\left(
\begin{array}{c}
{\bm u} \\
{\bm v}
\end{array}
\right)+\cdots 
\end{eqnarray}
defines the Majorana operator with $\gamma^{\dagger}=\gamma$ and
$\gamma^2=1$.
As a result, vortices in this system obey non-Abelian anyon statistics.

A challenge of this idea is how to realize
Dirac fermions in condensed matter systems.
%In 2003, 
%Indeed, 
In general, the Nielsen-Ninomiya theorem tells us that such a single species
of 2D Dirac fermions cannot exist in bulk periodic systems
like solid states.\cite{nielsen-ninomiya}
Indeed, no condensed matter realization of a single Dirac fermion was known in 2003.
To avoid this difficulty, an application to high-energy physics was
discussed in Ref.\citen{sato}.
In 2007, however, it was discovered that such a 2D Dirac fermion can exist
as a surface state of topological insulators.\cite{fu-kane-mele,moore-balents,roy3D}
Because the surface state is not a true 2D state but spreads to
the bulk as well, it can avoid the constraint imposed by
the Nielsen-Ninomiya theorem. 
Thus, if one places an $s$-wave superconductor on top of
a topological insulator, one can realize the Hamiltonian (\ref{eq:BdGDirac}).
This celebrated idea was first proposed by Fu and Kane.\cite{fu-kane}.  In the interface between an $s$-wave superconductor and a topological insulator, the chemical potential is not zero, and thus the BdG
Hamiltonian is slightly modified as
\begin{eqnarray}
{\cal H}=\left(
\begin{array}{cc}
-i\sigma_i\partial_i-\mu & \Phi^* \\
\Phi & i\sigma_i\partial_i+\mu
\end{array}
\right). 
\end{eqnarray}
The chemical potential, however, does not affect the existence of
the Majorana zero mode. 
The zero mode in the presence of the chemical potential is given by
\begin{eqnarray}
{\bm u}&=&N i^{3/2}
\left(
\begin{array}{c}
e^{-i\theta} J_1(\mu r)\\
i^{-1}J_0(\mu r)
\end{array}
\right)
\exp\left[-\int_0^r dr'\Phi(r')\right],
\nonumber\\
{\bm v}&=&i\sigma_y {\bm u}^* ,
\end{eqnarray}  
with the Bessel function $J_n(x)$.\cite{fukui-fujiwara}
Transport properties of Majorana fermions in superconductor-topological insulator junction systems have been
studied by several authors.\cite{tanaka-yokoyama,nori1,nori2}

\subsection{System with Rashba spin-orbit interaction}
\label{sec:system_rashba}

In the scenarios discussed above, unconventional systems such as
chiral $p$-wave superconductors or superconducting Dirac fermions are
needed for the realization of non-Abelian anyons.
If we take into account the spin-orbit interaction, however, we may realize
Majorana fermions and non-Abelian anyons even for ordinary electron
systems with an $s$-wave
pairing.\cite{fujimoto,sato-takahashi-fujimoto1,sato-takahashi-fujimoto2,sau,alicea} 
Below, we explain this realization scheme, following
Ref.\citen{sato-takahashi-fujimoto1},
which provided a complete topological analysis and established this scheme
for the first time.

The model BdG Hamiltonian we consider  is given by 
\begin{eqnarray}
{\cal H}=\frac{1}{2}\sum_{\bm k}
\left(c_{{\bm k}\uparrow}^{\dagger},
c_{{\bm k}\downarrow}^{\dagger},
c_{-{\bm k}\uparrow}, c_{-{\bm k}\downarrow}\right)
{\cal H}({\bm k})
\left(
\begin{array}{c}
c_{{\bm k}\uparrow} \\
c_{{\bm k}\downarrow}\\
c_{-{\bm k}\uparrow}\\
c_{-{\bm k}\downarrow}
\end{array}
\right) 
\end{eqnarray}
with
%\begin{widetext}
\begin{align}
{\cal H}({\bm k})=
\left(
\begin{array}{cc}
\varepsilon({\bm k})+{\bm g}({\bm k})\cdot{\bm \sigma}-\mu_{\rm
 B}H_z\sigma_z 
&  i\Delta\sigma_y\\
-i\Delta\sigma_y &
-\varepsilon({\bm k})+{\bm g}({\bm k})\cdot{\bm \sigma}^*+\mu_{\rm
 B}H_z\sigma_z 
\end{array}
\right). 
\label{eq:BdGRashba}
\end{align}
%\end{widetext}
Here $\varepsilon({\bm k})=\frac{k_x^2+k_y^2}{2m}-\mu$ is the energy of
electrons in the normal state, which is measured from the Fermi level, and 
$\mu_{\rm B}H_z\sigma_z$ is the Zeeman term. 
${\bm g}({\bm k})\cdot{\bm \sigma}$ is the antisymmetric spin-orbit
interaction which
may arise when spatial inversion symmetry is broken.
In particular, we consider here the Rashba spin-orbit interaction, ${\bm
g}({\bm k})=2\lambda(k_y, -k_x)$, which arises when the system
breaks the spatial reflection with respect to the $xy$ plane. 
As is shown below, from the Rashba spin-orbit interaction, the system
may have a similar property to chiral $p$-wave superconductors. 

To see the similarity between this system and a chiral $p$-wave
superconductor, we perform the following unitary transformation:
\begin{eqnarray}
{\cal H}^{\rm D}({\bm k})=D{\cal H}({\bm k}) D^{\dagger},
\quad
D=\frac{1}{\sqrt{2}}
\left(
\begin{array}{cc}
1 & i\sigma_y\\
i\sigma_y & 1
\end{array}
\right).
\end{eqnarray}
From this unitary transformation, the original BdG Hamiltonian is mapped
into the dual Hamiltonian,
\begin{eqnarray}
{\cal H}^{\rm D}=\left(
\begin{array}{cc}
\Delta-\mu_{\rm B}H_z\sigma_z & -i\varepsilon({\bm k})\sigma_y-i{\bm g}({\bm
 k})\cdot{\bm \sigma}\sigma_y\\
i\varepsilon({\bm k})\sigma_y+i{\bm g}({\bm k})\cdot\sigma_y{\bm \sigma} 
&-\Delta+\mu_{\rm B}H_z\sigma_z
\end{array}
\right).
\label{eq:dualHamiltonian} 
\end{eqnarray}
Interestingly, the dual Hamiltonian has the following gap function as
the off-diagonal elements,
\begin{eqnarray}
\Delta^{\rm D}({\bm k})=-i\varepsilon({\bm k})\sigma_y-i{\bm g}({\bm
 k})\cdot{\bm \sigma}\sigma_y. 
\end{eqnarray}
In particular, one should note that the Rashba spin-orbit interaction in
the original Hamiltonian is now become the spin-triplet component 
$-i{\bm g}({\bm k})\cdot {\bm \sigma}\sigma_y$ of the dual gap function.
Since the dual Hamiltonian is unitary equivalent to the original Hamiltonian,  
this suggests that the original Hamiltonian may have a similar property to
the spin-triplet superconductor owing to the Rashba spin-orbit
interaction. 
Whereas this is not always the case in reality because the diagonal term
of the dual Hamiltonian is not the standard electron energy but a
constant term $\Delta-\mu_{\rm B}H_z \sigma_z$, if we apply the Zeeman
field beyond a critical value, there appears a topological
superconducting state similar to chiral $p$-wave superconductors. 

\begin{figure}[t]
\includegraphics[width=8cm]{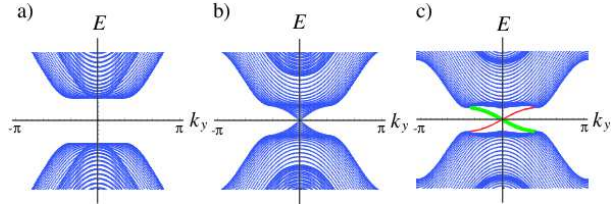}
\caption{(Color online) Energy spectra of a 2D $s$-wave pairing state
with the Rashba spin-orbit interaction in a ribbon geometry. a) $H_z=0$. 
No gapless edge mode appears as in the case of a conventional $s$-wave superconductor.
b) $\mu_{\rm
 B}H_z=\sqrt{\varepsilon({\bm 0})^2+\Delta^2}$. 
A topological quantum phase transition point emerges. The bulk gap closes.
c) $\mu_{\rm
 B}H_z>\sqrt{\varepsilon({\bm 0})^2+\Delta^2}$.
The red thin line indicates a chiral Majorana mode localized on 
 one edge of the ribbon, and the thick green  line represents an anti-chiral
 Majorana mode on the other edge of the ribbon.
[Reproduced from Fig.1 of Phys. Rev. Lett. ${\bm 103}$ (2009) 020401 by
 Sato et al.\cite{sato-takahashi-fujimoto1}]
}
\label{fig:edge_rashba}
\end{figure}

First, let us examine the edge states in this model.
Figure \ref{fig:edge_rashba} shows the quasiparticle spectra for the system
with an open boundary condition in the $x$-direction and a periodic boundary
condition in the $y$-direction.
In the absence of $H_z$ [Fig.\ref{fig:edge_rashba}(a)], the
quasiparticle spectrum shows a bulk gap as in the case of a conventional $s$-wave
superconductor, but if one turns on $H_z$, the bulk gap closes at the
critical value of $H_z$ [Fig.\ref{fig:edge_rashba}(b)], then when
the condition, 
\begin{eqnarray}
(\mu_{\rm B}H_z)^2>\varepsilon({\bm 0})^2+\Delta^2  
\label{eq:Hzcondition}
\end{eqnarray}
is satisfied, the bulk gap opens again and gapless edge states appear. 
[see  thin red and thick green lines in Fig. \ref{fig:edge_rashba}(c).]
The present system has two edges, and the one side of the edges has  
the edge state with the dispersion 
\begin{eqnarray}
E\sim c k_y, 
\end{eqnarray}
[ thin red line in Fig. \ref{fig:edge_rashba}(c))]
and the other has the edge state with
\begin{eqnarray}
E\sim -c k_y, 
\end{eqnarray}
[thick green line in Fig. \ref{fig:edge_rashba}(c)].
Therefore, each edge supports an edge state similar to that of the chiral
$p$-wave superconductors in Sect.\ref{subsec:chiralp}.

The existence of the edge states is also confirmed by the evaluation of
the TKNN number $\nu_{\rm TKNN}$.
While $\nu_{\rm TKNN}$ is zero in the absence of the Zeeman field, 
it becomes nonzero (i.e., $|\nu_{\rm TKNN}|=1$) when the condition
(\ref{eq:Hzcondition}) is satisfied.
Hence, the existence of edge states in Fig.\ref{fig:edge_rashba}(c) is
ensured by the bulk-boundary correspondence.
Since the TKNN number is the same as that of the spinless chiral $p$-wave
superconductor, the present system belongs to the same topological class 
as the spinless chiral $p$-wave superconductor.
When the condition (\ref{eq:Hzcondition}) is met, there also exists a
Majorana zero mode in a vortex.\cite{sato-takahashi-fujimoto1} 
 
Here one should note that the Zeeman field larger than the gap function
$\Delta$ is needed to satisfy Eq. (\ref{eq:Hzcondition}).
This means that it is difficult for usual bulk $s$-wave superconductors
to satisfy this condition.
In general, such a strong magnetic field may
destroy $s$-wave Cooper pairs by the Pauli depairing effect or the orbital
depairing effect.  
In the present case, the Pauli depairing can be avoided
because the Fermi surface is split by the Rashba spin-orbit
interaction; thus, the Zeeman splitting is strongly suppressed when the Rashba
spin-orbit interaction is strong enough.\cite{fujimoto1}
On the other hand, the orbital depairing effect cannot be avoided for
usual bulk superconductors.

\begin{figure}
\begin{center}
\includegraphics[width=4.5cm]{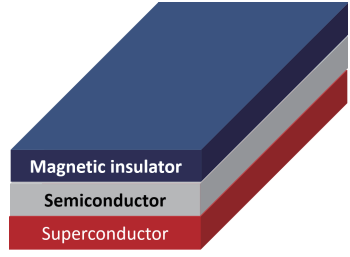}
\caption{(Color online) Realization of the Hamiltonian (\ref{eq:BdGRashba}) in a
 heterostructure. A proximity induced $s$-wave pairing state is
 realized in the semiconductor.} 
\label{fig:JDSau}
\end{center}
\end{figure}

However, it is known that this difficulty can be solved in
various ways:
\begin{enumerate}
 \item If one consider $s$-wave superfluids of cold atoms,
       instead of superconductors, one can avoid this
       problem.\cite{sato-takahashi-fujimoto1,Zhang-Sarma} 
This is because Cooper pairs in superfluids are charge neutral, and thus, no Lorentz
       force leading to the orbital depairing arises.
The spin-orbit interaction in cold atoms can be created
       artificially by using a sophisticated laser technique.\cite{spielman}

\item A superconducting state of heavy fermions may also not suffer from this
      problem.\cite{sato-takahashi-fujimoto2}
Because the effective mass of heavy fermions is about a hundred times the
      electron mass in the vacuum,
      the Lorentz force is strongly suppressed.  Thus, the orbital depairing
      effect does not work. In particular, heavy fermion noncentrosymmetric
      superconductors with the Rashba spin-orbit interaction are promising candidates.
      
\item A widely adopted solution is to use proximity induced
      superconductivity, instead of bulk superconductivity. By placing
      a semiconductor on top of an $s$-wave superconductor, $s$-wave
      superconductivity can be induced in the semiconductor.\cite{fujimoto,sau,alicea}
The Zeeman field in the semiconductor can be
      given in two different manners. 
The first one is to consider a heterostructure illustrated in
      Fig.\ref{fig:JDSau}.
%
%      and to induce the Zeeman term as a proximity of the exchange
%      interaction from the attached magneto. 
In this case, the electron energy in the semiconductor is
      Zeeman-split by proximity effects of the exchange interaction
      of the attached magnet.
% so the Zeeman term is induced effectively. 
Since no direct magnetic field is applied,
      no Lorentz force for the orbital depairing is induced.
The other way is to apply in-plane Zeeman fields, instead of Zeeman
      fields in the $z$-direction.\cite{alicea}
%The merit of this solution is that the
%      superconductivity is realized in the truly two-dimensional
%      system, in
%      which the bulk intrinsic superconductivity is impossible due to large
%      quantum fluctuations. 
In a 2D system, the motion of Cooper pairs in the perpendicular direction
      is severely restricted, and hence, the orbital depairing effect becomes ineffective if
      one applies Zeeman fields in the parallel direction.
In both cases, the topological superconductivity in a
      semiconductor with small $\varepsilon({\bm 0})$ is considered in
      order to satisfy the condition (\ref{eq:Hzcondition}) easily.
Note that the heterostructure breaks the inversion symmetry so it
      naturally induces the Rashba spin-orbit interactions.

The above-mentioned idea was generalized to 1D
nanowires on top of an $s$-wave superconductor.\cite{lutchyn,oreg} 
The Hamiltonian of this system is the same as  Eq.(\ref{eq:BdGRashba}) with $k_y=0$, 
and Majorana zero modes appear at the ends of the nanowires when the
condition (\ref{eq:Hzcondition}) is satisfied. 
\end{enumerate}

In these systems, one can control the topological superconductivity by
changing model parameters such as $\varepsilon({\bm 0})$
(namely, the chemical potential) and $H_z$.
Actually, the adiabatic operation by controlling the model parameters of topological
superconductivity was proposed in Ref.\citen{alicea2}.  

So far, we have considered $s$-wave pairing as the
superconducting gap $\Delta$. We should note that, however, Majorana
fermions may appear even for other unconventional pairings.
Indeed, the details of the gap function does not matter for the argument
using the dual Hamiltonian (\ref{eq:dualHamiltonian}) since the 
spin-triplet gap function is induced in the dual Hamiltonian even for
unconventional pairing. The only difference is that because the gap
function vanishes at ${\bm k}={\bm 0}$ for many unconventional pairings,   
the condition (\ref{eq:Hzcondition}) changes to 
\begin{eqnarray}
(\mu_{\rm B}H_z)^2>\varepsilon({\bm 0}). 
\end{eqnarray} 
Therefore, for systems with small $\varepsilon({\bm 0})$, one can
realize Majorana fermions even for weak Zeeman fields that do not cause
the orbital depairing. 
Majorana fermions in $p$-wave superconductors and $d$-wave
superconductors were discussed in
Refs. \citen{sato-fujimoto1, potter}, and
Refs. \citen{linder,sato-takahashi-fujimoto2, sato-fujimoto2,law1,Thomale2},
respectively.   

\subsection{Topological superconductor from coupling with non-collinear magnetic order}

It was recently proposed that a topological superconductor with Majorana zero modes
can be realized in an electron system with proximity-induced $s$-wave superconducting gap coupled with non-collinear magnetic order.
\cite{choy,martin,nadj,pascal,franz,loss,nakosai}
A simple example is a 1D $s$-wave superconductor coupled with helical magnetic order of localized spins via the exchange interaction $J\bm{s}(x)\cdot\bm{S}(x)$. 
Here $\bm{s}(x)=\frac{1}{2}c^{\dagger}_{\sigma}(x)\bm{\sigma}_{\sigma\sigma'}c_{\sigma'}(x)$ is the spin density operator of itinerant electrons,
and $\bm{S}(x)$ is the localized spin at position $x$. $J$ is the exchange coupling strength.
Let us consider a simple case of 1D helical order with of the localized spin configuration given by 
\begin{eqnarray}
\bm{S}(x)=S(\sin(k_Fx),0,\cos(k_Fx)),
\end{eqnarray}
which may be realized by the RKKY interaction.\cite{martin,pascal,franz,loss} 
Here $k_F$ is the Fermi wave number of the 1D itinerant electron system, and
$S$ is the magnitude of the localized spin.
Then, the unitary transformation, 
\begin{eqnarray}
U=\exp(i\frac{k_Fx}{2}\sigma_y), 
\end{eqnarray}
which rotates the spin axis as $U\bm{S}\cdot\bm{\sigma}U^{\dagger}=\sigma_z$, 
changes the kinetic energy term of the first-quantized Hamiltonian of the electron system $\frac{\hat{p}_x^2}{2m}$ into 
\begin{eqnarray}
\frac{1}{2m}\hat{p}_x^2-\frac{v_F}{2}\hat{p}_x\sigma_y+\frac{k_F^2}{8m},
\label{eq:helical}
\end{eqnarray}
where $\hat{p}_x=-i\partial_x$, $m$ is the effective mass of electrons, and $v_F=k_F/m$.
The second term of Eq.(\ref{eq:helical}) mimics the 1D version of the
spin-orbit interaction considered in Sect. \ref{sec:system_rashba}.
Thus, with the "Zeeman term" $JS\sigma_z$ and the $s$-wave pairing term $\Delta i\sigma_y$, 
the Hamiltonian of the system is equivalent to that of the 
1D $s$-wave superconductor with the Rashba SO interaction and the Zeeman coupling, which, indeed, belongs to
the same topological class as 
1D $p$-wave superconductor when $(JS)^2>\mu^2+\Delta^2$,
as discussed in the previous sections.
The case of 2D systems with skyrmion spin textures is discussed in Ref. \citen{nakosai}.
It should be noted that in the above scenario, we can deform the structure of the magnetic order without closing the bulk energy gap preserving the topological invariant which guarantees the existence of Majorana zero modes.
Then, the specific form of the magnetic order given by Eq. (\ref{eq:helical}) is not a necessary ingredient of this scenario. From a different viewpoint,
the Majorana zero mode found in this scenario can be associated with the Shiba state, which is
the bound state formed around a magnetic impurity in a superconductor.\cite{shiba}

\subsection{Other systems}
Varieties of hybrid systems with spin-orbit interaction may realize
Majorana fermions. For instance, quantum (anomalous) Hall
state/superconductor interfaces,\cite{qi-hughes-zhang1}
half-metal/superconductor heterostructures,\cite{chung} and quantum-dot-superconductor chains\cite{sau1} are proposed to
host Majorana fermions in $s$-wave superconducting states.

\section{Candidate Materials}
\label{sec:materials}

There are several candidate compounds for topological superconductors.
One of promising materials for time-reversal symmetry broken topological superconductors in class D is Sr$_2$RuO$_4$ which is a candidate of a quasi-2D chiral $p+ip$ superconductor.\cite{maeno1,maeno2}
A representative candidate for time-reversal symmetric topological superconductors in class DIII in three dimensions
is  Cu$_x$Bi$_2$Se$_3$, which is a carrier-doped topological insulator, and is expected to realize
an odd-parity pairing
state.\cite{CuBiSe1,CuBiSe2,CuBiSe3,CuBiSe4,CuBiSe5,hao-lee,CuBiSe6,yamakage,CuBiSe7,yamakage1,
hashimoto1,nagai,takami,mizushima,hashimoto2, nematic,matano}
Since there are already several excellent reviews on these materials,
\cite{maeno1,maeno2,ando,sasaki-mizushima} we will not discuss them here.
Instead, we discuss some other possible candidate materials in this section.
%Since there are already extensive reviews on this material, we will not present detailed discussions.
%The $\mu$SR

%\subsection{Sr$_2$RuO$_4$: Time-reversal-symmetry-broken (TRSB) superconductor in class D}

%\subsection{Cu$_x$Bi$_2$Se$_3$: Time-reversal-symmetric (TRS) superconductor in class DIII}

%Cu$_x$Bi$_2$Se$_3$ is a candidate of three-dimensional class DIII topological superconductors, which is
%topologically protected by time reversal symmetry.
%There are nice reviews on this material.\cite{ando,sasaki-mizushima} 
%We, here, briefly comment

\subsection{Proximity-induced superconductivity in nanowire:
%Experimental explorations for Majorana fermions in nanowire systems
 Time-reversal-symmetry-broken (TRSB) superconductor in class D}
\label{sec:nanowire}

In 2012, three experimental research groups independently reported the observation of a zero-bias conductance peak in nanowire-superconductor systems,
which may be associated with Majorana zero energy modes at open ends of the nanowire.\cite{mourik,das,deng}
In such nanowire systems, a class D topological superconducting state which harbors Majorana zero-energy states at its open ends is realized for $E_z>\sqrt{\Delta^2+\mu^2} $,
where $E_z$ is the Zeeman energy, $\Delta$ is the proximity-induced $s$-wave superconducting gap, and 
$\mu$ is the chemical potential of the nanowire.\cite{sato-takahashi-fujimoto1,sau,lutchyn,oreg} 
%\begin{figure}
%\includegraphics{fig01.eps}
%\caption{You can embed figures using the \texttt{\textbackslash includegraphics} command. EPS is the only format that can be embedded. Basically, figures should appear where they are cited in the text. You do not need to separate figures from the main text when you use \LaTeX\ for preparing your manuscript.}
%\label{f1}
%\end{figure}
These groups used InSb and InAs for the nanowire, as proposed in Refs.\citen{lutchyn,oreg}.
A zero-bias conductance peak appears when the Zeeman energy $E_z$ satisfies the above condition, which strongly indicates
the realization of Majorana end states.
According to previous theoretical studies,\cite{sato-takahashi-fujimoto1} 
a topological phase transition occurs at $E_z=\sqrt{\Delta^2+\mu^2}$, and the bulk energy gap closes at this critical value of $E_z$.
However, in the experimental data of these conductance measurements, gap-closing behaviors were not clearly observed. 
It was argued by Stanescu et al.\cite{stanes} that the absence of clear gap-closing behaviors is due to the dominant contributions from non-topological
superconducting bands which stem from multi-band character of the nanowires, and do not exhibit gap-closing at $E_z=\sqrt{\Delta^2+\mu^2}$.
Nevertheless, as discussed in several theoretical studies,\cite{stanes,liu-potter,rainis} 
the interpretation of these experimental observations is still controversial, and 
it has not yet been established whether or not the observed zero-bias peaks correspond to the signature of Majorana fermions.
For instance, Liu et al.\cite{liu-potter} and Rainis et al.\cite{rainis} argued that disorder effects give rise to low-energy peak of conductance which is due to sub-gap states originating from trivial bands, even in trivial superconducting states.
We need further experimental and theoretical investigations to establish the existence of Majorana zero modes in nanowire superconducting systems
with strong spin-orbit interactions.

\subsection{Li$_2$Pt$_3$B: time reversal symmetric (TRS) superconductor in class DIII}

It is expected that the noncentrosymmetric superconductor Li$_2$Pt$_3$B, which has the cubic crystal structure $P4_332$ and $T_c\sim 2.7$ K,
may be a  TRS topological superconductor similar to the B phase of the superfluid $^3$He.
It is known that the B phase of the superfluid $^3$He is a typical example of TRS topological superfluidity in class DIII.
The $\bm{d}$-vector of this spin-triplet $p$-wave pairing state which characterizes
the nontrivial topology is given by $\bm{d}(\bm{k})\propto \bm{k}$ with $\bm{k}$ a wave number.  
Since the crystal structure of Li$_2$Pt$_3$B breaks inversion symmetry, the pairing state cannot be classified by the parity of Cooper pairs.
However, if the spin-triplet pair component dominates the spin-singlet component, the system can be a TRS topological superconductor. 
The NMR measurement of this material suggests the existence of the dominant spin-triplet component.\cite{zheng}
Furthermore, the form of the $\bm{d}$-vector is constrained by the anti-symmetric spin-orbit interaction, the form of which is
$\sim \lambda_{\rm SO}\bm{\sigma}\cdot\bm{k}$, respecting the symmetry $P4_332$.
This spin-orbit interaction stabilizes the $\bm{d}$-vector similar to that of the B phase of $^3$He.
Unfortunately, single crystals of this material have not been synthesized so far, and thus, the topological properties of Li$_2$Pt$_3$B
have not yet been clarified.

\subsection{In$_x$Sn$_{1-x}$Te: TRS superconductor in class DIII or topological crystalline superconductor}

There have been extensive studies on crystalline topological insulators (CTIs) which are protected by crystalline symmetry in addition to time-reversal symmetry.\cite{CTI-fu}
A promising material is the semiconductor SnTe, for which surface Dirac cone bands protected by mirror symmetry
are observed via ARPES measurements.\cite{CTI-ex1} 
The partial substitution of Sn with In induces carrier-doping, rendering the system metallic. 
The doped CTI In$_x$Sn$_{1-x}$Te exhibits superconductivity with $T_c\sim 1$ K.
Remarkably, a zero energy bias peak of the conductance was observed for $x=0.045$, which implies a topological superconducting state
with surface zero energy Majorana modes.\cite{InSnTe}
The possible pairing state may be deduced by the group theoretical
argument.\cite{InSnTe}
Corresponding surface states have been investigated
theoretically.\cite{hashimoto}
The fermi surfaces of In$_x$Sn$_{1-x}$Te are located around $L$ points in the fcc BZ.
In the vicinity of the $L$ points, the band has $D_{3d}$ point group symmetry, which imposes a constraint on pairing symmetry.
There are two possible odd-parity pairing state allowing Majorana surface states; i.e., $A_{1u}$ and $A_{2u}$.
The $A_{1u}$ state has full energy gap, and thus, it is a TRS topological superconducting state in class DIII.
On the other hand, the $A_{2u}$ state has gap-nodes, and the bulk topological invariant can not be defined.
However, in this gapless superconducting state, a weak topological invariant can be defined in a 2D plane crossing
fully gapped region of the Fermi surfaces. Thus, this state is a weak topological superconductor.

For larger $x$, the superconducting state of In$_x$Sn$_{1-x}$Te is changed to a conventional state, which may be an $s$-wave pairing state.\cite{InSnTe2,InSnTe3}
If the surface Dirac bands protected by mirror symmetry still exist for this case, a vortex in the superconducting state can possess
the Majorana zero energy bound state at the intersection between the vortex core and the surface of the system.\cite{jackiw-rossi}
An interesting feature of these Majorana modes is that they are protected by magnetic group symmetry: 
the symmetry of the system under the combination of time-reversal symmetry operation and mirror reflection 
results in the enhancement of topological protection of the Majorana zero modes.\cite{fang}
According to the $K$-theory argument, zero modes in a vortex of a 2D superconductor in class D are classified as $Z_2$, which is protected
by the particle-hole symmetry as discussed in Sect. \ref{subsec:topodefect}.
Thus, multiple Majorana zero modes in the vortex core are not stable, and
there is only one topologically protected Majorana bound state.
However, a vortex penetrating the surface of In$_x$Sn$_{1-x}$Te also has the above-mentioned magnetic group symmetry when the vortex line
is parallel to the mirror plane of the system, the existence of which guarantees the topological protection of the surface Dirac cone.
Then, the mirror Chern number on this mirror plane leads to the $Z$ classification of the vortex core zero-energy bound states.
The existence of multiple Majorana zero-energy modes is allowed in this case.
This state is regarded as a topological crystalline superconducting vortex state.

\subsection{URu$_2$Si$_2$: Weyl superconductor}

The heavy fermion system URu$_2$Si$_2$ exhibits the superconducting
transition at $T_c=1.45$ K. 
According to thermal transport measurements, which detected the nodal structure of the gap function, a possible pairing state consistent with
the space group symmetry is the spin-singlet $d_{xz}+id_{yz}$ pairing state.\cite{URu2Si2-1}  (Also see Ref. \citen{kittaka})
The gap function of this state for small $|\bm{k}|$ is proportional to
$k_z(k_x+ik_y)$, which is regarded as the $p+ip$-wave symmetry multiplied by the $k_z$ factor.
This pairing state breaks time-reversal symmetry in the orbital degrees of freedom, and thus this material
is a promising candidate of a 3D chiral superconductor.
The gap function has point nodes for $k_x=k_y=0$ at the north and south poles of the Fermi surface, for which $k_z=k_0$ and $k_z=-k_0$,
respectively. Here $\pm k_0$ is the Fermi momentum at the $\bm{k}$-points with $k_x=k_y=0$, and for simplicity, we assume nearly spherical Fermi surface centered
around the $\Gamma$ point.
These points nodes are Weyl points, which possess monopole charge in the momentum space, and
thus this system is a Weyl
superconductor.\cite{sumiyoshi-berry-phase,URu2Si2-2,gos, goswami, kobayashi}
The basic properties of Weyl superconductors are reviewed in section \ref{sec:weyl}.
For $-k_0< k_z <0$ and $0<k_z<k_0$, the spin-singlet $d_{xz}+id_{yz}$ pairing state has 
the nonzero Chern number $C=2$ on the $xy$-plane for fixed value of $k_z$. Here the factor $2$ is due to spin degeneracy.
On the other hand, $C=0$ for $|k_z|>k_0$. This implies that the point-node of the gap function possesses the monopole charge $\pm 2$ in the momentum space.
Thus, the situation is similar to Weyl semimetals for which the Chern number on the 2D $\bm{k}$-plane between two Weyl points is nonzero.
For Weyl superconductors, it is expected that anomalous thermal Hall effect occurs because of the nonzero Chern number, as discussed in Sect.\ref{sec:thermal}. See section \ref{sec:weyl} for details.

The spin-singlet $d_{xz}+id_{yz}$ pairing gap has a line-node on the
$k_z=0$ plane, and correspondingly, it has a zero energy flat
band on the $(001)$ surface.\cite{gos, goswami, kobayashi}
However, its associated topological invariant is an accidental chiral
symmetry, and the surface flat band is fragile against the surface
misorientation.\cite{kobayashi} On other surfaces, 
topological surface arc associated with the coexistence of line and
point nodes exists.\cite{kobayashi}

Time-reversal symmetry breaking of the chiral pairing state of URu$_2$Si$_2$ was recently examined by the measurement of the Kerr effect.\cite{kapi} 
Furthermore, the recent measurement of the Nernst effect strongly supports the chiral pairing state of this material.\cite{URu2Si2-2,sumiyoshi-berry-phase}
The Nernst effect is a transverse thermoelectric  responses akin to the Hall effect, and a good probe for superconducting fluctuations. 
A colossal Nernst signal was observed for URu$_2$Si$_2$ above and close to the transition temperature $T_c$.\cite{URu2Si2-2} Surprisingly, this signal becomes larger for cleaner samples, which is inconsistent with
the prediction of conventional theories of superconducting fluctuations. Also, the strong enhancement of the Nernst signal appears in the temperature region where a vortex liquid is not realized. Thus, the dependence on the purity of samples is not due to pinning effect of vortices.
This remarkable experimental result can be understood by considering the chiral character of Cooper pair 
fluctuations.\cite{sumiyoshi-berry-phase}
Chiral superconducting fluctuations give rise to asymmetric skew-scattering of electrons, leading
to the giant Nernst effect, which is akin to the anomalous Hall effect.
This scenario successfully explains the above-mentioned experimental observation, confirming 
the realization of chiral Cooper pairs in URu$_2$Si$_2$.

\subsection{UPt$_3$: crystalline topological superconductor or Weyl superconductor}

%%%amend
UPt$_3$ is a heavy fermion superconductor with $T_c\sim 0.5 $ K.
The NMR measurement supports the realization of a spin-triplet pairing state.\cite{toh}
Under applied magnetic fields, three different superconducting phases appear.
The pairing symmetries of these phases are still controversial.
However, recent experimental studies strongly indicates that there are two promising candidates for the pairing state
in the low temperature and low field region, which is the so-called B phase.\cite{izawa,kerr}
One is expressed in terms of the $E_{1u}$ representation of the space group symmetry of this compound\cite{izawa} and the other one is the $E_{2u}$ representation.\cite{kerr}
Also, even within the $E_{1u}$ representation, there are two possible pairing states, 
the $E_{1u}$ planar $f$-wave state and the $E_{1u}$ chiral $f$-wave state.\cite{izawa2}
The $E_{1u}$ planar pairing state is characterized by the $\bm{d}$ vector of the form
$\bm{d}(\bm{k})\propto \lambda_a\bm{b}+\lambda_b\bm{c}$ where
$\bm{b}$ and $\bm{c}$ are the lattice vectors of the hexagonal lattice of UPt$_3$, and
$\lambda_{a(b)}=k_{a(b)}(5k_c^2-k^2)$ with $a$, $b$, and $c$ the axes of the hexagonal lattice.
Also, for higher magnetic fields parallel to the $c$-axis, the $\bm{d}$-vector rotates, and changes
to $\bm{d}(\bm{k})\propto \lambda_a\bm{b}+\lambda_b\bm{a}$ within the $B$ phase.
On the other hand, the $E_{1u}$ chiral pairing state and the $E_{2u}$ pairing state break the time-reversal symmetry, 
and their $\bm{d}$ vectors are, respectively, given by
$\bm{d}(\bm{k})\propto \bm{c}(k_a+ik_b) (5k_c^2-k^2)$, 
%On the other hand, the $\bm{d}$ vector of the $E_{2u}$ pairing state is given by
and $\bm{d}(\bm{k})\propto \bm{c}(k_a+ik_b)^2k_c$.
%which breaks time reversal symmetry.
%%%%%

We, first, consider the $E_{1u}$ planar $f$-wave state.
The gap function of this state has line and point nodes, which give rise to gapless excitations, and thus, it is not a bulk topological superconducting state in the usual sense. However, as elucidated by Tsutsumi et al.\cite{tsutsumi}, 
for fixed $k_b=0$,
the $E_{1u}$ pairing state has chiral symmetry $\Gamma$ defined by the product of the time-reversal symmetry operation $\mathcal{T}$,
the particle-hole symmetry operation $\mathcal{C}$, and the mirror symmetry operation with respect to $ac$-plane $\mathcal{M}_{ac}$: i.e. $\Gamma=\mathcal{T}\mathcal{C}\mathcal{M}_{ac}$ and the BdG Hamiltonian $\mathcal{H}(\bm{k})$ satisfies
$\{\Gamma,\mathcal{H}(\bm{k})\}=0$ for $k_b=0$.
Then, using this chiral symmetry, we can define a 1D winding number $w(k_c)$ for $k_b=0$
which characterizes the $Z$ non-triviality. For the $E_{1u}$ state, we have $w(k_c)=\pm 2$ for $k_c\leq k_F$, and zero
for other $k_c$.
That is, Majorana modes appear on a surface normal to the $a$-axis, with the energy dispersion $\sim \pm vk_b$.
These surface Majorana modes are protected by the combination of time-reversal symmetry,
particle-hole symmetry, and mirror symmetry, and thus the $E_{1u}$ pairing state is a crystalline topological superconducting state.

On the other hand, the $E_{1u}$ and $E_{2u}$ chiral pairing states
%the pairing state with 
%$\bm{d}(\bm{k})\propto \bm{c}(k_a+ik_b)^2k_c$ is a chiral $f$-wave pairing state 
break time-reversal symmetry.
Since there are point nodes of the gap at $k_a=k_b=0$, and the Fermi surface of UPt$_3$ is three dimensional,\cite{upt3-fermi} a single-particle excitation from these point nodes behaves as a Weyl fermion, as in the case of URu$_2$Si$_2$.
The point node of the $E_{1u}$ chiral state, $\bm{d}(\bm{k})\propto \bm{c}(k_a+ik_b) (5k_c^2-k^2)$, has the monopole charge $q_m=\pm 1$, while the $E_{2u}$ state, $\bm{d}(\bm{k})\propto \bm{c}(k_a+ik_b)^2k_c$, has $q_m=\pm 2$.
Then, the Chern number on the $k_a-k_b$ plane for fixed $k_c$ between the two point nodes of the $E_{1u}$ ($E_{2u}$) chiral state
is equal to $2$ ($4$). Here,  the factor $1$  ($2$) from the momentum dependence of $\bm{d}$-vector, and the other factor $2$ from the spin degeneracy.  Because of the nonzero Chern number, 
there are Majorana arcs on the surface BZ for surfaces parallel to the $c$-axis.
As in the case of URu$_2$Si$_2$, the anomalous thermal Hall effect, the Kerr effect, and the Nernst effect due to chirality fluctuation are expected to occur.

For the $E_{2u}$ state, there exists a line node on the $k_c$ plane,
which may result in a flat band in the $(001)$ surface.
A line node in an odd parity superconductor is, however, topologically
unstable.\cite{kobayashi-shiozaki}
Correspondingly, the flat band on the (001) surface is fragile
against surface misorientation and the surface Rashba spin-orbit
interaction.\cite{kobayashi}  

\subsection{Cd$_3$As$_2$: Superconducting Dirac semimetal}
Dirac semimetals are 3D materials
that possess gapless Dirac points in the bulk BZ, whose low-energy excitations are effectively
described as Dirac fermions.
Cd$_3$As$_2$ is one of recently discovered Dirac
semimetals,\cite{cd3as21, cd3as22, cd3as23, cd3as24, cd3as25, cd3as26,
cd3as27, cd3as28} 
in which a superconducting phase transition has been
reported.\cite{supercd3as21, supercd3as22, supercd3as23}
Because of the experimental observation of a zero bias conductance peak in the tunneling
conductance,\cite{supercd3as22} topological
superconductivity is expected.
Actually a recent theoretical analysis indicates that unique orbital
texture of Dirac points favors topological superconductivity with a
quartet of surface Majorana fermions.\cite{kobayashi-sato, hashimoto3}

\section{Non-Abelian Statistics}
\label{sec:non-abelian}

\subsection{Exchange operation of Majorana zero-energy modes}
\label{sec:exch}

One of the most intriguing and remarkable features of Majorana zero-energy modes in superconductors is that they obey non-Abelian statistics, which is significantly different from Fermi statistics and Bose statistics.\cite{moore-read,nayak-wilczek,read-green,ivanov}
This novel quantum statistics is a variant of anyon statistics which is realized in fractional quantum Hall effect states, and
basically realized in 2D systems.
However, we note that it is also possible to generalize the non-Abelian statistics to three dimensions, which will be discussed in the next subsections.
As is well known, the conventional quantum statistics is characterized by the change in the $U(1)$ phase of a many-body wave function arising from
the exchange of identical particles,
and hence the many-body state does not depend on the order of the exchange processes.
In other words, the exchange operations constitute an Abelian group.
In contrast, for non-Abelian statistics, exchange operations of particles are non-commutative, i.e.
a different order of exchange operations of particles leads to different many-body states.
As a matter of fact, the particle-exchange operations (or more precisely, braiding operations\cite{com1}) 
are described by non-Abelian unitary operators that act on the (topological) ground state space.
Originally, it was proposed by Read, Green, and Ivanov that a vortex with a Majorana zero mode in its core obeys
non-Abelian statistics.\cite{read-green,ivanov} 
However, we would like to stress that non-Abelian statistics is a more general property of Majorana zero modes in topological superconductors,
irrespective of whether they are realized in vortex cores or at open-boundary edges of a sample.

Let us consider $N$ Majorana zero-energy modes located at spatial positions denoted by $1$, $2$, $3$, ... , $N$ in a topological superconductor, whose fields are respectively given by
$\gamma_1$, $\gamma_2$, $\gamma_3$, ...,  $\gamma_N$.
Here, all Majorana fields satisfy,
\begin{eqnarray}
\gamma_i^2&=&1, \\
\gamma_i\gamma_j&=& -\gamma_j\gamma_i \quad\mbox{for} \quad i\neq j.
\end{eqnarray}
A key factor of the non-Abelian statistics of Majorana modes is the following rule for the exchange (braiding) of any pairs of two Majorana fields, $\gamma_i$
and $\gamma_j$.
\begin{eqnarray}
\gamma_i\rightarrow\gamma_j, \quad \gamma_j\rightarrow -\gamma_i.
\label{eq:rule1}
\end{eqnarray}
We will see in the next subsection that the exchange (braiding) rule (\ref{eq:rule1}) gives rise to the non-Abelian statistics of Majorana zero-energy modes.
Note that in Eq. (\ref{eq:rule1}) one Majorana field changes its sign, while the other does not.
This peculiar behavior is often understood as an effect of the phase-winding of a vortex where a Majorana fermion exists: i.e., 
vortices with Majorana modes accompany a branch cut at which the phase jumps by $2\pi$, and thus, the braiding of two vortices leads
to a phase change of $\pi$ in one of the Majorana fields when the Majorana fermion traverses the branch cut.\cite{ivanov,stone-chung}
However, as mentioned before, non-Abelian statistics is realized even for Majorana edge states in a nanowire, 
and Eq.(\ref{eq:rule1}) also holds in such cases without vortices.\cite{clarke,halperin-alicea}
To see this, let us consider the braiding of $\gamma_1$ and $\gamma_2$ more precisely.
To proceed with the argument, we assume that the topological ground state is separated from the first excited state by a finite energy gap,
and the exchange is carried out adiabatically within this ground state manifold. Then, the Majorana zero modes are
still Majorana zero modes after the exchange operation. This exchange operation is described by a unitary operator $U_{1,2}$.
We recall that all Majorana fields are expressed by linear combinations of fields of electrons with coefficients given by
the wave function of the BdG Hamiltonian.
Thus, according to the adiabatic theorem, after the braiding operation, $\gamma_{1(2)}$ is moved to $\gamma_{2(1)}$ with an additional phase factor 
denoted by $s_{2(1)}$.
Then, we obtain,
\begin{eqnarray}
s_2\gamma_2=U_{1,2}\gamma_1U_{1,2}^{\dagger}, \quad 
s_1\gamma_1=U_{1,2}\gamma_2U_{1,2}^{\dagger}.
\label{eq:u12}
\end{eqnarray}
Since $s_1^2\gamma_1^2=1$, and $\gamma_1^2=1$, we have $s_1=1$ or  $-1$. Similarly, $s_2=1$ or  $-1$.
Combining the Majorana zero modes $\gamma_1$ and $\gamma_2$, we can construct a complex fermion field 
$\psi_{12}=(\gamma_1+i\gamma_2)/2$.
The occupation number of the complex fermion $n_{12}=\psi^{\dagger}_{12}\psi_{12}$ is $1$ or $0$.
Thus, the state that consists of $\gamma_1$ and $\gamma_2$ only is doubly-degenerate.
More generally, if $N=2m$ with $m$ an integer, we can construct $m$ complex fermion fields from $2m$ Majorana fields,
and each complex fermion state is occupied or unoccupied, which leads to a total degeneracy $2^m$. 
This is so-called topological degeneracy.
We note that this degeneracy can not be lifted by local perturbations as long as the system is isolated. This property is important
in connection with the application to topological quantum computation, which we will mention later.
Since Majorana fields are given by the superposition of original electron fields, 
the total number of the complex fermions is equal to the number of quasiparticles
that arise from the breaking of Cooper pairs in a superconductor. 
This implies that the parity of the total occupation number must be preserved for an isolated system, since the destruction or creation of a Cooper pair changes
the number of quasiparticles only by $2$.
Hence, the degeneracy is reduced to $2^{m-1}$ for each parity sector.
Keeping this in mind, we, now, consider how the occupation number,
\begin{eqnarray}
n_{12}=\psi^{\dagger}_{12}\psi_{12}=\frac{1}{2}(1+i\gamma_1\gamma_2),
\label{eq:n12}
\end{eqnarray}
is affected by the exchange of $\gamma_1$ and $\gamma_2$. Operating $U_{1,2}$ on $n_{12}$, we obtain,
\begin{eqnarray}
U_{1,2}n_{12}U_{1,2}^{\dagger}=\frac{1}{2}+\frac{i}{2}U_{1,2}\gamma_1\gamma_2U_{1,2}^{\dagger}=\frac{1}{2}-\frac{i}{2}s_1s_2\gamma_1\gamma_2.
\label{eq:un12}
\end{eqnarray}
Here, we used Eq.(\ref{eq:u12}).
In the case of $N=2$, i.e. there are only two Majorana zero modes $\gamma_1$ and $\gamma_2$, the parity of the occupation number $n_{12}$ must not be changed by the exchange operation. Then, we have $s_1s_2=-1$. On the other hand, in the case of $N>2$, it is not so trivial whether or not the parity of $n_{12}$
is changed by $U_{12}$. However, if each Majorana zero modes are spatially well separated, 
the exchange of $\gamma_1$ and $\gamma_2$ can not affect the occupation numbers of complex fermion states composed of the other
Majorana zero modes, i.e.
$n_{ij}=\psi_{ij}^{\dagger}\psi_{ij}$ with $\psi_{ij}=(\gamma_i+i\gamma_j)/2$ for $i$, $j=3,4,5, ..., N$, and $i\neq j$.
%$\gamma_3$, $\gamma_4$, $\gamma_5$, ... , $\gamma_N$.
Otherwise, the exchange of $\gamma_1$ and $\gamma_2$ can affect fermion states infinitely far away from them! 
Thus, $n_{12}=U_{1,2}^{\dagger}n_{12}U_{1,2}$ holds, and we obtain $s_1s_2=-1$.
This leads to the rule (\ref{eq:rule1}). Depending on the gauge and the choice of left-handed or right-handed rotation, 
we have $s_1=1$, $s_2=-1$, or $s_1=-1$, $s_2=1$.
 Note that in the above argument, the presence or absence of vortices does not play any role, and hence
 Eq.(\ref{eq:rule1}) holds for any Majorana zero modes realized in topological superconductors.\cite{clarke,halperin-alicea}
 
 The unitary operator $U_{i,j}$ for the exchange operation of $\gamma_i$ and $\gamma_j$ can be conveniently 
 expressed as,\cite{ivanov}
 \begin{eqnarray}
 U_{i,j}=\exp\left(-\frac{\pi}{4}\gamma_i\gamma_j\right)=\frac{1}{\sqrt{2}}(1-\gamma_i\gamma_j).
 \label{eq:uij}
 \end{eqnarray} 
 It is easy to verify that Eq. (\ref{eq:uij}) leads to the following exchange rule:
 \begin{eqnarray}
\gamma_j=U_{i,j}\gamma_iU_{i,j}^{\dagger}, \quad 
-\gamma_i=U_{i,j}\gamma_jU_{i,j}^{\dagger}.
\label{eq:uij2}
\end{eqnarray}

\begin{figure}
\includegraphics[width=9cm]{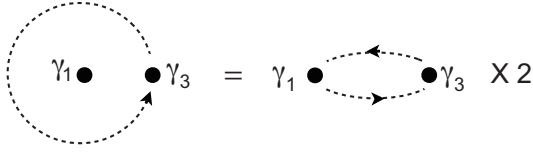}
\caption{Moving the Majorana fermion $\gamma_3$ around the Majorana fermion $\gamma_1$.
}
\label{fig:non-abe1}
\end{figure} 
 
\subsection{Non-Abelian statistics} 
\label{sec:non-abe}

As mentioned above, non-Abelian statistics is characterized by non-commutativity of the exchange operation of identical particles.\cite{moore-read,ivanov,read-green}
To demonstrate this, we consider a simple system that consists of four Majorana fermions, $\gamma_1$, $\gamma_2$,
$\gamma_3$, and $\gamma_4$. 
Let us see what happens when the Majorana fermion $\gamma_3$ is moved adiabatically around the Majorana fermion $\gamma_1$, and returns
to the initial position; i.e., the trajectory encircles $\gamma_1$, which amounts to twice the exchange operation of $\gamma_1$ and $\gamma_3$.(Fig.\ref{fig:non-abe1})
 An important observation is that the Majorana fields $\gamma_1$, $\gamma_2$, and
$\gamma_3$ constitute the Pauli matrices,
\begin{eqnarray}
-i\gamma_1\gamma_2=\sigma_z, \quad -i\gamma_2\gamma_3=\sigma_x, \quad -i\gamma_3\gamma_1=\sigma_y.
\end{eqnarray}
We can verify that $\sigma_{x,y,z}$ defined above actually satisfy the commutation relation for the Pauli matrices,
$[\sigma_{\mu}, \sigma_{\nu}]=2i\epsilon_{\mu\nu\lambda}\sigma_{\lambda}$ with $\mu,\nu,\lambda=x,y,z$.
It is also noted that from Eq. (\ref{eq:n12}), the occupation number of the complex fermion $\psi_{12}=(\gamma_1+i\gamma_2)/2$ 
is expressed by the eigenvalue of $\sigma_z$, i.e. $n_{12}=\frac{1}{2}(1-\sigma_z)$.
We denote the state vector with the occupation number $n_{12}$ as $| n_{12} \rangle$, which is also the eigenstate of $\sigma_z$; i.e.
$\sigma_z=1$ for $|0\rangle$ and $\sigma_z=-1$ for $|1\rangle$. 
Then, since exchanging $\gamma_1$ and $\gamma_3$ twice is expressed as $(U_{3,1})^2=e^{-\frac{\pi}{2}\gamma_3\gamma_1}=-\gamma_3\gamma_1=-i\sigma_y$,
the resulting states after the operation are
\begin{eqnarray}
(U_{3,1})^2|0\rangle =|1 \rangle, \quad
(U_{3,1})^2|1\rangle =-|0 \rangle.
\label{eq:u31-2}
\end{eqnarray}
Thus, this operation changes the parity of the occupation number of $\psi_{12}$; i.e.
an electron state $|1 \rangle $ is transformed into a hole state $| 0\rangle$ and vice versa. 
%by twice operation of the exchange of the Majorana fermions $\gamma_1$ and $\gamma_3$. 
In the next section, we will see that this property can be utilized for experimental detection of the non-Abelian statistics.
As mentioned before, the parity of the total number of quasiparticles should not change in the superconducting state.  
In fact, under the above operation, the parity of the occupation number $n_{34}$ of $\psi_{34}=(\gamma_3+i\gamma_4)/2$  
is also changed, and the total change of the parity cancels. 
The state composed of the four Majorana fermions $\gamma_1$, $\gamma_2$, $\gamma_3$, and $\gamma_4$ is completely specified by the occupation numbers $n_{12}$ and $n_{34}$.
We denote this state as $|n_{12}, n_{34}\rangle=(\psi^{\dagger}_{12})^{n_{12}}(\psi^{\dagger}_{34})^{n_{34}}|0,0\rangle$.
Since the parity of $n_{12}+n_{34}$ must be preserved, there are two sets of doubly-degenerate states, i.e.
$\{|1, 1\rangle,  |0, 0\rangle \}
$, and
$ \{|1, 0\rangle,  |0, 1\rangle \}
$. 
The exchange operation of any two of the four Majorana fermions acts on these doubly-degenerate states.
For instance, the exchange of $\gamma_1$ and $\gamma_3$ results in,
\begin{eqnarray}
U_{3,1}|1,1\rangle =\frac{1}{\sqrt{2}}(|1,1\rangle -|0,0\rangle), \label{eq:exch1}
\end{eqnarray}
\begin{eqnarray}
U_{3,1}|0,0\rangle =\frac{1}{\sqrt{2}}(|1,1\rangle +|0,0\rangle),\label{eq:exch2}
\end{eqnarray}
\begin{eqnarray}
U_{3,1}|1,0\rangle =\frac{1}{\sqrt{2}}(|1,0\rangle -|0,1\rangle), \label{eq:exch3}
\end{eqnarray}
\begin{eqnarray}
U_{3,1}|0,1\rangle =\frac{1}{\sqrt{2}}(|1,0\rangle +|0,1\rangle).\label{eq:exch4}
\end{eqnarray}
These transformation rules are obtained as follows.
We, first, note that from $\psi_{12}|0,0\rangle=\psi_{34}|0,0\rangle=0$, and $\psi_{12}|0,1\rangle=\psi_{34}|1,0\rangle=0$,
it follows that $\gamma_1|0,0\rangle=-i\gamma_2|0,0\rangle$, $\gamma_3|0,0\rangle=-i\gamma_4|0,0\rangle$, 
$\gamma_1|01\rangle=-i\gamma_2|0,1\rangle$, and $\gamma_3|1,0\rangle=-i\gamma_4|1,0\rangle$.
Then, we have
$|1, 1\rangle=\psi_{12}^{\dagger}\psi^{\dagger}_{34}|0,0\rangle=\gamma_1\gamma_3|0,0\rangle$,   $|0, 0\rangle=-\gamma_1\gamma_3|1,1\rangle$,
$|1, 0\rangle=\psi_{12}^{\dagger}|0,0\rangle=\gamma_1\gamma_3|0,1\rangle$, and 
$|0, 1\rangle=\psi^{\dagger}_{34}|0,0\rangle=-\gamma_1\gamma_3|1,0\rangle$.
Using these relations and Eq.(\ref{eq:uij}),
%$U_{13}^{\dagger}=(1+2\gamma_1\gamma_3)/\sqrt{2}$, 
we obtain Eqs.(\ref{eq:exch1})-(\ref{eq:exch4}).

As seen in Eqs.(\ref{eq:exch1})-(\ref{eq:exch4}), the exchange operation $U_{31}$  is the unitary transformation in the 2D degenerate spaces.
More generally, the exchange operations of Majorana fermions defined by (\ref{eq:uij}) are non-Abelian unitary transformation.
The non-commutativity  of $U_{ij}$ can be easily verified as
\begin{eqnarray}
U_{i,j}U_{j,k}-U_{j,k}U_{i,j}=-\gamma_i\gamma_k=i(2n_{ik}-1). \label{eq:noncom}
\end{eqnarray}
Since $n_{ik}=1$ or $0$, the operation of Eq.(\ref{eq:noncom})  on any state leads to a nonzero eigenvalue.
Thus, Majorana zero-energy states in superconductors obey non-Abelian statistics characterized by the noncommutativity of particle exchange.

%%%%added on 2016 Jan 4
There have been several proposals on how to realize the braiding operation of Majorana fermions.
We will discuss some of these ideas briefly at the end of Sect. \ref{sec:TQC}.
%%%%%%

%\subsection{Application to quantum computation}
  
%The unitary transformation generated by the exchange of Majorana zero modes discussed above can be utilized for building up quantum gates. This is an idea of topological quantum computation based on the manipulation of Majorana fermions.  
  
\subsection{Non-Abelian statistics in three dimensions}

In standard textbooks of quantum mechanics, it is explained that
in three spatial dimensions, exchanging the positions of two identical point-like particles twice results in the same state as the original one.
In other words, in three dimensions, the exchange of two particles is trivial, and only Bose or Fermi statistics is possible.
However, as explained in the section \ref{sec:exch}, the exchange rule (\ref{eq:rule1}), which is the basis of the non-Abelian statistics,
holds for Majorana zero modes in superconductors irrespective of spatial dimensions.
In fact, it was clarified by Teo and Kane that the non-Abelian statistics of Majorana zero modes is possible even in three-dimensional systems,
when there are point-like defects with Majorana zero modes in a superconductor.\cite{teo-kane1} 
They demonstrated that such a defect is realized in heterostructure systems such as the intersection of a vortex of an $s$-wave pairing and the interface between a topological insulator and
a trivial superconductor.  Suppose that there is a vortex line perpendicular to the interface between the superconductor and
the topological insulator. The vortex line is terminated at the interface, and the zero-energy Majorana mode appears at this end point of the vortex, behaving as a non-Abelian anyon.
An important point here is that the point defect is not really a point-like object, but is spatially extended. This is because of the texture structure composed 
of the spatially-varying order parameter of the superconductor and the band gap which spatially varies from a negative value in the topological insulator region to a positive value in the trivial superconductor region.
Because of such additional degrees of freedom, the exchange of two particles can be non-trivial, in contrast to conventional quantum statistics.
It was also elucidated by Freedman et al. that the non-Abelian statistics of spatially-extended defects with zero modes is associated with 
projective ribbon permutation statistics, where a texture accompanying a zero mode is intuitively regarded as a ribbon, at the open edge of which
there is a localized zero mode.\cite{freedman}
Also, the non-Abelian statistics in three dimensions is possible in three-dimensional network of nanowires in which Majorana zero modes appear
at boundaries between a topological sector and a trivial sector.

\subsection{Proposal for experimental detection of non-Abelian statistics}

There are several proposals for experimental schemes of detection of non-Abelian statistics.
A remarkable feature of non-Abelian statistics is the non-commutativity of particle exchanges.
Thus, an important issue is how one can detect the non-commutativity in experimentally observable quantities.
Most of these proposals utilize certain kinds of interferometers, in which the exchange of Majorana fermions can be achieved when
a Majorana edge mode travels encircling a vortex Majorana zero mode inside the bulk of a superconductor.\cite{law,akhmerov,fu-kane-majorana,butiker} (see Fig.\ref{fig:non-abe2})
The signature of the noncommutativity appears in transport properties of the interferometers.
A simple example is a class D topological superconductor attached to a single metallic lead (Fig.\ref{fig:non-abe2}).
A more sophisticated approach proposed by Grosfeld and Stern\cite{grosfeld} is to utilize the interplay between the Aharonov-Bohm effect and Aharonov-Casher effect.

\subsubsection{One-lead conductance measurement}
\label{sec:one-lead}

\begin{figure}
\includegraphics[width=9cm]{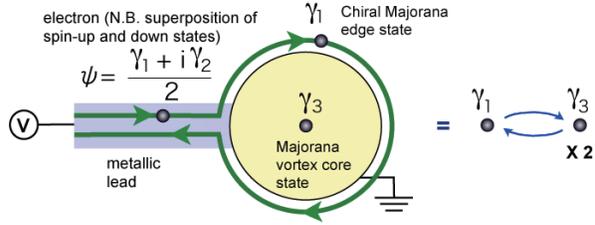}
\caption{(Color online) Setup of an interferometer for the detection of the non-Abelian statistics. A disk-shaped topological superconductor in class D is attached to a metallic lead. 
There is a vortex in the superconductor with a Majorana zero mode $\gamma_3$ in the core.
A chiral Majorana edge mode $\gamma_1$ propagates along the circumference of the disk,
encircling the Majorana fermion $\gamma_3$.
This setup realizes the braiding of the Majorana fermions $\gamma_1$ and $\gamma_3$.
}
\label{fig:non-abe2}
\end{figure}

We consider the setup shown in Fig. \ref{fig:non-abe2}, in which the bulk is a class D topological superconductor in two dimensions.\cite{law} 
For simplicity, we assume that 
there is only one chiral Majorana gapless mode on the 1D edge of the system.
There is a finite tunneling amplitude between
electrons (or holes) in the lead  and the gapless edge state.
The Hamiltonian for this tunneling process with amplitude $t$ is given by\cite{flensberg}
\begin{eqnarray}
H_{\rm tun}=i\frac{t}{2}\gamma(0)\sum_{\sigma}[c_{\sigma}(0)+c_{\sigma}^{\dagger}(0)], \label{eq:tun}
\end{eqnarray}
where $\gamma(0)$ is the Majorana field for the edge state at $\bm{r}=0$ where the lead is attached,
and $c_{\sigma}(0)$ and $c_{\dagger}^{\dagger}(0)$ are electron annihilation and creation operators with spin $\sigma$ at the same point, respectively.
Equation (\ref{eq:tun}) implies that only one of the real fields, $\gamma_1$, that constitute the complex field $\psi=(\gamma_1+i\gamma_2)/2$
couples with the Majorana edge state, and the other field $\gamma_2$ is decoupled.
Let us consider the situation that there is one vortex inside the bulk superconductor, which contains one Majorana zero mode.
We denote the Majorana field for the vortex core state as $\gamma_3$.
The tunneling between the lead and the superconductor induces the injection of $\gamma_1$ into the Majorana edge state. 
The injected Majorana fermion travels along the circumference of the superconductor encircling the Majorana zero mode $\gamma_3$ in the vortex core, and returns to the lead, constituting the electron (or hole) field $\psi^{\dagger}$ ($\psi$) with $\gamma_2$ again.
In this process, twice the exchange operation of $\gamma_1$ and $\gamma_3$ occurs, which is expressed by $(U_{31})^2$ as explained in the section \ref{sec:non-abe}. Then, from Eq.(\ref{eq:u31-2}), the occupation number of an electron $n=\psi^{\dagger}\psi$ in the lead is changed by
this process; i.e. the electron state in the lead is completely converted into the hole state and vice versa.
We can generalize this consideration to the case that there are multiple vortices inside the bulk superconductor.
If the number of vortices with a single Majorana zero mode in each vortex core is odd,  
the conversion between the electron state $|1\rangle$ and the hole state $|0\rangle$ occurs with probability of unity
after the above process.
Hence, we arrive at the noteworthy consequence that the conductance for the attached lead is quantized as $2e^2/h$ when
the number of enclosed vortices is odd.\cite{law}

The quantized conductance can be derived explicitly by using scattering theory.
Here, we follow a sophisticated argument presented by Li, Fleury, and B\"uttiker.\cite{butiker}
For simplicity, we consider the case that there is only one chiral Majorana zero mode on the edge of the superconductor 
and one channel of conduction electrons in the lead, and that there are also $n_v$ vortices with Majorana zero modes in the core in the bulk of the superconductor.
Hereafter, it is assumed that the vorticity of each vortex is unity.
$n_v$ must be odd, since the total number of Majorana fermions including the edge state and the vortex core states, which emerge
from the splitting of the original electrons into two parts, must be even.
Then, the scattering matrix for electrons and holes at the junction connects
the incoming state of electrons and holes $(\psi^{\dagger}_{\rm in},\psi_{\rm in} )$ and their outgoing state 
$(\psi^{\dagger}_{\rm out},\psi_{\rm out} )$:
\begin{eqnarray}
%\left(
\left(\begin{array}{c}
\psi_{\rm out} \\
\psi^{\dagger}_{\rm out}
\end{array}
\right)
=S
\left(\begin{array}{c}
\psi_{\rm in} \\
\psi^{\dagger}_{\rm in}
\end{array}
\right).
\end{eqnarray}
Here, spin indices are omitted.
In fact, for any realistic proposals to realize non-Abelian statistics in
class D topological superconductors, 
the spin-orbit interaction and the Zeeman effect split the Fermi surface into two parts, and only one of them is relevant
to the topological superconducting state.\cite{sato-fujimoto1,sato-takahashi-fujimoto1,sau,alicea} 
Thus, there is only one type of fermions which is the superposition of an up-spin state and a down-spin state.
$\psi$ and $\psi^{\dagger}$ can be expressed in terms of Majorana fields $\gamma_1$ and $\gamma_2$ via,
\begin{eqnarray}
\left(\begin{array}{c}
\psi \\
\psi^{\dagger}
\end{array}
\right)
=
\frac{1}{\sqrt{2}}
\left(\begin{array}{cc}
1 & i \\
1 & -i
\end{array}
\right)
\left(\begin{array}{c}
\gamma_1/\sqrt{2} \\
\gamma_2/\sqrt{2}
\end{array}
\right).
\end{eqnarray}
Thus, transforming the electron-hole basis into the Majorana basis, we obtain the scattering matrix for Majorana modes as,
\begin{eqnarray}
S_M=\frac{1}{2}
\left(\begin{array}{cc}
1 & 1 \\
-i & i
\end{array}
\right)S
\left(\begin{array}{cc}
1 & i \\
1 & -i
\end{array}
\right).
\label{eq:sm-s}
\end{eqnarray}
As mentioned above, only $\gamma_1$ couples to the chiral Majorana edge mode, leaving $\gamma_2$ unaffected.
Then, $S_M$ is expressed as
\begin{eqnarray}
S_M=
\left(
\begin{array}{cc}
r_{M1} & 0 \\
0 & 1
\end{array}\right),
\end{eqnarray}
where $r_{M1}$ is the reflection amplitude for the Majorana fermion $\gamma_1$.
$r_{M1}$ can be derived in the following manner.
At the junction between the lead and the superconductor, the Majorana state $\gamma_1$ tunnels into
the chiral Majorana edge state $\gamma$ with tunneling amplitude $\sqrt{1-r_0^2}$, where
$r_0$ is the bare reflection amplitude of $\gamma_1$ at the junction.
Traveling around the superconducting region, the Majorana state acquires
the phase change $\theta=n_v\pi+\pi+kL$ with $k$ the wave number of the chiral Majorana mode, and $L$ the circumference of 
the superconductor.
The second term of $\theta$, $\pi$, arises from
the Berry phase due to the rotation of the spin.
Note that for realistic proposals mentioned before, the spin-orbit interaction combined with the Zeeman effect gives rise to
a spin texture structure on the Fermi surface.
For instance, in the case of a superconductor with the Rashba spin-orbit interaction $\lambda\bm{\sigma}\cdot(k_y,-k_x,0)$, the direction of spin is in the $xy$-plane and perpendicular to
the propagating direction.
Hence, the motion of a Majorana fermion along the circumference accompanies the rotation
of the spin by $2\pi$, resulting in the Berry phase $\pi$.
After traveling around the edge, the Majorana mode tunnels again into the $\gamma_1$ state in the lead with probability $\sqrt{1-r_0^2}$.
Then,  the amplitude for this single-turn process is
$-(1-r_0^2)e^{i\theta}$, where the first minus sign arises from backward scattering.
In a similar manner, the amplitude for the Majorana mode moving around the circumference of the superconductor $n$-times with successive
tunneling into the lead is given by $-(1-r_0^2)r_0^{n-1}e^{in\theta}$.
Then, the total amplitude for the reflection is
\begin{eqnarray}
&&r_{M1}=r_0-(1-r_0^2)e^{i\theta}-(1-r_0^2)r_0e^{i2\theta}-(1-r_0^2)r_0^2e^{i3\theta}-\cdot\cdot\cdot \nonumber \\
&&\qquad =\frac{r_0-e^{i\theta}}{1-r_0e^{i\theta}}.\label{eq:rm1}
\end{eqnarray}
The scattering matrix in the electron-hole basis $S$ is obtained from Eq.(\ref{eq:sm-s}),
\begin{eqnarray}
S=\left(
\begin{array}{cc}
s_{ee} & s_{eh} \\
s_{he} & s_{hh}
\end{array}\right)=\frac{1}{2}
\left(
\begin{array}{cc}
r_{M1}+1 & r_{M1}-1 \\
r_{M1}-1 & r_{M1}+1
\end{array}\right).
\label{eq:s-rm}
\end{eqnarray}
The current due to the Andreev reflection at the junction in the zero temperature limit is given by
\begin{eqnarray}
I=\frac{2e}{h}\int^{eV}_0dE|s^{he}|^2,
\end{eqnarray}
\begin{eqnarray}
|s^{he}|^2=\frac{(r_0-1)^2}{2}\cdot \frac{1+\cos\theta}{1+r_0^2-2r_0\cos\theta},
\label{eq:she}
\end{eqnarray}
where $V$ is the bias-voltage applied to the lead, and $E=v_Mk$ with $v_M$ velocity of the chiral Majorana edge state. 
Note that when $n_v$ is odd and $\theta=2\pi\times\mbox{integer}$, $|s^{eh}|^2=1$ in the limit of $E\rightarrow0$, irrespective of the value of $r_0$,
as seen from Eq.(\ref{eq:she}).
Thus, we obtain the quantized conductance $G=dI/dV|_{V\rightarrow 0}=2e^2/h$.
On the other hand, when $n_v$ is even, i.e. $\theta=\pi\times\mbox{(odd integer)}$, $|s^{eh}|^2=0$, and hence, $G=0$ irrespective of $r_0$.
In fact, in the case of even $n_v$, there is no chiral Majorana zero mode on the edge of the superconductor,
because, as mentioned before, the total number of Majorana fermions including the edge state and the vortex core states, which emerge
from splitting of the original electrons into two parts, must be even.
The vanishing zero-bias conductance $G=0$ for even $n_v$ is consistent with this physical picture.
The quantized conductance $G=2e^2/h$ irrespective of the coupling strength between the lead and the superconductor in the case of odd $n_v$
is remarkable, characterizing the non-Abelian character that leads to the perfect conversion between an electron and a hole, i.e. Eq.(\ref{eq:u31-2}).
The above argument can be generalized to more realistic cases with multiple conducting channels in the lead.
In such cases, one of the Majorana fermion fields from electrons (or holes) in the lead couples to the chiral Majorana edge mode, if there is only one chiral edge mode, leading to the quantized conductance for this channel.
However, because of contributions from other channels that are not involved with perfect Andreev reflection mediated via a Majorana zero mode, 
the total conductance is not quantized generally. Thus, the observation of the non-Abelian character of Majorana fermions using the one-lead conductance measurement is rather difficult for realistic experimental setup.
We need more sophisticated interference experiments for the detection of the non-Abelian statistics, which will be discussed in the next 
subsection.

It is noted that the above calculation for odd $n_v$ can also be applicable to the case of a nanowire, for which a Majorana edge state is localized at
the junction between the normal lead and the superconducting wire.
In this case, $\theta=0$, and thus we obtain the quantized conductance $G=2e^2/h$.
However, underlying physics for the origin of the quantized conductance here is slightly different from that in the case of 2D systems.
In the 2D cases considered above, the quantized conductance is a result of the braiding of Majorana fermions, featuring the non-Abelian character,
while in the case of nanowires, it is due to the perfect conversion of electrons and holes caused by the Andreev reflection mediated via a Majorana zero mode.

\subsubsection{Aharonov-Bohm effect and Aharonov-Casher effet}

\begin{figure}
\includegraphics[width=8.5cm]{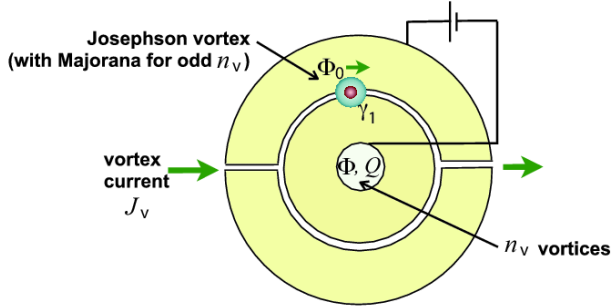}
\caption{(Color online) Setup for the detection of the non-Abelian statistics based on the AC effect.
The system consists of three part of topological superconductors, which are connected  via Josephson junction regions.
The Josephson vortices with Majorana zero modes can propagate along the junction, encircling 
the charge $Q$ at the center of the system, which leads to the AC effect.
}
\label{fig:non-abe3}
\end{figure}

As seen in the previous section, the conductance does not depend on the number of vortices $n_v$ encircled by the trajectory of the propagating chiral Majorana edge mode when $n_v$ is odd.
This implies that in the case of odd $n_v$, there is no Aharonov-Bohm (AB) effect with respect to the magnetic flux threaded in the superconductor.
We note that this remarkable feature is not due to the charge neutrality of Majorana fermions.  In fact,
there is a sign change of a Majorana field, when it encircles a single vortex, as can be seen from the expression for a Majorana field in  terms of electron fields:
\begin{eqnarray}
\gamma_i=\sum_{\sigma}\int d\bm{r} [u_{0\sigma i}(\bm{r})e^{i\frac{\phi}{2}}c_{\sigma}^{\dagger}(\bm{r})+v_{0\sigma i}(\bm{r})e^{-i\frac{\phi}{2}}c_{\sigma}(\bm{r})],
\label{eq:mf-f}
\end{eqnarray}
%eq.(\ref{eq:mf-f}).
where $\phi$ is the phase of the superconducting gap.
Instead, however, the absence of the AB effect is understood as a result 
of the non-commutativity of Majorana fermions, i.e. the non-Abelian character.\cite{stern,bonderson}
When a Majorana fermion travels along the surface edge of the superconductor, the braiding of the Majorana edge mode and a Majorana bound state
in a vortex core in the superconductor occurs, and the Majorana edge state acquires a phase. 
Another incoming quasiparticle gives rise to another braiding. However, this does not commute with the previous one because of Eq.(\ref{eq:noncom}), and hence
the resulting state cannot be an eigenstate of both of these two braiding operators. This results in dephasing of the interference.
Thus, the absence of the AB effect signifies non-Abelian statistics.
A scheme utilizing this property for the detection of non-Abelian statistics was proposed by Grosfeld and Stern.\cite{grosfeld}
They considered Josephson-coupled superconductors threaded by a magnetic flux, in which there is a Josephson vortex at the junction, and furthermore
an electric charge $Q$ is placed in a hole of the superconductor as well as the magnetic flux $\Phi$ (see Fig.\ref{fig:non-abe3}). 
A key idea is to use the Aharonov-Casher (AC) effect associated the Josephson vortex  which carries a Majorana zero mode. 
The AC effect is realized by changing the role of a charged particle and a magnetic flux in the AB effect; i.e., a quantum mechanical particle with a magnetic dipole moving around a charge flux (an electric field) acquires a phase due to the gauge field associated with the electric field.
Thus, the vortex current $J_v$ carried by the conventional Josephson vortex without a Majorana fermion exhibits
a periodic dependence on the charge $Q$ in the hole of the system, $J_v\sim J_{v0}+J_{v1}\cos(2\pi \frac{Q}{2e})$. 
However, in the setup considered here, the Josephson vortex harbors a Majorana zero mode, and hence
the AC effect of the moving Josephson vortex disappears when the number of vortices in
the center hole $n_v$ is odd; i.e., there are an odd number of Majorana fermions in the inner part of the junction system encircled by the trajectory of the Josephson vortex.
The absence of the AC effect is due to the above-mentioned mechanism of dephasing arising from the noncommutativity of Majorana fermions in the Josephson vortex and in the inner part
of the junction.

\section{"Fractionalization"-- $4\pi$-periodic Josephson Effect}
\label{sec:4pi-josephson}

\begin{figure}
\includegraphics[width=7cm]{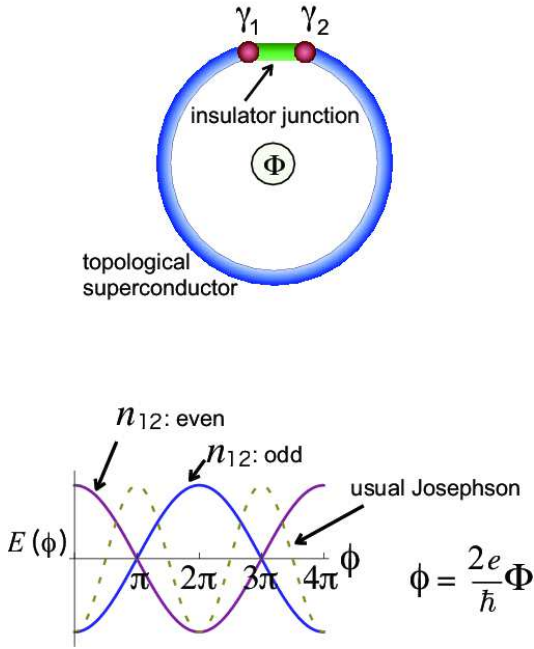}
\caption{(Color online) (Upper panel) Setup for the $4\pi$-periodic Josephson effect. Majorana zero modes $\gamma_1$ and $\gamma_2$ exist
at the edges of the topological superconducting wire, which are connected via an insulator, constituting the Josephson junction. The flux $\Phi$ penetrates the ring composed of the superconducting wire and the insulating junction. (Lower panel) Energy as a function of $\Phi$ for the $4\pi$-periodic Josephson junction.
}
\label{fig:frac-jose}
\end{figure}

As seen in previous sections, a Majorana fermion field of superconductors emerges from separating the real and imaginary parts of a complex fermion field of an electron.
In this sense, an electron fractionalizes into two Majorana fermions.
It should be noted that this is not fractionalization arising from intrinsic topological order which is realized in the fractional quantum Hall effect states,
and leads to fractionalized quasiparticles.
In fact, there is no intrinsic topological order in topological superconductors.
However, the "fractionalization" into Majorana fermions gives rise to some interesting electromagnetic properties.
One of the most remarkable phenomena is the $4\pi$-periodic Josephson 
effect.\cite{kitaev-majorana-chain,yakovenko,jiang}
To explain this, let us consider the Josephson junction system with a ring geometry shown in Fig.\ref{fig:frac-jose}, where a magnetic flux $\Phi$ threads the hole of the ring.
This is a superconductor-insulator-superconductor junction, and the phase difference between the two junction separated via the small insulating region
is $\phi=\frac{2e}{\hbar}\Phi$.
Then, the single-electron tunneling Hamiltonian is
$H_{1t}=-te^{i\phi}\psi_{\sigma 1}^{\dagger}\psi_{\sigma 2}+h.c.$, where $\psi_{\sigma 1}$ and $\psi_{\sigma 2}$ are electron fields at two junctions.
Also, in the case with Majorana zero-energy modes, the mode expansion of the electron fields is given by
\begin{eqnarray}
\psi_{\sigma i}=u_{\sigma i}(x)\gamma_i+\mbox{(non-zero energy modes)},
\label{eq:psimode0}
\end{eqnarray}
with $i=1$, $2$.
Thus, the single-electron tunneling Hamiltonian reads,
\begin{eqnarray}
H_{1t}=J_Mi\gamma_1\gamma_2\cos \frac{\phi}{2},
\label{eq:h1t}
\end{eqnarray}
where $J_M$ is a real constant.
In addition, there is also a usual Josephson coupling term for Cooper pair tunneling, which is expressed as $H_{2t}=J\cos\phi$ with $J$
the Josephson tunneling amplitude.
Since the system is isolated, the total charge number is conserved, and hence, it
is an eigenstate of the fermion parity given by Eq.(\ref{eq:n12}), i.e., $i\gamma_1\gamma_2=1$ or $-1$.
As seen from Eq.(\ref{eq:h1t}), this eigenstate of the parity exhibits $4\pi$-periodicity of the Josephson current as
a function of $\phi$. This implies that the Josephson current is not carried by Cooper pairs with charge $2e$, but rather by
particles with charge $e$; i.e., the "fractionalization" of Cooper pairs occurs. 
%\begin{eqnarray}
%\gamma_i=\sum_{\sigma}\int d\bm{r} [u_{0\sigma i}(\bm{r})e^{i\frac{\phi}{2}}c_{\sigma}^{\dagger}(\bm{r})+v_{0\sigma i}(\bm{r})e^{-i\frac{\phi}{2}}c_{\sigma}(\bm{r})]
%\label{eq:mf-f}
%\end{eqnarray}

%\begin{eqnarray}
%H=J_Mi\gamma_1\gamma_2\cos \frac{\Phi}{2}+J_0\cos \Phi
%\end{eqnarray}

The experimental exploration of the $4\pi$-periodic Josephson effect was achieved by Rokhinson et al.\cite{rokhinson}
They used a quasi-1D nanowire with proximity-induced superconductivity for the realization of a topological superconductor.
In this case, there are two additional Majorana zero-energy end states at the two open edges as well as those at the junction.
Thus, the fermion-parity at the junction is  not conserved, but the mutual exchange of the fermion parity between the junction and
the open edges  occurs.
In this situation, the transition between the two parity-eigenstates occurs, and the periodicity of the Josephson current is changed to $2\pi$,
which is the same as that in the conventional Josephson effect.
However, as clarified by San-Jose, Prada, and Aguado,\cite{sanjose} the $4\pi$-periodicity is partially recovered in the case of the ac Josephson effect,
for which nonadiabatic transitions between states with the same fermion-parity are allowed.
The measurement of Shapiro steps, which demonstrates the ac Josephson effect, was carried out by Rokhinson's group\cite{rokhinson} 
and double the height of the Shapiro steps was observed. This observation may be a signature of the $4\pi$-periodic Josephson effect. 
However, to this date, no other experimental groups have succeeded in confirming this observation.

It is noted that the "fractionalization" character of Majorana fermions is also detectable via thermodynamic measurements of topological entanglement entropy. For details, see, e.g., Refs. \citen{cooper-stern,EE1,EE2}.

\section{Non-local Correlation Between Majorana Fermions and Teleportation}
\label{sec:non-local}

\begin{figure}
\includegraphics[width=7cm]{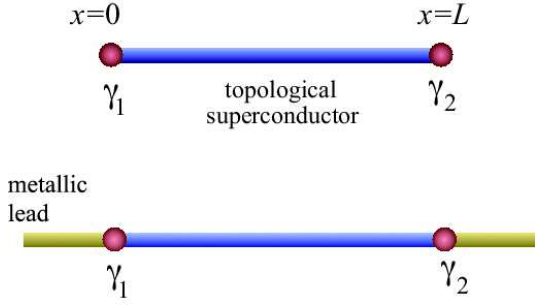}
\caption{(Color online) 1D topoogical superconductor with Majorana fermions $\gamma_1$ and $\gamma_2$ at 
the edges $x=0$ and $L$, respectively. Metallic leads are attached to the open edges in the lower panel.
}
\label{fig:non-local}
\end{figure}

As mentioned before, we can interpret the emergence of a Majorana fermion in a superconductor as the splitting of a complex fermionic field of an electron into real and imaginary parts. From this viewpoint, one may expect a non-local correlation between two spatially separated Majorana fermions.
This is indeed the case for certain situations.\cite{ bolech, tewari,semenoff,nilsson,fu-majorana}
To deal with this issue, we consider a simple toy model of a 1D topological superconductor with a Majorana zero-energy mode localized at
the open boundaries of the system. (See Fig.\ref{fig:non-local})
We denote Majorana fields for these edge states at $x=0$ and $L$ as $\gamma_1$ and $\gamma_2$, respectively.
For this system, the mode expansion of an electron operator with spin $\sigma$ at $x$, $\psi_{\sigma}(x)$, is given by
\begin{eqnarray}
\psi_{\sigma}(x)=\sum_{i=1,2} u_{\sigma i}(x)\gamma_i+\mbox{(non-zero energy modes)}.
\label{eq:psimode}
\end{eqnarray}
The non-local correlation can be immediately seen from the long-distance behaviors of the correlation function $\langle\psi_{\sigma}(x)\psi_{\sigma}^{\dagger}(y)\rangle$.  
Since $u_{\sigma 1(2)}$ is localized around $x=0$ $(L)$, and there are no low-energy excitations other than the Majorana edge states, we have, 
for $x\sim 0$ and $y\sim L$,\cite{tewari,semenoff,nilsson}
\begin{eqnarray}
\langle\psi_{\sigma}(x)\psi_{\sigma}^{\dagger}(y)\rangle \sim u_{\sigma 1}(x)u_{\sigma 2}^{*}(y)\langle \gamma_1\gamma_2 \rangle.
\label{eq:psi-cor}
\end{eqnarray}
As mentioned in the previous sections, the topological superconducting phase is the eigenstate of the parity operator
 $i\gamma_1\gamma_2=2n-1$ with $n_{12}=\psi_{12}^{\dagger}\psi_{12}$,
 as long as the system is isolated.
Thus, the correlation function (\ref{eq:psi-cor}) is non-zero even for $|x-y|\rightarrow \infty$, which signifies non-local correlation resulting from
the existence of Majorana zero modes.
This non-local correlation affects the tunneling probability of electrons mediated via Majorana zero modes in an interesting way.
Let us consider electron tunneling from one end at $x=0$ to the other end at $x=L$.
The tunneling amplitude is proportional to the correlation function (\ref{eq:psi-cor}), which is 
independent of the distance between the two edges $L$.
This remarkable $L$-independent behavior of the tunneling  is referred to as "teleportation".
However, the experimental detection of this "teleportation" is a tricky issue, because of the following reason.
To distinguish whether an electron detected at $x=L$ comes from the opposite edge at $x=0$ via tunneling or is created from
breaking up a Cooper pair in the superconductor, we need to detect the parity state of $\psi_{12}$ simultaneously.
Also, we note that this "teleportation" does not break causality, because the information of the one end state, say at $x=0$, must be transferred
to the observer at $x=L$ in a classical way to detect the above-mentioned tunneling process.
Furthermore, to probe the Majorana end state, one needs to couple the boundary edge with a lead that results 
in the failure of parity conservation.\cite{budich}  Then, the systems is no longer in the eigenstate of $\gamma_1\gamma_2$, and the average of this operator leads
to the vanishing of the correlation Eq.(\ref{eq:psi-cor}).\cite{bolech}
Concerning the last point, however, the situation is changed, when there is a finite overlap of the wave functions of two Majorana edge modes, 
and the topological degeneracy associated with the fermion number parity $i\gamma_1\gamma_2=\pm 1$ is lifted.
In this case, the value of $\langle \gamma_1\gamma_2\rangle$ is determined by the lowest of the two states with $n_{12}=0$ and $1$.
If the energy splitting between these levels is sufficiently larger than the typical excitation energy of electrons in the attached lead,
the transition between the states with $n_{12}=0$ and $1$ is suppressed, and the non-local correlation Eq.(\ref{eq:psi-cor}) survives.\cite{nilsson}
We note that the correlation Eq.(\ref{eq:psi-cor}) is still independent of the distance between the two edges $L$, though the overlap between the two Majorana end states depends on $L$.

The non-local correlation may be detected by tunneling experiments. The Hamiltonian for tunneling processes between 
the Majorana end state of the 1D topological superconductor and external leads is expressed by,\cite{bolech,flensberg}
\begin{eqnarray}
H_T&=&\sum_{k,\sigma}\int dx[t_k(x)c^{\dagger}_{k\sigma}\psi_{\sigma}(x)+h.c.] \nonumber \\
&=&\sum_{k,\sigma}\sum_{i=1,2}[V_{k\sigma i}c^{\dagger}_{k\sigma}-V^{*}_{k\sigma i}c_{k\sigma}]\gamma_i, 
\label{eq:tun-g}
\end{eqnarray}
where $c_{k\sigma}$ ($c_{k\sigma}^{\dagger}$) is an annihilation (creation) operator for an electron in the leads, 
$V_{k\sigma i}=\int dx t_{k}(x)u_{\sigma i}(x)$, and $t_k(x)$ is the tunneling amplitude.
We have used Eq.(\ref{eq:psimode}) to derive Eq.(\ref{eq:tun-g}).
$V_{k\sigma 1}$ and  $V_{k\sigma 2}$ are the effective tunneling amplitudes between electrons in the leads and Majorana fermions in the superconductor at $x=0$  and $x=L$, respectively. 
Using $\psi_{12}=(\gamma_1+i\gamma_2)/2$, we rewrite the above equation in the form,
\begin{eqnarray}
H_T&=&\sum_{k,\sigma}[V_{k\sigma}^{(-)}c^{\dagger}_{k\sigma}\psi_{12}+V_{k\sigma}^{(-)*}\psi_{12}^{\dagger}c_{k\sigma} \nonumber \\
&&+V_{k\sigma}^{(+)}c^{\dagger}_{k\sigma}\psi_{12}^{\dagger}+V_{k\sigma}^{(+)*}\psi_{12}c_{k\sigma}],
\label{eq:tunneling2}
\end{eqnarray}
where 
$V_{k\sigma}^{(\pm)}=V_{k\sigma 1}\pm i V_{k\sigma 2}$.
The first line of the right-hand side of Eq.(\ref{eq:tunneling2}) corresponds to the normal tunneling processes between electrons in the leads and
the non-local fermion $\psi_{12}$.
On the other hand, the second line of (\ref{eq:tunneling2}) expresses the Andreev scattering processes mediated via the non-local fermion. 
In the case that the two Majorana fermions are well separated, and the the level splitting $E_M$ of the two Majorana modes induced by finite overlapping
is  sufficiently smaller than bias voltage $V$ applied to the lead, local Andreev reflection at each junction dominates, and the non-local correlation does not
appear.\cite{bolech} 
In fact, by integrating out the $\psi_{12}$ fermion, we obtain a term of the form $(V_{k\sigma 1}V_{-k\sigma'1}+V_{k\sigma 2}V_{-k\sigma'2})c^{\dagger}_{k\sigma}c^{\dagger}_{-k\sigma'}$ which corresponds to the local Andreev reflection, and dominates  
the crossed Andreev reflection term of the form $V_{k_{\sigma}1}V_{-k\sigma' 2}c_{k\sigma}^{\dagger}c^{\dagger}_{-k\sigma'}$.
However, in the case of the opposite situation, i.e. $E_M \gg V$, crossed Andreev reflection mediated via $\psi_{12}$ occurs;
an electron injected to one end of the superconducting wire is converted to a hole emitted from the other end, characterizing
the non-local correlation.\cite{nilsson}

\begin{figure}
\includegraphics[width=8cm]{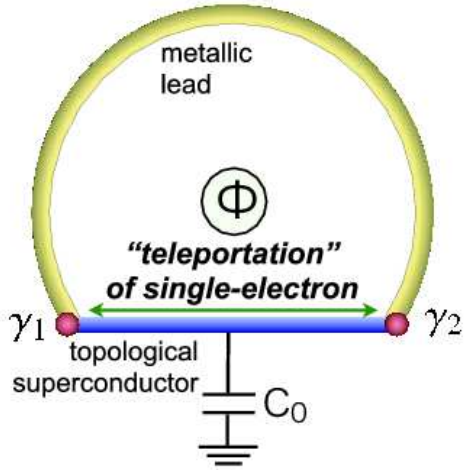}
\caption{(Color online) Setup for "teleportation" of electrons via Majorana fermions. The ring-shaped system consists of 
a topological superconductor nanowire and a metallic lead. Flux $\Phi$ penetrates the ring.
There are Majorana fermions $\gamma_1$ and $\gamma_2$ at the edge of the topological superconductor.
}
\label{fig:tele}
\end{figure}

A more remarkable non-local correlation effect appears in a mesoscopic topological superconductor in which the charging energy effect is not negligible.
This sophisticated idea was first proposed and examined by Fu.\cite{fu-majorana,egger1,egger2} An important point of this idea is that 
for sufficiently large Coulomb energy, the charge conservation induces
the phase fluctuation of the superconducting gap, which suppresses the local Andreev reflection mentioned above. 
%We, here, present a bit modified version of this argument.
Let us consider a 1D mesoscopic-size superconductor which is not grounded, but has a capacitor inserted between the system and the ground.
Note that the zero energy states with $n_{12}=0$ and $n_{12}=1$ are still degenerate, 
though the total number of Cooper pairs $_{\rm pair}$ are constrained by the charging energy.
Thus, for sufficiently large charging energy, and with a suitable bias voltage, which controls
the total number of electrons in the superconductor, 
the lowest energy states are doubly degenerate corresponding to $n_{12}=0$ and $1$.
There are two possibilities for the doubly-degenerate states; i.e.,
(i) $(N_{\rm pair}, n_{12})=(N,0)$ and $(N,1)$, 
(ii)  $(N_{\rm pair},n_{12})=(N-1,1)$, and $(N,0)$. 
In the case (i), the number of Cooper pairs is not changed, while, in the case (ii), the change of $n_{12}$ 
from $0$ to $1$ accompanies
breaking of a Cooper pair.
To take into account the phase fluctuation explicitly, we include the phase operator $\phi$ in the mode expansion of an electron field, 
Eq. (\ref{eq:psimode}),
\begin{eqnarray}
\psi_{\sigma}(x)=\sum_{i=1,2} u_{\sigma i}(x)\gamma_ie^{-i\frac{\phi}{2}}+\mbox{(non-zero energy modes)},
\label{eq:psimode2}
\end{eqnarray}
Then, for the case (i), the tunneling Hamiltonian between the normal metal leads and the superconductor, 
which couple at $x=0$ and $L$
is,\cite{egger1}
\begin{eqnarray}
H_T=\sum_{k,\sigma}[V_{k\sigma 1}c^{\dagger}_{k\sigma}\psi_{12}+iV_{k\sigma 2}^{*}\psi_{12}^{\dagger}c_{k\sigma}+h.c.]
%H_T=\sum_{k,\sigma}[V^{(-)}_{k\sigma}c^{\dagger}_{k\sigma}\psi_{12}+V_{k\sigma}^{(-)*}\psi_{12}^{\dagger}c_{k\sigma}].
\label{eq:tun-tele}
\end{eqnarray}
To derive Eq.(\ref{eq:tun-tele}), we have applied a gauge transformation to the $\psi_{12}$ field.
Equation (\ref{eq:tun-tele}) implies that the single-electron tunneling from terminal 1 to terminal 2 mediated via the non-local fermion $\psi_{12}$ occurs, in spite of the absence of single-particle excitations in the bulk of the
superconductor at sufficiently low temperatures. 
Remarkably, this tunneling process is independent of the length of the superconductor.
It may be legitimate to call this process "teleportation" of single electron mediated via Majorana fermions.
This effect is a result of the non-local correlation of Majorana fermions $\gamma_1$ and $\gamma_2$.
On the other hand, for case (ii), we have,\cite{egger1}
\begin{eqnarray}
H_T=\sum_{k,\sigma}[V_{k\sigma 1}c^{\dagger}_{k\sigma}\psi_{12}^{\dagger}e^{-i\phi}+iV_{k\sigma 2}^{*}\psi_{12}c_{k\sigma}e^{i\phi}+h.c.].
%H_T=\sum_{k,\sigma}[V_{k\sigma}c^{\dagger}_{k\sigma}\psi_{12}^{\dagger}e^{i\phi}+V_{k\sigma}^{*}\psi_{12}c_{k\sigma}e^{-i\phi}].
\label{eq:tun2}
\end{eqnarray}
Since $e^{i\phi}$ ($e^{-i\phi}$) is the raising (lowering) operator for the number of Cooper pairs,
the first (second) term of Eq.(\ref{eq:tun2}) creates (annihilates) the $\psi_{12}$ fermion accompanying 
the breaking (formation) of Cooper pairs, preserving the total electron charge.
Equation (\ref{eq:tun2}) also implies the single-electron tunneling from terminal 1 to terminal 2 mediated
via $\psi_{12}$, causing "teleportation" of electrons.
As a matter of fact, after integrating over the $\psi_{12}$ field in Eq.(\ref{eq:tun-tele}) or (\ref{eq:tun2}), we obtain
a term of the form $V_{k\sigma 1}V_{k'\sigma' 2}^{*}c_{k\sigma}^{\dagger}c_{k'\sigma'}$, and there is no
Andreev scattering term. 
The above results can be understood as follows.
The phase fluctuation suppresses the Andreev reflection mediated via Majorana bound states at open edges, 
in contrast to the case without charging energy expressed by Eq.(\ref{eq:tunneling2}). 
As a result, only normal single-particle tunneling processes survive because of the charge conservation, leading to the tunneling between
electrons at the two open edges mediated via the Majorana modes. 
%This implies that electrons in the normal metal leads coupled with Majorana modes at $x=0$ via $V_{k\sigma 1}$ and at $x=L$ via $V_{k\sigma2}$ tunnel through the superconductor. This tunneling is mediated by the $f$ fermion, which is akin to resonant tunneling. But an important difference is
%that the $f$ fermion, which is independent of spatial coordinates and time is not a point particle, but an spatially extended object spreading over the superconducting region. This feature is a result of non-local correlation of Majorana fermions at $x=0$ and $L$.
%Since the tunneling amplitude does not depend on the length of the system $L$ explicitly, we can call this electron transfer process as "teleportation".
We stress that this "teleportation" does not break causality because of the above-mentioned reason.
This effect can be experimentally detected as the AB effect.
Let us consider the setup shown in Fig.\ref{fig:tele}, 
which consists of a topological superconducting wire with Majorana zero modes at boundary edges
attached to a normal metal lead, constituting a ring geometry threaded by a magnetic flux.
Because of the single-electron resonant tunneling mediated via Majorana modes, at sufficiently low temperatures for which
the quantum coherence of electrons in the normal metal lead is retained, 
the AB effect with the periodicity $2\phi_0$ with $\phi_0=h/2e$ appears.
This is in strong contrast with the case without Majorana zero modes, for which the AB effect is absent, since
the single-electron coherence is suppressed by the superconducting gap, and the coherence of Cooper pairs does not survive in the normal metal lead.

The above result also implies that the charge transport of the topological superconducting nanowire
via the two Majorana end states leads to the quantized conductance $G=e^2/h$, which is a half of 
the conductance in the case of the Andreev scattering mediated via the Majorana zero mode discussed in Sect.\ref{sec:one-lead}. This is due to the suppression of the Andreev scattering at the two ends as mentioned above.
The conductance $G=e^2/h$ also characterizes the "teleportation" of an electron through the superconducting nanowire
via the Majorana end states.\cite{fu-majorana,egger1}

We close this section by mentioning that non-local correlation also appears as the entanglement of spin states of Majorana fermions.\cite{flensberg-spin,pascal-spin}
For instance, for a topological superconducting wire in the class D, 
the spin density of a Majorana end state is correlated with that at the other open end. 
The detection of the spin correlation at two open ends may signify the existence of Majorana fermions.

\section{Thermal Responses of Majorana Fermions}
\label{sec:thermal}

In some cases,
topological invariants characterizing topological phases can appear in experimentally observable quantities.
A famous and well-established example is the quantum Hall effect in a 2D electron gas under a strong magnetic field, for which
the Hall conductivity is expressed by the first Chern number (the so-called TKNN number).
An analogous effect is possible in topological superconductors.
However, since charge is not conserved in superconducting states, the Hall conductivity is not quantized.
Instead, energy conservation still holds, which implies that thermal transport can be an effective probe for bulk topological invariants.
For topological superconductors in which Majorana fermions are only low-energy excitations,
Majorana fermions carry the heat current, resulting in the quantum thermal Hall effect.
In this section, we discuss this phenomenon in the case of chiral superconductors with broken time-reversal symmetry 
in two dimensions (class C and D) and time-reversal-symmetric topological 
superconductors in three dimensions (class DIII).

\subsection{Case of chiral superconductors}

In the case of chiral superconductors, there are chiral gapless edge states, which break time-reversal symmetry by carrying
spontaneous heat currents without external magnetic fields.
The spontaneous edge currents result in the thermal Hall effect without magnetic fields in chiral superconductors.
This effect was first discussed by Read and Green,\cite{read-green} who exploited the edge theory of class D superconductors in two dimensions.
The chiral gapless edge state of this system, which propagates only in one direction along the open boundary edge, 
is a 1D free Majorana fermion, which, via the Jordan-Wigner transformation, is equivalent to the 1+1D Ising model.
Thus, this gapless edge state is described by 1+1D conformal field theory (CFT) of the Ising universality class with the central charge $c=1/2$.
Since the specific heat of this CFT is given by $c\frac{\pi k_{\rm B}^2T}{6\hbar}$, % (in the unit of $k_{\rm B}$=1, $\hbar=1$), 
and the velocity of the Majorana edge state and its density of states cancel with each other,
the thermal Hall conductivity is given by  $\kappa=c\frac{\pi k_{\rm B}^2T}{6\hbar}$. 
%Based on this effective field theory, Read and Green obtained the formula for the thermal Hall conductivity $\kappa=c\frac{\pi T}{6}$ (in the unit of $k_{\rm B}$=1, $\hbar=1$).
It is noted that this expression is the same as that for the integer quantum Hall effect, but with $c=1$.
In the case of the integer quantum Hall effect, the edge state is a chiral 1D free electron system which corresponds to the Gaussian universality class of CFT
with the central charge $c=1$. 
In contrast, a Majorana fermion has half the degree of freedom of a complex fermion field of an electron, and hence the central charge is equal to $c=1/2$.

One can also derive 
the thermal Hall conductivity from the Kubo formula for bulk heat current correlation functions.\cite{sumiyoshi}
In fact, the bulk topological invariant, i.e. the first Chern number, can be directly related to the Hall conductivity 
via the expression for the current-current correlation function.
Furthermore, although the derivation based on the CFT analysis can not determine the sign of the thermal Hall conductivity,
the bulk calculation based on realistic band structures enables us to obtain the complete expression including its sign.
However, in contrast to the case of charge transport, the calculation of the thermal transport coefficient based on the Kubo formula 
is much more involved because of the following reason.
First, we need to introduce a mechanical force that couples to heat currents to apply linear response theory, since a temperature gradient 
that practically induces thermal transport cannot be included in microscopic Hamiltonian as a potential term.
This difficulty is circumvented by introducing a fictitious gravitational potential $\Phi$ that couples to the energy density of electrons, but not to
their real mass.
Then, heat currents are induced by a gravitational potential gradient $\nabla\Phi$ which plays the same role as the temperature gradient $\nabla T/T$.
This potential gradient is referred to as a gravitoelectric field, which we denote as $\bm{E}_g=-\nabla\Phi$ in analogy with a usual electric field.
Second, to obtain a correct expression for the thermal Hall conductivity, one needs to extract 
contributions from energy magnetization currents, since circulating magnetization currents induced by breaking time-reversal symmetry do not participate into the transport currents generated by an applied temperature gradient.
The above two points were carefully examined by Qin, Niu and Shi, which led to an expression for the thermal Hall conductivity in terms of the Berry curvature. From their analysis, 
the thermal Hall conductivity for transport currents reads as\cite{niu,cooper,streda}
\begin{align}
\kappa^{tr}_{xy}=\kappa^{Kubo}_{xy}+\frac{2M^z_E}{TV}. \label{thc}
\end{align}
The first term is given by the usual Kubo formula for a heat current correlation function, and $M^z_E$ is the gravitomagnetic energy (heat) magnetization, which characterizes the circulation of the energy (heat) flow. $V$ is the volume of the system.
The second term arises from the extraction of the magnetization current from the total heat current.
Equation (\ref{thc}) can be expressed in terms of the Berry curvature:
\begin{align}
\kappa^{tr}_{xy}=-\frac{1}{T V}\int dE E^2\sum_{\substack{kn\\E_{kn} \leq E}} \mathrm{Im} \Braket{ \frac{\partial u_{kn}}{\partial k_x} | \frac{\partial u_{kn}}{\partial k_y }}f'(E),
\label{thermalHall}
\end{align}
where $|u_{kn}\rangle$ is the periodic part of the Bloch wave function for an electron with momentum $k$ in the $n$-th band,
and $\Braket{ \frac{\partial u_{kn}}{\partial k_x} | \frac{\partial u_{kn}}{\partial k_y }}$ is the Berry curvature in the momentum space, which arises
from a  topologically non-trivial structure of the pairing state.
$f(E)$ is the fermi distribution function.
By using the Sommerfeld expansion, we obtain the following expression in the low-temperature limit:\cite{sumiyoshi}
\begin{align}
\kappa^{tr}_{xy}=\frac{C_1(0)}{2}\frac{\pi k_{\rm B}^2T}{6\hbar},
\end{align}
where 
$C_{1}(E) \equiv \sum_{n} \int \frac{d^2 k}{\pi} \mathrm{Im} \Braket{ \frac{\partial u_{kn}}{\partial k_x} | \frac{\partial u_{kn}}{\partial k_y }}  \Theta(E-E_{kn}),$
with $\Theta(x)$ the Heaviside step function.
When the energy $E$ lies within the energy gap, $C_1(0)$ is the first Chern number $C$ (the TKKN number)
defined in Sect. \ref{subsec:PHS}.
For a single-band spinless $p+ip$ superconductor, $C_1(0)=\pm 1$, and hence, we obtain $\kappa_{xy}^{tr}=\pm \frac{\pi k_{\rm B}^2T}{12\hbar}$, which coincides with
the CFT result mentioned above up to the sign. The sign of $\kappa_{xy}^{tr}$ is determined by the details of the band structure, i.e., the Berry curvature.

The above result is directly applicable to various systems showing a chiral superconducting state.
For instance. in the case of Sr$_2$RuO$_4$,  which is one of the most promising candidate of a $p+ip$ chiral superconductor,
there are three bands with spin degeneracy, and the Fermi surfaces of two of them have positive curvature leading to positive Berry curvature,
while the other one has negative Berry curvature.
Thus, the total contributions to the thermal Hall conductivity is
$\kappa_{xy}^{tr}=\frac{\pi T}{12}(2+2-2)=\frac{\pi T}{6}\sim 10^{-4} ~T$ (W/K$\cdot$m).
Although this magnitude is quite small, rendering its experimental detection highly challenging, we believe that it is still possible
to observe this effect by precise measurements.
Recently, a detailed analysis of the gap function for Sr$_2$RuO$_4$ suggested
the existence of higher harmonics of the gap function which leads to large Chern number,
and the thermal Hall conductivity $\kappa_{xy}=\frac{7}{12}\pi T$.\cite{s-simon}

\subsection{Case of TRS topological superconductors}

For class DIII time-reversal-symmetric topological superconductors in three dimensions, 
there are Majorana surface states, which carry heat currents. 
For this class, 
the bulk topological invariant is defined as a winding number which takes any integer value, as explained in section \ref{subsec:TRS},
and the winding number $w_{\rm 3D}$ is precisely equal to the total number of topologically protected surface gapless  Majorana modes. 
This topological invariant appears in thermal transport coefficient, as in the case of chiral superconductors.
However, an important difference is that, in this case, one needs to break the time-reversal symmetry to realize the quantum thermal Hall effect
by adding external perturbations to the system.
Perturbations that break the time-reversal symmetry (more precisely chiral symmetry, as will be elucidated later),
induce mass gap of the Majorana surface state, and hence, the 2D surface state exhibits the quantum thermal Hall effect,
as in the case of 2D chiral superconductors. 
An effective theory for the surface state is described by $M$-species of 2D free Majorana fermions with mass gaps $m_i$ ($i=1, 2,..., M$).
We can apply Eq.(\ref{thermalHall}) to these Majorana fermion systems.
Since the first Chern number for a massive Majorana fermion %(with proper regularization adding $k^2$-term to the kinetic energy) 
is given by the sign of the mass ${\rm sgn}(m_i)/2$,
the thermal Hall conductivity of this system is given by,\cite{nomura}
\begin{eqnarray}
\kappa_{xy}^{tr}=\sum_i{\rm sgn}(m_i)\frac{\pi k_{\rm B}^2T}{24\hbar}=(N_{+}-N_{-})\frac{\pi k_{\rm B}^2T}{24\hbar},
\end{eqnarray}
where $N_{+(-)}$ is the number of Majorana surface modes with positive (negative) mass.
An important question is how $N_{+}-N_{-}$ is related to the bulk topological invariant $w_{\rm 3D}$.
This is by no means trivial, because the signs of mass gaps ${\rm sgn}(m_i)$ crucially depend on characters of perturbations
that generate them.
%the existence of non-topological surface states, which depends sensitively to
%energy gap structures is generally allowed, and may contribute to the above heat current.
Nevertheless, it is possible that for an appropriate perturbation, $N_{+}-N_{-}=w_{\rm 3D}$ holds.
Indeed, for class DIII topological superconductors in three dimensions,
such a perturbation is realized as a chiral-symmetry breaking field.\cite{shiozaki,wang-zhang}
The Hamiltonian of this perturbation is given by,
\begin{eqnarray}
\mathcal{H}_{\Gamma}=\frac{\gamma}{2}\int d\bm{r}\Psi^{\dagger}(\bm{r})\Gamma\Psi(\bm{r}),
\end{eqnarray}
where $\Psi(\bm{r})$ is the Nambu spinor, and $\Gamma$ is the chiral symmetry operation, which acts 
on the BdG Hamiltonian $\mathcal{H}_{\rm BdG}$ as
$\Gamma\mathcal{H}_{\rm BdG}\Gamma^{-1}=-\mathcal{H}_{\rm BdG}$; i.e. 
$\Gamma$ anticommutes with the Hamiltonian. $\gamma$ is a coupling constant.
It is obvious that since $\Gamma$ commutes with itself, $\mathcal{H}_{\Gamma}$ breaks chiral symmetry.
For class DIII topological superconductors, chiral symmetry is realized as a combination of time-reversal symmetry and particle-hole symmetry.
In the representation of the Nambu spinor, $\Psi=(\psi_{\uparrow}, \psi_{\downarrow},\psi^{\dagger}_{\downarrow},-\psi^{\dagger}_{\uparrow})^T$,
$\mathcal{H}_{\Gamma}=-i\gamma \int d{\bm{r}} (\psi_{\uparrow}\psi_{\downarrow}-\psi^{\dagger}_{\downarrow}\psi^{\dagger}_{\uparrow})$; 
 i.e., the imaginary part of a spin-singlet pairing field.
%This implies that Majorana surface states with opposite sign mass are associated with Majorana edge states which propagate in the opposite direction, leading to backward scattering due to impurities; i.e. Majorana edge states associated with opposite sign mass are not topologically protected.
%Hence, when the numbers of surface Majorana modes with positive and negative mass are, respectively, $N_{+}$ and $N_{-}$,  
%the winding number $N$, i.e. the number of topologically protected Majorana modes, is equal to $N=N_{+}-N_{-}$.
%Since $\sum_i{\rm sgn}(m_i)=N_{+}-n_{-}$, 
Hence, for a heterostructure system composed of a topological superconductor with a surface on which 
a conventional spin-singlet superconductor with a nonzero imaginary gap is placed, 
the bulk topological invariant $N$ appears in 
the quantized thermal Hall conductivity given by,
\begin{eqnarray}
\kappa_{xy}^{tr}=w_{\rm 3D}\frac{\pi k_{\rm B}^2T}{24\hbar}.
\end{eqnarray}
An important point here is that the signs of the 
mass gaps of Majorana surface states generated by the $\Gamma$ perturbation exactly coincide with those of the winding number
corresponding to the Majorana surface states, which guarantees the relation $\sum_i{\rm sgn}(m_i)=w_{\rm 3D}$.\cite{shiozaki,wang-zhang}

\begin{figure}
\includegraphics[width=9cm]{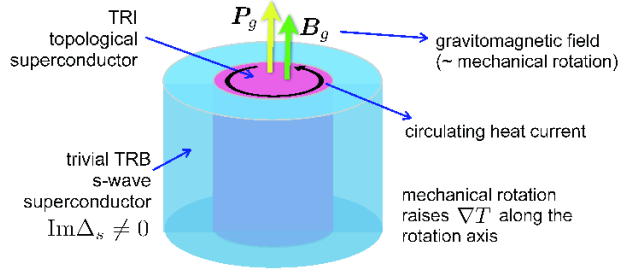}
\caption{(Color online) Setup for thermal analogue of magnetoelectric effect. A topological superconductor with cylindrical geometry
is surrounded by a spin-singlet trivial superconductor with an imaginary part of the gap function.
}
\label{fig:gra}
\end{figure}

%Thus, the windingg number appears in the coefficient of the quantized thermal Hall conductivity.
%In the above argument, perturbations which generate mass gap of surface Majorana fermions are essentially important.
Thermal responses characterized by the winding number also appear in the thermal analogue of magnetoelectric effects, 
which are referred to as gravitomagnetoelectric effects.\cite{shiozaki}
To explain such effects, let us consider a system with cylindrical geometry composed of a topological superconductor and a trivial spin-singlet superconductor with the nonzero imaginary part of the gap function,
as shown in Fig.\ref{fig:gra}.
We here introduce a gravitomagnetic field $\bm{B}_g$, which is associated with a circulating heat (or energy) current, in analogy with a magnetic field
related to a circulating charge current. We note that $\bm{B}_g$ is related to the existence of circulating particle flow, and thus,
it is expected that the rotation of the system around the symmetric axis can induce $\bm{B}_g$ parallel to the axis.
Then, $\bm{B}_g$ gives rise to the gravitomagnetoelectric effect, inducing heat polarization $\bm{P}_g$ along the axis.
Furthermore, the response function of this effect is characterized by the winding number:\cite{shiozaki}
\begin{eqnarray}
\bm{P}_g=\pm \frac{\pi k_{\rm B}^2T^2}{24\hbar v}w_{\rm 3D}\bm{B}_g,
\end{eqnarray}
where $v$ is the Fermi velocity of the surface Majorana fermions.
The heat polarization $\bm{P}_g$ can be detected as the temperature difference between the top and bottom surfaces of the system shown in Fig.\ref{fig:gra}.

%\subsection{Gravitational effective field theory}

\section{Topological Quantum Computation}
\label{sec:TQC}

It was proposed by Kitaev, Freedman and coworkers that non-Abelian anyons can be utilized 
for the construction of fault-tolerant quantum computation.\cite{TQC-rev,bravyi-kitaev,bravyi-kitaev2,bravyi,freedman2,georg,nori-QC1,nori-QC2}
As mentioned before, a topological state with $2N$ Majorana zero modes has $2^{N-1}$-fold topological degeneracy,
which cannot be lifted by local perturbations.
The braiding operations of non-Abelian anyons give rise to unitary operations in these degenerate states, which constitute
quantum gates.
In this proposal, a qubit is made from a set of several non-Abelian anyons such as Majorana zero-energy modes in topological superconductors.
Thus, the qubit is defined as a non-local object, which ensures stability against decoherence effects caused by local perturbations.

Universal quantum gates can be constructed from one-qubit operations including the Hadamard gate, the phase gate, 
and the $\pi/8$-gate and the two-qubit
controlled NOT (CNOT) gate.
Let us first consider one-qubit operations.
One qubit is constructed from four localized Majorana zero modes represented by Majorana fields $\gamma_1$, $\gamma_2$, $\gamma_3$,
and $\gamma_4$. The state is represented by $|n_{12}n_{34}\rangle$ with $n_{12}$ and $n_{34}$, respectively, the occupation numbers of the fermion operators $\psi_{12}=\frac{1}{2}(\gamma_{1}+i\gamma_{2})$  and $\psi_{34}=\frac{1}{2}(\gamma_{3}+i\gamma_{4})$.
For the even (odd) parity sector, the doubly-degenerate states are $|11\rangle$ and $|00\rangle$ ($|10\rangle$ and $|01\rangle$).
In the 2D space spanned by these states, the braiding operators (\ref{eq:uij}) are represented by using the Pauli matrices:
$U_{1,2}=(1-i\sigma_z)/\sqrt{2}$, $U_{2,3}=(1-i\sigma_x)/\sqrt{2}$, $U_{3,1}=(1-i\sigma_y)/\sqrt{2}$.  
Thus, the braidings of two Majorana fermions lead to $\pi /2$-rotations around the $x$-, $y$-, and $z$-axes.
The NOT gate is given by $(U_{2,3})^2=-i\sigma_x$, and the phase gate is,
\begin{eqnarray}
U_{1,2}=
e^{-i\frac{\pi}{4}}
\left(
\begin{array}{cc}
1 & 0 \\
0 & i
\end{array}
\right)
\end{eqnarray}
up to a phase factor.
Also, the Hadamard gate $H$ is 
\begin{eqnarray}
H\equiv (U_{2,3})^2U_{3,1}=
\frac{-i}{\sqrt{2}}
\left(
\begin{array}{cc}
1 & 1 \\
1 & -1
\end{array}
\right).
\end{eqnarray}
All one-qubit gates except the $\pi/8$-gate can be constructed from the above operations. 
%and the $\pi/8$-gate expressed by the matrix ${\rm diag}(1, e^{i\pi/4})$.
%However, the topologically protected braiding operations of Majorana fermions can not generate the $\pi/8$-gate.
%To circumvent this drawback, Bravyi and Freedman et al. proposed to use tropologically unprotected operations.\cite{bravyi,freedman}
The $\pi/8$-gate may be realized by topologically unprotected operations such as the fusion of two Majorana fermions which yields a dynamical phase
change, \cite{bravyi,freedman2}  The dynamical phase can be tuned to be equal to $\pi/8$. 
An alternative method for realizing the $\pi/8$-gate is to change the genus of the shape of the system during the braiding of Majorana modes,
though the implementation in real systems  is technically challenging.\cite{freedman2}

We now consider two-qubit operations. 
The following argument is based on the seminal papers by Bravyi and Kitaev.\cite{bravyi-kitaev,bravyi-kitaev2,bravyi} 
For brevity, we change the notation of the one qubit state as $|00\rangle$  (or $|01\rangle$)  $\rightarrow |0\rangle$ and $|11\rangle$  
(or $|10\rangle$)  $\rightarrow |1\rangle$.
The basis of the two-qubit state space is $\{ |0\rangle |0\rangle, |0\rangle |1\rangle,|1\rangle |0\rangle,|1\rangle |1\rangle\}$,
where the first qubit consists of Majorana fermions $\gamma_1$, $\gamma_2$, $\gamma_3$, and $\gamma_4$, and
the second qubit consists of $\gamma_5$, $\gamma_6$, $\gamma_7$, and $\gamma_8$.
The CNOT gate for this two-qubit space is expressed in terms of the Hadamard gate $H$ and the controlled $\sigma_z$:
\begin{eqnarray}
\Lambda(\sigma_x)&\equiv&
\left(
\begin{array}{cc}
\mathbb{I} & 0 \\
0 & \sigma_x
\end{array}
\right) \nonumber \\
&=&
\left(
\begin{array}{cc}
H & 0 \\
0 & H
\end{array}
\right)
\Lambda(\sigma_z)
\left(
\begin{array}{cc}
H^{\dagger} & 0 \\
0 & H^{\dagger}
\end{array}
\right),
\label{eq:CNOT}
\end{eqnarray}
with the controlled $\sigma_z$ gate,
\begin{eqnarray}
\Lambda(\sigma_z)\equiv \left(
\begin{array}{cc}
\mathbb{I} & 0 \\
0 & \sigma_z
\end{array}
\right),
\end{eqnarray}
where $\mathbb{I}$ is the $2\times 2$ identity matrix. 
Note that $|0\rangle$ and $|1\rangle$ for the first (second) qubit are the eigenstates of $i\gamma_1\gamma_2$ 
($i\gamma_5\gamma_6$) with the eigenvalues $-1$ and $1$, respectively.
Then, the controlled $\sigma_z$ gate is expressed as,
 \begin{eqnarray}
 &&\Lambda (\sigma_z)=\exp(i\frac{\pi}{4}(1-i\gamma_1\gamma_2)(1-i\gamma_5\gamma_6)) \nonumber \\
&&=e^{i\frac{\pi}{4}}\exp(-i\frac{\pi}{4}\gamma_1\gamma_2\gamma_5\gamma_6)U_{1,2}U_{5,6}, \label{eq:Csigmaz}
\end{eqnarray}
where $U_{1,2}$  and $U_{5,6}$ are the braiding operators for $\gamma_1$, $\gamma_2$, and $\gamma_5$, $\gamma_6$,
respectively.
Furthermore, the operation $\exp(-i\frac{\pi}{4}\gamma_1\gamma_2\gamma_5\gamma_6)$ can be implemented 
in the following way. We prepare two ancillary Majorana fermions $\gamma_9$ and $\gamma_{10}$
in addition to the two Majorana qubits. We also prepare a state satisfying $(\gamma_9+i\gamma_{10})|\psi\rangle=0$;
i.e., the vacuum state for the complex fermion operator $\psi_{9,10}=(\gamma_9+i\gamma_{10})/2$. 
Then, it follows that,
\begin{eqnarray}
&&\exp(-i\frac{\pi}{4}\gamma_1\gamma_2\gamma_5\gamma_6)|\psi\rangle=2U_{10,6}
\Pi^{(2)}_{+}\Pi^{(4)}_{+}|\psi\rangle   \nonumber \\
&&=2i(U_{2,1})^2(U_{5,6})^2
U_{10,6}
\Pi^{(2)}_{+}\Pi^{(4)}_{-}|\psi\rangle \nonumber \\
&&=2i(U_{2,1})^2(U_{5,6})^2
U_{6,10}
\Pi^{(2)}_{-}\Pi^{(4)}_{+}|\psi\rangle \nonumber \\
&&=2U_{6,10}
\Pi^{(2)}_{-}\Pi^{(4)}_{-}|\psi\rangle,
\label{eq:exp1256}
\end{eqnarray}
%\begin{eqnarray}
%&&\exp(-i\frac{\pi}{4}\gamma_1\gamma_2\gamma_5\gamma_6)|\psi\rangle=2\exp(\frac{\pi}{4}\gamma_6\gamma_{10})
%\Pi^{(2)}_{+}\Pi^{(4)}_{+}|\psi\rangle   \nonumber \\
%&&=2i\exp(\frac{\pi}{2}\gamma_1\gamma_2)\exp(\frac{\pi}{2}\gamma_6\gamma_5)
%\exp(\frac{\pi}{4}\gamma_6\gamma_{10})
%\Pi^{(2)}_{+}\Pi^{(4)}_{-}|\psi\rangle \nonumber \\
%&&=2i\exp(\frac{\pi}{2}\gamma_1\gamma_2)\exp(\frac{\pi}{2}\gamma_6\gamma_5)
%\exp(-\frac{\pi}{4}\gamma_6\gamma_{10})
%\Pi^{(2)}_{-}\Pi^{(4)}_{+}|\psi\rangle \nonumber \\
%&&=2\exp(-\frac{\pi}{4}\gamma_6\gamma_{10})
%\Pi^{(2)}_{-}\Pi^{(4)}_{-}|\psi\rangle,
%\end{eqnarray}
where $\Pi^{(2)}_{\pm}=\frac{1}{2}(1\mp i\gamma_6\gamma_9)$ and $\Pi^{(4)}_{\pm}=\frac{1}{2}(1\pm \gamma_1\gamma_2\gamma_5\gamma_9)$.
$\Pi^{(2)}_{+(-)}$ is the projection operator for the measurement of the eigenvalue of $-i\gamma_6\gamma_9$
which results in the eigenvalue $+1$ ($-1$).
$\Pi^{(4)}_{+(-)}$ is the projection to the state with $\gamma_1\gamma_2\gamma_5\gamma_9=+1$ ($-1$).
Thus, the measurement of the eigenvalues of $i\gamma_6\gamma_9$ and $\gamma_1\gamma_2\gamma_5\gamma_9$
combined with the braiding operations in Eq.(\ref{eq:exp1256}) realizes
the operation $\exp(-i\frac{\pi}{4}\gamma_1\gamma_2\gamma_5\gamma_6)$.
For instance, if the measurement of $-i\gamma_6\gamma_9$ and $\gamma_1\gamma_2\gamma_5\gamma_9$
gives the eigenvalues $-i\gamma_6\gamma_9=1$ and $\gamma_1\gamma_2\gamma_5\gamma_9=1$, respectively,
then the braiding operation $U_{10,6}$ combined with $U_{1,2}$ and $U_{5,6}$ leads to  
the controlled $\sigma_z$ gate $\Lambda(\sigma_z)$, Eq.(\ref{eq:Csigmaz}). 
This completes the realization scheme of the CNOT gate (\ref{eq:CNOT}).
There is also a  different approach for the realization of the CNOT gate, which
utilizes two-qubit states composed of six Majorana fermions.\cite{georg}
%In this alternative scheme, the first qubit consists of $\gamma_1$, $\gamma_2$, $\gamma_3$, and $\gamma_4$,
%while the second qubit consists of $\gamma_3$, $\gamma_4$, $\gamma_5$, and $\gamma_6$, sharing
%the two Majorana fermions with the first qubit. Then, the CNOT gate is implemeted by the operation ....

In the above scenarios, it is crucial for the readout scheme of the Majorana qubit to carry out the non-demolition measurement of the fermion parity. 
The direct measurement of the Majorana qubit state via the coupling with electronic probes such as STM results in decoherence of the qubit state, \cite{budich} since, as mentioned before, the coupling with external electron leads breaks the conservation of the fermion parity.
A promising approach for the non-demolition measurement of the fermion parity is 
to use the Aharonov-Casher effect.\cite{hassler}
When vortex currents flow encircling an isolated superconducting island that contains Majorana qubits, 
the interference of the vortex currents is affected by the fermion parity of Majorana qubits. 
This detection scheme does not destroy the fermion parity in the island.

%%%%amended on Jan4 2016
The braiding operation of Majorana fermions that leads to non-Abelian statistics
is the most important factor in all the above schemes.
This operation may be carried out by using the interferometer-type setup considered in Sect.4.4, and also 
by controlling the bias voltage in nanowire superconductors, which transport the position of the Majorana end state of the wire.\cite{alicea2}%, braiding1,braiding2} 
Another promising proposal is to utilize Coulomb-blockaded quantum dots coupled to Majorana end states of superconducting nanowires.\cite{flensberg-non-abelian} In this approach, a change in the occupation number
of electrons in the dot induces the flip of Majorana qubits. Although this operation is not topologically protected, and 
requires tuning of the parameters of the system, its feasibility and controllability are fascinating.
The braiding operation can also be carried out by controlling the tunneling amplitude between two Majorana fermions.\cite{braiding1,braiding2} For instance, when there are three Majorana fermions $\gamma_1$, 
$\gamma_2$, and $\gamma_3$ with the tunneling Hamiltonian, $\mathcal{H}=t_{12}\gamma_1\gamma_2+t_{23}\gamma_2\gamma_3$,
 ($t_{12}>0$, $t_{23}>0$),
the Majorana fermion $\gamma_1$ is transferred to $\gamma_3$ by changing the parameter $t_{12}/t_{23}$ 
from $0$ to $\infty$. The tuning of the tunneling amplitude can be implemented by controlling the gate voltage,\cite{braiding1} or also by controlling the charging energy via the change in the magnetic flux penetrating a Josephson junction system.\cite{braiding3,braiding4}
 %%%%%%%%%%%%%%%%% 

\section{Interaction Effects}
\label{sec:interaction}

To this date, most of the studies on topological phases have been focused on non-interacting fermion systems, 
though there have been some important research works on quantum spin systems and boson systems, 
where strong interactions play an important role for the realization of topological phases.\cite{pollmann,wen} 
Effects of electron-electron interaction on topological insulators and superconductors are not yet fully understood.
However, significant progress has recently been achieved for topological classification of
interaction systems in certain symmetry classes. 
The interaction between electrons can change the topological classification obtained for free-fermion systems.
An important example is a time-reversal symmetric 1D topological superconductor in class BDI.\cite{fidkow1,fidkow2} 
In the absence of interactions, this class allows multiple numbers of Majorana zero modes, signifying
the $Z$ nontriviality
characterized by the winding number.
Fidkowski and Kitaev showed that mutual interactions between fermions preserving time-reversal symmetry
can change the $Z$ non-triviality of the 1D BDI class to the $Z_8$ non-triviality,
gapping out eight Majorana modes.
Another important example is a class DIII superconductor in three dimensions, which is also characterized by the winding number taking any integer values.
Interactions between electrons can change the $Z$ non-triviality to the $Z_{16}$ non-triviality.\cite{fidkow3,wang,metlitski}

\subsection{1D class BDI : $Z \rightarrow Z_8$ }

The topological phase of the 1D BDI class is characterized by a winding number that takes integer values; 
i.e. it has $Z$ non-triviality. 
Thus, it is possible that there are multiple Majorana zero modes at the open edges of the system.
The essential features of this class are ingeniously described by the Kitaev Majorana chain model,  which
is a simple model for 1D spinless $p$-wave superconductor.\cite{kitaev-majorana-chain}
Let us consider an $N$-channel Kitaev chain model, the Hamiltonian of which is,
\begin{eqnarray}
\mathcal{H}&=&\sum_{\alpha=1}^N\sum_{j=1}^L[-w(a^{\dagger}_{j\alpha}a_{j+1\alpha}+h.c.)
-\mu (a^{\dagger}_{j\alpha}a_{j\alpha}-\frac{1}{2}) \nonumber \\
&&+\Delta a_{j\alpha}a_{j+1\alpha}
+\Delta a^{\dagger}_{j+1\alpha}a^{\dagger}_{j\alpha}].
\label{eq:kitaev-chain}
\end{eqnarray}
where $a_{j\alpha}$ ($a^{\dagger}_{j\alpha}$) is the annihilation (creation) operator for a spinless fermion at $j$-site and in the $\alpha$-th channel,
and $L$ is the total number of lattice sites along the chain direction.
Introducing Majorana fields $\gamma_{2j-1\alpha}$ and $\gamma_{2j\alpha}$ satisfying
\begin{eqnarray}
a_{j\alpha}=\frac{1}{2}(\gamma_{2j-1\alpha}+i\gamma_{2j\alpha}), 
\end{eqnarray}
we can rewrite the Hamiltonian in the form,
\begin{eqnarray}
\mathcal{H}&=&\frac{i}{2}\sum_{\alpha=1}^N\sum_{j-1}^{L-1}[-\mu \gamma_{2j-1\alpha}\gamma_{2j \alpha}+ (w+\Delta)\gamma_{2j\alpha}\gamma_{2j+1\alpha}
\nonumber \\
&&+(-w+\Delta)\gamma_{2j-1\alpha}\gamma_{2j+2\alpha}].
\end{eqnarray}
We can continuously deform the parameters of the non-trivial phase with $\Delta\neq 0$ 
to the case with $\Delta=w>0$ and $\mu=0$ without closing the energy gap. The topological character is maintained by this deformation. 
The deformed Hamiltonian is,
\begin{eqnarray}
\mathcal{H}_{\rm D}=iw\sum_{\alpha=1}^{N}\sum_{j-1}^{L-1}\gamma_{2j\alpha}\gamma_{2j+1\alpha}.
\end{eqnarray}
Then, the Majorana fermions $\gamma_{1\alpha}$ at the site $j=1$ and $\gamma_{2L\alpha}$ at the site $j=L$ are decoupled.
There are $N$ Majorana zero modes at each end of the chain, characterizing the $Z$ non-triviality of the 1D BDI class.
Time-reversal symmetry characterized by the invariance under the complex conjugate operation $K$ prohibits terms such as $i \gamma_{1\alpha_1}\gamma_{1\alpha_2}$, which generates a mass gap for the Majorana modes. 
However, mutual interactions between fermions change the situation drastically.
For the case of $1\leq N \leq 3$, $N$ Majorana zero modes are still protected by time-reversal symmetry.
For the case of $N=4$, there are four Majorana zero modes at one end in the absence of interactions.
We denote the four Majorana fermions at the $j=1$ site as $\gamma_1$, $\gamma_2$, $\gamma_3$, and $\gamma_4$
omitting the site index for brevity. 
Then,  the time-reversal symmetry allows the interaction term
\begin{eqnarray}
\mathcal{H}_{\rm int}=U\gamma_{1}\gamma_{2}\gamma_{3}\gamma_{4}.
\label{eq:majo-int} 
\end{eqnarray} 
In the absence of the interaction $U=0$, there is fourfold ($2\times 2=4$) degeneracy at the $j=1$ site,
expressed by the pseudo-spin operators $\sigma_{z1}=-i\gamma_{1}\gamma_{2}$ and
$\sigma_{z2}=-i\gamma_{3}\gamma_{4}$.
The interaction (\ref{eq:majo-int}) lifts this fourfold degeneracy to the double degeneracy, 
$(\sigma_{z1},\sigma_{z2})=(1,-1), (-1,1)$ or $(-1,-1), (1,1)$.
The case with $5\leq N \leq 7$ can be understood as the combination of the $N=4$ case and the $N-4$ case.
Thus, the cases of $1\leq N \leq 7$ are still non-trivial, though the topological degeneracy is changed by
the interaction effect.
In contrast, for the case with $N=8$, the topological degeneracy is completely lifted by mutual interactions, and the system becomes trivial.
To see this, we consider eight Majorana fermions at the site $j=1$, $\gamma_{\alpha}$ with $\alpha=1, 2, 3, ..., 8$.
We divide them into two groups $\alpha=1- 4$ and $\alpha=5 - 8$, and introduce the interaction term,
\begin{eqnarray}
\mathcal{H}_{\rm int}=U_1\gamma_{1}\gamma_{2}\gamma_{3}\gamma_{4}+U_2\gamma_{5}\gamma_{6}\gamma_{7}\gamma_{8},
\end{eqnarray}
which lifts the 16-fold degeneracy in the absence of interactions to the four-fold degeneracy (two sets of the doubly-degenerate state).
Let us assume $U_{1,2}<0$, and restrict the argument within a lower energy state $|0\rangle$ that satisfies
$\gamma_1\gamma_{2}\gamma_{3}\gamma_{4}|0\rangle=|0\rangle$ and
$\gamma_{5}\gamma_{6}\gamma_{7}\gamma_8|0\rangle=|0\rangle$. 
(For the case of the opposite signs of $U_{1,2}$, we can use a lower energy state satisfying
$\gamma_1\gamma_{2}\gamma_{3}\gamma_{4}|0\rangle=-|0\rangle$ and
$\gamma_{5}\gamma_{6}\gamma_{7}\gamma_8|0\rangle=-|0\rangle$)
We, now, introduce two sets of auxiliary spin operators,\cite{kitaev-model}
\begin{eqnarray}
\sigma_{x}^{(1)}=i\gamma_1\gamma_4, \quad \sigma_{y}^{(1)}=i\gamma_2\gamma_4,
\quad \sigma_{z}^{(1)}=i\gamma_3\gamma_4,
\end{eqnarray}
and
\begin{eqnarray}
\sigma_{x}^{(2)}=i\gamma_5\gamma_8, \quad \sigma_{y}^{(2)}=i\gamma_6\gamma_8,
\quad \sigma_{z}^{(2)}=i\gamma_7\gamma_8.
\end{eqnarray}
We can easily see that $\sigma_{\mu}^{(1)}$ and $\sigma_{\mu}^{(2)}$ satisfy 
the commutation relations $\sigma^{(1)}_{\mu}\sigma^{(1)}_{\nu}=i\epsilon_{\mu\nu\lambda}\sigma_{\lambda}^{(1)}$
and $\sigma^{(2)}_{\mu}\sigma^{(2)}_{\nu}=i\epsilon_{\mu\nu\lambda}\sigma_{\lambda}^{(2)}$ ($\mu,\nu,\lambda=x,y,z$)
within the lower energy state $|0\rangle$.
The eigenstates of $\sigma_z^{(1)}$ and $\sigma_z^{(2)}$ represent the four-fold degeneracy of
$|0\rangle$.
Thus, introducing the "exchange interaction" between these spins,
\begin{eqnarray}
\mathcal{H}_{\rm ex}&=&J\bm{\sigma}^{(1)}\cdot\bm{\sigma}^{(2)} \nonumber \\
&=&-J(\gamma_1\gamma_4\gamma_5\gamma_8+\gamma_2\gamma_4\gamma_6\gamma_8+\gamma_3\gamma_4\gamma_7\gamma_8),
\end{eqnarray}
we can lift the fourfold degeneracy.
For $J>0$, the lowest-energy state is "spin-singlet", i.e. a trivial state with no topological degeneracy.
This implies that the $Z$ non-triviality for the non-interacting 1D BDI class is reduced to the $Z_8$ non-triviality
for interacting systems.
There are only seven different topological phases. 
The above argument is rather heuristic. More sophisticated argument is presented in Refs.\citen{fidkow1,fidkow2}

\subsection{3D class DIII : $Z\rightarrow Z_{16}$ }

Interaction effects on the 3D topological superconductors in class DIII have been studied by Fidkowski et al., Wang and Senthil, and Metlitski et al.\cite{fidkow3, wang, metlitski}
They showed that an electron-electron interaction changes the $Z$ non-triviality for non-interacting systems to the $Z_{16}$ non-triviality, and thus, 16 Majorana surface states are gapped out by the interaction effect without breaking the symmetry of the systems, which leads to a trivial phase.
To see this, let us consider 16 surface Majorana states on the (001)-surface of a 3D topological superconductor. 
We assume that energies of all the Majorana fermions cross zero at $\bm{k}=0$ for simplicity. 
Then, in the absence of the mutual interaction between electrons, the Hamiltonian is
\begin{eqnarray}
\mathcal{H}=\int d{\bm{r}}\sum_{i=1}^{16}\chi_i^{T}(\bm{r})(\hat{k}_x\sigma_z+\hat{k}_y\sigma_x)\chi_i(\bm{r}),
\end{eqnarray}
where $\hat{k}_{\mu}=-i\partial_{\mu}$, $\bm{r}=(x,y)$, and $\chi_i^T=(\gamma_{i\uparrow},\gamma_{i\downarrow})$. 
$\gamma_{i\sigma}$ is the $i$-th Majorana field with spin $\sigma$. 
Using the complex fermion fields $\psi_i(\bm{r})=\chi_{2i-1}(\bm{r})+i\chi_{2i}(\bm{r})$,
we rewrite the Hamiltonian into that of Dirac fermions,
\begin{eqnarray}  
\mathcal{H}=\int d\bm{r}\sum_{i=1}^8\psi^{\dagger}_i(\bm{r})(\hat{k}_x\sigma_z+\hat{k}_y\sigma_x)\psi_i(\bm{r}).
\end{eqnarray}
However, the Dirac fermion field $\psi_i$ obeys an unusual time-reversal symmetry operation.
Since, under time-reversal $\mathcal{T}$,
\begin{eqnarray}
\chi_{i\uparrow}\rightarrow \chi_{i\downarrow}, \qquad
\chi_{i\downarrow}\rightarrow -\chi_{i\uparrow},
\end{eqnarray}
the $\psi_i$ field transforms as,
\begin{eqnarray}
&&\psi_{i\uparrow}\rightarrow \psi^{\dagger}_{i\downarrow}, \qquad
\psi_{i\downarrow}\rightarrow -\psi^{\dagger}_{i\uparrow}, \nonumber \\
&&\psi_{i\uparrow}^{\dagger}\rightarrow \psi_{i\downarrow}, \qquad
\psi_{i\downarrow}^{\dagger}\rightarrow -\psi_{i\uparrow}.
\label{eq:psi-tr}
\end{eqnarray}
We now add a spin-singlet $s$-wave pairing interaction term for the Dirac fermion to the free-fermion Hamiltonian,
\begin{eqnarray}
\mathcal{H}_{\rm pair}=\sum_{i=1}^8\int d\bm{r}\Delta(\bm{r})\psi^{\dagger}_{i\uparrow}(\bm{r})\psi^{\dagger}_{i\downarrow}(\bm{r})+h.c.
\end{eqnarray}
%It is noted that this pairing term does not imply the true BCS pairing of electrons on the surface, but 
This pairing term breaks time-reversal symmetry as well as $U(1)$ symmetry, because, as seen from (\ref{eq:psi-tr}), under $\mathcal{T}$,
\begin{eqnarray}
\Delta \rightarrow -\Delta^{*}.
\end{eqnarray}
Thus, the surface Majorana fermions protected by time-reversal symmetry are gapped out by the pairing term.
To recover time-reversal symmetry, we introduce many vortices of the phase of $\Delta$, which are not in the Abrikosov lattice state, but fluctuate; i.e. the spatial correlation of the phase decays exponentially. 
We consider the situation that vortices exist only for the surface pairing state, and not in the bulk of the 3D topological superconductor.
It is known that there is no thermodynamics phase transition between the normal state and a vortex liquid state.
Thus, in this surface vortex liquid state, time-reversal symmetry as well as $U(1)$ symmetry is preserved, though
the surface Dirac fermion $\psi_{i}$ acquires an energy gap.
However,  we need to be careful regarding vortex core states, because there is a Majorana zero mode 
in a vortex core of the superconducting state of Dirac fermions.\cite{jackiw-rossi}
For our system with eight Dirac fermions, there are eight Majorana zero modes in a single vortex core.
This situation is similar to the case discussed in the previous subsection.
We can apply the same argument as that for the 1D BDI model to the eight Majorana zero modes in the vortex core,
and we find that the eight Majorana fermions are gapped out by interaction effects, and are not topologically protected.
Thus, there is no gapless mode in vortex cores in the surface vortex liquid state.
The surface state with 16 Majorana cones is fully gapped by interaction effects, preserving symmetry of the system; i.e. 
the bulk and the surface states are topologically equivalent to a trivial superconducting state.
This implies that the $Z$ non-triviality of the 3D class DIII superconductors for non-interacting systems
is reduced to the $Z_{16}$ non-triviality for interacting systems.

\section{Weyl Superconductors}
\label{sec:weyl}

\subsection{Basic features}

In this section, we briefly review a novel topological superconducting state referred to as a Weyl superconductor, which 
is different from, but closely related to topological superconductors. 
Weyl superconductors are characterized by nodal excitations from point-nodes of the superconducting gap, where the amplitude of the gap function is
zero.\cite{volovik,balents1} 
To illustrate the basic idea, we use a simple toy model of a spinless chiral $p+ip$ wave superconductor in three dimensions.
The Hamiltonian is,
\begin{eqnarray}
\mathcal{H}=
\left(
\begin{array}{cc}
\frac{\bm{k}^2-k_F^2}{2m} & \Delta_0 (k_x-ik_y) \\
\Delta_0 (k_x+ik_y) & -\frac{\bm{k}^2-k_F^2}{2m}
\end{array}
\right),
\label{eq:weyl0}
\end{eqnarray}
with $k_F$ the Fermi wavenumber, and $\Delta_0 >0$.
There are two point nodes of the superconducting gap at $(0,0,k_F)$ and $(0,0,-k_F)$.
The Bogoliubov quasiparticle excitations from the point nodes have $\bm{k}$-linear energy dispersion which mimics 
massless Dirac fermions in three dimensions. 
A low-energy effective Hamiltonian $\mathcal{H}_{+}(\bm{k})$ for the nodal excitation from the point node at $(0,0,k_F)$ can be written in the form
\begin{eqnarray}
\mathcal{H}_{+}(\bm{k})=v_xk_x\tau_x+v_yk_y\tau_y+v_z(k_z-k_{F})\tau_z, 
\label{eq:weyl1}
\end{eqnarray}
where $\bm{\tau}=(\tau_x,\tau_y,\tau_z)$  are the Pauli matrices for the particle-hole space, $v_{x,y,z}$ are the velocities in the $x,y$, and $z$ directions. %and $\bm{k}_p=(k_{px},k_{py},k_{pz})$ is the Fermi wave-number vector for the position of the point node on the Fermi surface.  
It noted that $v_{x,y,z}>0$.
Thus, we can adiabatically change the values of $v_{x,y,z}$ to satisfy $v_{x}=v_y=v_z=1$ without changing the topological features as explained below. 
Then, Eq.(\ref{eq:weyl1}) is recast into 
\begin{equation}
\mathcal{H}_{+}(\bm{k})=\bm{\tau}\cdot(\bm{k}-\bm{k}_p),
\label{eq:h-weyl1}
\end{equation}
with $\bm{k}_p=(0,0,k_F)$.
This form represents the fictitious ``Zeeman interaction" between the pseudospin $\bm{\tau}$ and ``a magnetic field" $\bm{k}-\bm{k}_p$.
``The magnetic field" in the momentum space $\bm{k}-\bm{k}_p$ 
has a hedgehog-type configuration originating from the point  $\bm{k}=\bm{k}_p$.
Thus, we can consider that 
there is a monopole at $\bm{k}=\bm{k}_p$ with the monopole charge $q_m=+1$.
It is noted that the adiabatic change of the parameters $v_{x,y,z}$ mentioned above, which does not change the signs of $v_{x,y,z}$,
preserves the monopole charge $q_m$. 
%As explained below, the existence of the monopole charge at the nodal point characterizes the Weyl superconducting state.
On the other hand, the nodal excitations from the point $(0,0,-k_F)$ 
is expressed by the Hamiltonian, $\mathcal{H}_{-}(\bm{k})=v_xk_x\tau_x+v_yk_y\tau_y-v_z(k_z+k_{F})\tau_z$.
We can change the sign of $v_x$ and $v_y$ by the gauge transformation $\Delta \rightarrow -\Delta$.
Also, we can adiabatically change the values of $v_{x,y,z}$ without changing their signs to satisfy $|v_x|=|v_y|=|v_z|=1$.
Then, the effective Hamiltonian for the point node at $(0,0,-k_F)$ is,
\begin{equation}
\mathcal{H}_{-}(\bm{k})=-\bm{\tau}\cdot(\bm{k}+\bm{k}_p).
\label{eq:h-weyl2}
\end{equation}
In this form, the direction of ``the magnetic field" in the momentum space is reversed compared with Eq.(\ref{eq:h-weyl1}), and then,
the monopole charge at $\bm{k}=-\bm{k}_p$ is $q_m=-1$.
We  stress again that the monopole charge, which is a topological quantity, is not changed as long as the signs of the velocity
$v_{x,y,z}$ are not changed.
If these point nodes are not degenerated, and the point nodes have the definite monopole charge, we refer to the nodal excitations from the point nodes
as Weyl fermions. A Weyl fermion is a gapless Dirac fermion with a definite chirality. The monopole charge mentioned above 
corresponds to the chirality of the Weyl fermion.
Weyl superconductors are characterized by the existence of Weyl fermions as nodal excitations, which carry monopole charges
in the momentum space.
This state is a superconducting analogue of the Weyl semimetal state, for which 
Weyl fermons are realized by closing the energy gap of a topological insulator via time-reversal-symmetry breaking 
or inversion-symmetry breaking fields.\cite{balents2,balents3,yang}

Generally, Weyl fermions with opposite chirality (i.e. opposite sign of the monopole charge) should appear in pair in Weyl superconductors,
because, in the periodic BZ, the "magnetic flux" generated from the monopole with $q_m=+1$ must be absorbed
into the anti-monopole with $q_m=-1$.\cite{nielsen-ninomiya}

\begin{figure}
\includegraphics[width=8.5cm]{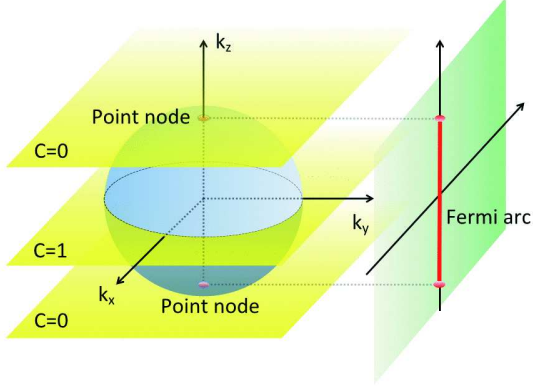}
\caption{(Color online) 
The point nodes at the north and south poles on the spherical Fermi surface are Weyl points, which possess
the monopole charges $q_m=+1$ and $-1$, respectively. 
A Majorana Fermi arc appears in the surface BZ that is parallel to $k_z$-axis. 
The Fermi arc is terminated at
two points that are the projection of the bulk Weyl points onto the surface BZ.
In the bulk, the first Chern number on the $k_x$-$k_y$ plane for fixed $k_z$ satisfying $-k_F < k_z <k_F$ is nonzero, $C=1$.
On the other hand, that for fixed $k_z$ with $|k_z|>k_F$ is zero, $C=0$.
}
\label{fig:weyl}
\end{figure}

The nonzero monopole charge associated with the Weyl point leads to a remarkable consequence, the existence of a Fermi arc on 
the surface of a sample.
%To be concrete, we consider the case that there are two Weyl points with opposite chirality.
%The effective Hamiltonian $H_s$ for the Weyl fermons with the chirality $s=\pm 1$ is given by,
 %\begin{eqnarray}
 %\mathcal{H}_s(\bm{k})=s\bm{\tau}\cdot\bm{k}-\bm{\tau}\cdot\bm{k}_p.
 %\end{eqnarray}
%Here, the Weyl fermion with the chirality $+1$( $-1$ ) is located on $\bm{k}\sim \bm{k}_p$ ( $-\bm{k}_p$ ).
To explain this phenomenon, we note again that
the  "magnetic flux" generated from the monopole at $\bm{k}=\bm{k}_p$ is absorbed into the anti-monopole at $\bm{k}=-\bm{k}_p$.
This implies that there is a "magnetic field" oriented in the direction $-\bm{k}_p$ in the momentum space. 
The "magnetic field" in the momentum space is the Berry curvature, and its integral over the 2D BZ on
a plane perpendicular to  $\bm{k}_p $ is the first Chern number (TKNN number). 
The nonzero Chern number in the 2D momentum space results in
the existence of gapless chiral edge states on the surface of the system. Thus, there is a gapless chiral excitation on the surface of
the sample parallel to $\bm{k}_p $  (see Fig. \ref{fig:weyl}). This chiral zero-energy state exists on a line segment connecting two $\bm{k}$-points on
the surface BZ, which are the projections of the points $\pm \bm{k}_p$ onto the surface BZ.
The line segment of the surface zero-energy state is called a Fermi arc.
This chiral state consists of Bogoliubov quasiparticles, and hence the zero-energy states are Majorana fermions.
Since the chiral state is in analogy with the chiral edge state of the quantum Hall effect state and 2D chiral superconductors, 
it gives rise to distinct transport phenomena such as the thermal Hall effect without an applied magnetic field,
as discussed in section \ref{sec:thermal}. 
The anomalous thermal Hall conductivity for the above model of a Weyl superconductor, (\ref{eq:weyl0}), is given by the 3D version of Eq.(\ref{thermalHall});
the sum $\sum_{k}$ is taken over the 3D BZ.
In the low temperature limit $k_{\rm B}T \ll \Delta$, it is given by,
\begin{eqnarray}
\kappa_{xy}^{tr}=\sum_{k_z}C_1(k_z)\frac{\pi k_{\rm B}^2T}{12\hbar},
\end{eqnarray}
where $C_1(k_z)$ is the first Chern number defined on the $k_x$-$k_y$ plane for a fixed $k_z$.
As mentioned in section \ref{sec:thermal}, its magnitude is very small $\kappa_{xy}^{tr}\sim 10^{-4} T$ (W/Km).

\subsection{Particle-hole symmetry}

Weyl superconductors are different from Weyl semimetals in that they possess particle-hole symmetry. Here, we will discuss
how the particle-hole symmetry imposes constraints on Weyl fermions.
Let us suppose that a Weyl fermion is realized as a nodal excitation of a spin-triplet superconductor.
The particle-hole transformation operator for spin-triplet pairing states is $\mathcal{C}=\tau_xK$ with $K$ the complex conjugate operation.
Then, we see that,
\begin{eqnarray}
-\mathcal{C}\mathcal{H}_{+}(-\bm{k})\mathcal{C}^{-1}=\mathcal{C}\bm{\tau}\mathcal{C}^{-1}\cdot (\bm{k}+\bm{k}_p).
\end{eqnarray}
Note that $\mathcal{C}\bm{\tau}\mathcal{C}^{-1}\cdot \bm{k}=-\bm{\tau}\cdot\bm{k}$ up to the
gauge transformation changing the phase of the superconducting gap by $\pi$, which flips the sign of the off-diagonal component of 
$\bm{\tau}\cdot\bm{k}$.
Thus, the particle-hole tranformation changes $\mathcal{H}_{+}(\bm{k})$ to $\mathcal{H}_{-}(\bm{k})$; i.e.
the particle-hole symmetry partners of the monopole in Weyl superconductors is the anti-monopole.

In the case of spin-singlet superconductors, the particle-hole transformation operator is changed to $i\tau_y K$.
However, the above result also holds for the spin-singlet pairing case.

\subsection{Breaking time-reversal symmetry is necessary}

We now consider the role of time-reversal symmetry for the realization of Weyl superconductors.
Applying the time-reversal symmetry operation $\mathcal{T}=i\sigma_y K$ to $\mathcal{H}_{+}(\bm{k})$, which possesses
the monopole charge $qm=+1$ at $\bm{k}=\bm{k}_p$, we obtain,
\begin{eqnarray}
\mathcal{T}\mathcal{H}_{+}(-\bm{k})\mathcal{T}^{-1}=-\bm{\tau}^{*}\cdot (\bm{k}+\bm{k}_p).
\end{eqnarray}
Note that the right-hand side has the monopole charge $q_m=+1$ at $\bm{k}=-\bm{k}_p$,  because 
the complex conjugate operation of $\bm{\tau}$
change the sign of the monopole charge. 
Also, as shown above, the particle-hole symmetry ensures the existence of Weyl fermions with the monopole charge $qm=-1$ at $\bm{k}=-\bm{k}_p$,
which is expressed by $\mathcal{H}_{-}(\bm{k})$.
Thus, there are two Weyl fermions with opposite chiralities at $\bm{k}=-\bm{k}_p$; i.e.
they form a Dirac fermion represented by a four-component spinor.
Similarly, the time-reversal symmetry operation of $\mathcal{H}_{-}(\bm{k})$
leads to a Weyl fermion with the monopole charge $q_m=-1$ at $\bm{k}=\bm{k}_p$.
Hence,
the nodal excitations from the point nodes are not Weyl but Dirac fermions.
It is necessary to break time-reversal symmetry for the realization of Weyl superconductors.

It is noted that in the case with time-reversal symmetry, there is a helical Fermi arc on the surface of the system, which consists of
two chiral Fermi arc with opposite chiralities.  
This state may be referred to as a Dirac superconductivity.\cite{fang}

\subsection{Candidate materials}

As already discussed in section \ref{sec:materials}, there are several candidate systems for Weyl superconductors. 

{\it (i) A-phase of $^3$He:}  The model Hamiltonian (\ref{eq:weyl0}) is realized in the A-phase of the superfluid Helium 3 up to the spin degeneracy; i.e., $\Delta_{\uparrow\uparrow}(\bm{k})=\Delta_{\downarrow\downarrow}(\bm{k})=\Delta_0 (k_x+ik_y)$.
 The spin degeneracy is trivial in this case, and the monopole charge at each Weyl points is simply doubled.
 The chiral anomaly associated with Weyl fermions in the A-phase of $^3$He was verified by  the measurement of forces acting on vortices.\cite{bevan,volovik}

{\it (ii) URu$_2$Si$_2$: } As discussed in Sect. \ref{sec:materials}, the heavy fermion superconductor  URu$_2$Si$_2$ has the gap
function with chiral $d_{zx}+id_{yz}$ wave symmetry;  $\Delta (\bm{k})=\Delta_0 k_z(k_x+ik_y)$.
It has point nodes at the north and south poles on the Fermi surface. 
On the surface of the system parallel to the $z$-axis, a Majorana arc appears, which connects the projected points
of the two Weyl points at the north and south poles on the Fermi surface.
Although there is a horizontal line node at $k_z=0$, it does not affect the Majorana arc except just at $k_z=0$.
Since this is a spin-singlet pairing state, there is spin degeneracy, which gives a trivial $2$ factor to the monopole charge.
Up to the spin degeneracy, the monopole charges at the nodal points are $q_m=\pm 1$.\cite{URu2Si2-2,gos}

{\it (iii) B-phase of UPt$_3$: } As mentioned in Sect. \ref{sec:materials}, there is controversy regarding the pairing state of
the B-phase of the heavy fermion superconductor UPt$_3$. There are three candidate pairing states; the $E_{1u}$ planar state, $E_{1u}$ chiral state, and $E_{2u}$ chiral state. Among them, the $E_{1u}$ and $E_{2u}$ chiral states realize Weyl superconducting states.
The $E_{1u}$ chiral state is characterized by the $\bm{d}$-vector, $\bm{d}(\bm{k})\propto \bm{c}(k_a+ik_b) (5k_c-k^2)$, 
the chiral part of which is the same as the chiral $p$-wave pairing symmetry.
On the other hand,  the $\bm{d}$-vector of the $E_{2u}$ chiral state is $\bm{d}(\bm{k})\propto \bm{c}(k_a+ik_b)^2k_c$,
the chiral part of which is equivalent to the chiral $d$-wave pairing symmetry.
There is also trivial spin degeneracy in both cases.
Up to the spin degeneracy, the point nodes of the $E_{1u}$ ( $E_{2u}$ ) chiral state have the monopole charges $q_m=\pm 1$ ($2$).
In the case of the $E_{2u}$ chiral state, there are two Majorana arcs on the surface of the system parallel to the $c$-axis.\cite{goswami}

{\it (iv) UCoGe: } In the heavy fermion system UCoGe, spin-triplet superconductivity coexists with ferromagnetism below $T_c\sim 0.6 $ K.\cite{UCoGe1,UCoGe2,UCoGe3,UCoGe4}
According to a group-theoretical argument, there are two possible spin-triplet pairing states in the ferromagnetic phase of UCoGe: 
(a) the $A$-phase with the $\bm{d}$-vector $\bm{d}\sim (a_1k_a+ia_2k_b, a_3k_b+ia_4 k_a,0)$, which possesses
point nodes of the superconducting gap at $k_a=k_b=0$, (b) the $B$-phase with the $\bm{d}$-vector $\bm{d}\sim (b_1k_c+ib_2k_ak_bk_c, ib_3k_c+b_4 k_ak_bk_c,0)$ which possesses a line node of the gap.\cite{mineev}
The NMR measurement carried out by Hattori et al. revealed that the pairing glue of this system is the ferromagnetic spin fluctuation.\cite{hattori}
A microscopic calculation of the upper critical field for the $A$-phase successfully explains the unusual behaviors of the upper critical fields
of this system, strongly supporting the realization of the $A$-phase in UCoGe.\cite{hattori}
The $A$-phase of this system is indeed the Weyl superconducting state, in which time-reversal symmetry is broken by the $\bm{d}$-vector; i.e., 
the nonunitary pairing state. The spin degeneracy is completely lifted by the exchange splitting of the Fermi surface.
Thus, the point-nodal quasiparticle excitations of the $A$-phase behave as genuine Weyl fermions.

{\it (v) SrPtAs :} $\mu$SR measurements suggest that time-reversal symmetry is broken in the superconducting phase of SrPtAs. \cite{biswas} 
According to microscopic model calculations, a chiral $d+id$ pairing state is a promising candidate of this superconductor state. \cite{Thomale1} The superconducting gap function possesses point-nodes characterizing the Weyl superconducting state.

\subsection{Superconducting doped Weyl semimetals}
 
An interesting realization of a Weyl superconductor is a superconducting
doped Weyl semimetal.\cite{cho,bolu,yi} 
Upon slight doping, Weyl semimetals have disconnected Fermi
surfaces, each of which surrounds one of the band-touching
Weyl points.
For time-reversal breaking Weyl semimetals,
a uniform superconducting state on the Fermi surface is found to support
bulk gap point nodes even for a constant $s$-wave pairing state.\cite{cho}
The existence of point nodes in an $s$-wave state is due to the nonzero
Chern number of Weyl points in the normal state.\cite{bolu,yi,murakami}
For time-reversal symmetry breaking Weyl semimetals, 
a surface Fermi arc does not have a partner arc with opposite surface
momentum; thus, it cannot participate in a Cooper pair even in an $s$-wave
superconducting state. As a result, the Fermi arc remains gapless in the
superconducting state; thus there must be point nodes at which the Fermi
arc terminates.\cite{bolu}  
The Fermi arc in the normal state make it possible to support an exotic
surface state with symmetry-protected crossed flat bands in the
superconducting state.\cite{bolu} 

\section{Concluding Remarks}
 
We make some remarks on recent developments that are not discussed in the main text.      
The experimental exploration for Majorana fermions in topological superconductors is still an important open issue.
In particular, the experimental realization of non-Abelian statistics, which will open the door to future applications to topological quantum computation, is one of the most crucial and challenging subjects.
Toward this goal, we still need to develop our theoretical understanding of the distinct phenomena
 arising from Majorana fermions in topological superconductors, which can be utilized for experimental detection.
From this perspective, it is important to investigate nonadiabatic
dynamics of Majorana fermions and related decoherence phenomena of
Majorana qubits. Recently, there have been some advances in this
direction.\cite{majorana-dynamics1,majorana-dynamics2,majorana-dynamics3,majorana-dynamics4,majorana-dynamics5,amorim,
majorana-dynamics6,martin1,martin2,martin3,martin4}

The effects of superconducting fluctuations on Majorana fermion states have also been studied recently.\cite{fid-fisher,halperin,altman,kuei1}
It was revealed that the topological degeneracy associated with Majorana fermions is maintained even for the case of quasi-long-range order 
with power-law decay of the superconducting correlation, for which the total electron charge is conserved, and the phase of the superconducting gap fluctuates.

Another important issue is the investigation of interaction effects discussed in Sect. \ref{sec:interaction}.
In particular, it is possible that novel topological phases emerge on the surface of topological superconductors because of interaction effects.\cite{fidkow3,wang, metlitski}
Exploration for such novel states arising from interaction effects, which cannot be realized
in non-interacting systems, is an important future issue.

In Sect. \ref{sec:thermal}, we discussed thermal responses of topological superconductors.
This subject is related to the issue of effective field theories that describe topological responses of superconductors.
 It is known that 
for 3D topological insulators, topological field theory is the Axion electromagnetism action.\cite{Qi-Hughes-Zhang}
It was proposed that the thermal responses of topological superconductors are described by gravitational field theory analogous to the Axion action.\cite{ryu-ludwig,wang-zhang}
However, it has turned out that the bulk gravitational field theory cannot describe the thermal Hall effect of topological superconductors
correctly, and that thermal responses characterizing topological features are due to the existence of gapless edge states.
This implies that a bulk-edge correspondence is not apparent in thermal responses\cite{stone,read-grav}
Also, it was recently clarified that thermal responses are successfully described by gravitational theory with torsion,
which is not included in the conventional theory of gravity.\cite{shitade,abanov}

Majorana surface states appear at various interfaces between a topological superconductor and a topologically trivial system.
For an interface between a topological superconductor and a ferromagnet, odd-frequency Cooper pairs appear in addition to Majorana zero energy states generically. The relation between these surface states was discussed by several authors.\cite{asano,ebisu}

\begin{acknowledgment}

%\acknowledgment
The authors thank Y. Ando, L. Fu, N. Kawakami, Y. Maeno, Y. Matsuda, T. Mizushima, T. Shibauchi, K. Shiozaki, H. Sumiyoshi, Y. Tada, R. Takashima, and Y. Tanaka for invaluable discussions.
This work was supported by Grants-in-Aids for Scientific
Research from MEXT of Japan [Grants No. 25287085, No. 23540406, No. 25220711, No.22103005, No. 25103714 (KAKENHI on Innovative Areas "Topological Quantum Phenomena''), No. 15H05852, No. 15H05855 (KAKENHI on Innovative Areas "Topological Materials Science'')]. 

\end{acknowledgment}

%\appendix
%\section{}

%Use the \verb|\appendix| command if you need an appendix(es). The \verb|\section| command should follow even though there is no title for the appendix (see above in the source of this file).

%For authors of Invited Review Papers, the \verb|profile| command is prepared for the author(s)' profile.  A simple example is shown below.

%\begin{verbatim}
%\profile{Taro Butsuri}{was born in Tokyo, Japan in 1965. ...}
%\end{verbatim}


\begin{thebibliography}{9}
\bibitem{com1} The ground state with several zero-energy Majorana fermions depends on whether the exchange processes of two Majorana fermions are carried out by moving them in a clockwise direction or anticlockwise direction. Thus, the exchange operation of Majorana fermions is a braiding operation.
\end{thebibliography}
\end{document}